**Study of Selectivity and Permeation in Voltage-Gated Ion Channels**

By

Janhavi Giri, Ph.D.
Visiting Research Faculty
Division of Molecular Biophysics and Physiology
Rush University Medical Center
Chicago, IL, USA
Email: janhavigiri@yahoo.com

# TABLE OF CONTENTS

**TABLE OF CONTENTS (CONTINUED)**

# LIST OF TABLES

# LIST OF FIGURES

# LIST OF FIGURES (CONTINUED)

**LIST OF FIGURES (CONTINUED)**

# LIST OF FIGURES (CONTINUED)

**LIST OF FIGURES (CONTINUED)**

**LIST OF FIGURES (CONTINUED)**

## LIST OF FIGURES (CONTINUED)

## LIST OF FIGURES (CONTINUED)

**LIST OF FIGURES (CONTINUED)**

## LIST OF FIGURES (CONTINUED)

**LIST OF FIGURES (CONTINUED)**

# SUMMARY

Ion channels are proteins with holes down their middle that regulate access to biological cells and are responsible for an enormous range of biological functions. A substantial fraction of the drugs used in clinical medicine are targeted directly or indirectly on channels. Ion selectivity and permeation are based on simple laws of physics and chemistry. Ion channels are therefore ideal candidates for physical investigation.

A reduced model generates the selectivity of voltage-gated L-type calcium channel (crystal structure not known) under a wide range of ionic conditions using only two parameters with unchanging values. The reasons behind the success of this reduced model are investigated since chemical intuition suggests that more detailed models are needed. Monte Carlo simulations are performed investigating the role of flexibility of the side chains in the selectivity of calcium channels under a wide range of ionic conditions. Results suggest that the exact location and mobility of oxygen ions have little effect on the selectivity behavior of calcium channels. The first order determinant of selectivity in calcium channels is the density of charge in the selectivity filter. Flexibility seems a second order determinant.

## SUMMARY (CONTINUED)

Single channel planar lipid bilayer experiments are performed to determine the selectivity, permeation, and sialic acid specificity of a bacterial outer-membrane channel NanC of *Escherichia Coli* with a known crystal structure. Measurements show that NanC exhibits a large unit conductance varying with the applied voltage polarity, anion over cation selectivity, and voltage-dependent gating. Unitary conductance of NanC decreases significantly in presence of the buffer HEPES. Alternate buffers are tested. Results suggest that the sialic acid specificity of NanC should be performed in low concentration salt solutions (250 mM) without pH buffers adjusted to neutral pH 7.0. Neu5Ac (a major sialic acid) is observed to change the gating and considerably increase the ionic conductance of NanC. The effect of Neu5Ac on the unitary current through NanC saturates at higher Neu5Ac concentrations. Interestingly, Neu5Ac reduces the ionic conductance of OmpF: frequent, long closures are seen. Therefore, in *E.coli* sialic acid translocation is specifically facilitated by NanC, and not by the general porin OmpF.

# CHAPTER 1. INTRODUCTION

## 1.1. <u>Background</u>

Cells are the functional and structural unit of all living organisms. The interior of a cell is isolated from the external environment by the plasma membrane. The cell membrane serves variety of functions; these include homeostasis, transport, communication and excitability. The cell membrane is selectively permeable, permitting the free passage of some materials and restricting the passage of others, thus regulating the passage of materials into and out of the cell. In the membrane there exist specialized pathways which allow the transport of ions (for example, $Na^+$, $K^+$, $Cl^-$ and $Ca^{2+}$) into the cell. These are water filled pores known as ion channels (1, 2). Ion channels selectively conduct ions across an otherwise impermeable cell membrane. The transport of ions through these proteins is facilitated by the ionic concentration gradient maintained by the pumps and other ATPases also present in the cell membrane.

Ion channels are critical for normal physiological functions such as nerve and muscle excitation, hormonal secretion, learning and memory, cell proliferation, sensory transduction, the control of water and salt balance and the regulation of blood pressure (3, 4). Any defect in their normal function can lead to medical conditions categorized as "channelopathies" (5). For example, mutations in the gene encoding the α-subunit of the human skeletal muscle $Na^+$ channel gives rise to a group of diseases known collectively as the periodic paralyses, while mutations in the cardiac muscle $Na^+$ channel gene result in long QT syndrome. Cystic fibrosis results from the complete absence of or dramatic reduction in, the epithelial $Cl^-$ conductance, as a consequence of mutations in CFTR (cystic fibrosis transmembrane conductance regulator). Thus, defects in ion channel functions have profound physiological effects. In order to understand how these defects cause diseases and find an appropriate cure, we need to investigate in detail the biophysical properties of ion channels.

The three main biophysical properties defining function of ion channels are selectivity, permeation and gating. Selectivity determines the ion specificity of the ion channel, permeation defines the conductance or the current (I)–voltage (V) relation and gating describes the opening and closing of the ion channels.

In order to describe these properties, some knowledge of the structure of the ion channel is required. These integral membrane proteins are extremely small in size (pore diameter ~ 7-10 Å) and are often difficult to crystallize because these proteins require detergent or other means to solubilize them in isolation and such detergents interfere with the crystallization process. The crystal structures for the $Ca^{2+}$ and $Na^+$ channels are still not known.

However, there are alternate approaches which are adopted to understand the structure-function characteristics. In order to explain selectivity of the channel, the amino acids that form the selectivity filter (from the primary sequences) are identified by performing site-directed mutagenesis. This approach has been successful in determining the amino acids that form selectivity filter of the voltage-gated L-type $Ca^{2+}$ (6-9) channel found in cardiac muscle cells and the DEKA $Na^+$ channel (10, 11) found in nerve cells.

The functional properties of ion channels namely selectivity and permeation must be based on basic principles of physics and chemistry. Selectivity is a phenomenon that describes the chemical interaction of an ion at a particular reactive site of the protein. Permeation on the other hand describes the current-voltage relation through the channel of interest. Selectivity and permeation are often intertwined while performing experimental measurements of these phenomena. Gating is a property of an ion channel to open or close and therefore, can be separated from selectivity and permeation.

Selectivity of ion channels has been a central issue in the study of cells and tissues at least since Hodgkin, Huxley, and Katz discovered the role of $Na^+$ and $K^+$ currents in the action potential (12-14). The L-type $Ca^{2+}$ channels play an important physiological role in skeletal, cardiac and smooth muscle, endocrine cells and other tissues. These channels open when the cell membrane depolarizes allowing $Ca^{2+}$ ions to enter the cytosol. The entry of $Ca^{2+}$ ions in the cytosol is sensed by the Ryanodine Receptor (RyR) present in the SR (intracellular $Ca^{2+}$ storage organelle) membrane which opens and releases a huge amount of $Ca^{2+}$ in the cytosol, resulting in muscle contraction. The L-type $Ca^{2+}$ channel therefore, plays a crucial physiological role in the excitation-contraction coupling of muscles. These channels are highly selective for $Ca^{2+}$ and have been studied in great detail (15-22). Few channels are more important than the calcium channel because calcium concentration inside cells is used as a signal in almost every tissue of an animal. Thus, the mechanism of selectivity has been studied for a long time. However, the unavailability of crystal structure for these channels has always been the bottleneck.

Mutagenesis experiments of these channels (20) indicate that only a few amino acids that form the selectivity filter, actually contribute to their ion selectivity. A quick literature survey indicates that selectivity of ion channels is interpreted in a variety of ways, ranging from structural discussions of selectivity (23-37) to scaling models (38), kinetic models (4, 39) and electrostatic models (40-43). However, all these different interpretations of selectivity fail to reproduce experimental measurements over a wide range of conditions.

On the contrary, reduced models or oversimplified models using only physical variables, which include the radius of the pore ($R$) and the dielectric constant of the protein ($\varepsilon_p$), are successful in explaining the mechanisms that govern selectivity of the L-type $Ca^{2+}$, $Na^+$ channels and also their mutations (44-49). It has come as a surprise that a model of selectivity that includes only a few features of the atomic structure has been able to describe the selectivity properties of calcium and sodium channels very well, in all solutions over a wide range of conditions, with only two adjustable parameters using crystal radii of ions.

In these reduced models, channel protein is represented as a cylinder and the crucial side chains as spherical charges free to move within the central region of the selectivity filter which is in equilibrium with surrounding bathing solutions containing spherical ions in a dielectric region. Ion radii are Pauling crystal radii (50) taken from the literature of physical chemistry and are never adjusted. The Monte Carlo simulations in these reduced models produce the distribution of ions and side chains that minimizes the free energy of the system. The structure of the overall selectivity filter is the mean location of the side chains and the permeating ions. The rest of the protein (assumed to be "unimportant" in this simplified model) is treated as a fixed structure. Water is not present as molecules in this calculation. It is represented as a dielectric, as is the protein. Ion selectivity is therefore understood by building a reduced description of the structure of the channel such that the selective properties are outputs of the model that arise from the balance between electrostatic and steric forces in the confined space of the channel (29, 30, 32, 33, 35, 37, 40).

Reduced models dealt quantitatively with many properties of other channels including the RyR (51-54) and bacterial channel OmpF (Outer membrane protein F) porin (55, 56) by calculating the current – voltage curves over a wide range of ionic conditions. In RyR, such models successfully predicted an anomalous mole fraction effect (a signature phenomenon of $Ca^{2+}$ channels observed in mole fraction experiments where conductance is measured in mixture of two species, as their relative amounts i.e.

mole fractions vary, indicated by a minimum even before it was measured (52, 54). These models also explained RyR mutations that reduced the structural charge density (of side chains with permanent charge) from 13 M to zero (51). Similar models also produced a successful plan for the conversion of a nonselective bacterial channel OmpF porin into a decent $Ca^{2+}$ channel (55-57).

In this work, the factors that account for success of the reduced model of equilibrium binding selectivity of the $Ca^{2+}$ channels are investigated. Monte Carlo simulations of this reduced model indicate the role of flexibility and confinement of side chains under a variety of ionic conditions. These simulations determine the structural relations that the reduced model calculates accurately and therefore, predicts selective properties of these channels in agreement with the experiments.

The model of selectivity is actually an induced fit model, in which the structure of the binding site is induced by all the forces in the system, depending on the ion concentrations in the bath, and the other parameters of the problem. The fit of the side chains to the ions is induced by the minimization of the free energies of the system. The model is in fact, a self-organized model where the only energies that determine the structure are electrostatic and excluded volume. No energies with a more chemical flavor are incorporated in the present version of the model. Monte Carlo simulations are particularly well suited to compute induced fit, self-organized structures because the simulations yield an ensemble of structures near their free energy minimum. The exact location and mobility of oxygen ions have little effect on the selectivity behavior of calcium channels. Seemingly, nature has chosen a robust mechanism to control selectivity in calcium channels: the first order determinant of selectivity is the density of charge in the selectivity filter. The density is determined by filter volume along with the charge and excluded volume of structural ions confined within it. Flexibility seems a second order determinant.

Besides numerical simulations, single channel planar lipid bilayer experiments are carried out for investigating the biophysical (ion selectivity, permeation, and gating) and biological role of a bacterial outer membrane sialic acid specific porin NanC (N-acetylneuraminic acid inducible Channel) with a known high resolution crystal structure (58).

Porins act as molecular filters in the outer membrane of the gram negative bacterium *Escherichia coli* allowing the passive diffusion of solutes (ions, sugars and amino acids) into the periplasm. Porins participate in a variety of functions ranging from the transport of solutes to contributing to the virulence of infection by pathogenic bacteria (59-65). Porins are usually classified as the general non-specific porins, and solute-specific porins.

The general porins OmpF, OmpC, and PhoE of *E. coli* are channels that allow passive diffusion of small solutes (under 600 Da) down their concentration gradient. The general porins transport a variety of substances from the external environment into the bacterial periplasm and do not demonstrate a marked preference for particular solutes. Other ('specific') porins show preference for particular solutes, for example, the maltoporin LamB (66) of *E. coli*. Specific porins are known to have a definite function important to bacterial survival. Specific porins are often induced under special growth conditions along with the components of an inner membrane uptake system, and enzymes necessary for degradation and fermentation of the solutes. These specific porins presumably are needed for the bacteria to thrive under these special conditions.

The functional properties of a sialic acid specific outer membrane porin NanC of *E. coli* are characterized using electrophysiological methods. NanC is named because it is a N-acetylneuraminic acid (Neu5Ac) inducible channel. N-acetylneuraminic acid is one of the family of naturally occurring sialic acids that include more than 40 nine carbon negatively charged sugar moieties (67). The ability of the bacteria to colonize, persist and cause disease (68, 69) depends on their ability to use sialic acids in many cases. The most abundant and well studied sialic acid is N-acetylneuraminic acid, Neu5Ac. If general porins OmpF/OmpC are not expressed by the *E. coli* bacteria (as is often the case (70)), NanC must be present (i.e., it must be induced) to promote efficient uptake of Neu5Ac across the outer membrane.

Recently, a high resolution structure of NanC has been reported (58). The structure of NanC shares many of the characteristic features of other outer membrane channel proteins. NanC is a monomer, a 12-stranded β-barrel with a relatively narrow pore of average radius 3.3Å. However, NanC is also different in many ways from other porin structures. In most of the porins, 'loop' partially occludes the pore region and is the most obvious location where specific interactions and gating might occur. Interestingly, NanC has no 'loop' occluding the pore. The pore region of NanC is predominately decorated by positively charged residues that are arranged to form two positively charged tracks facing each other across the pore. This particular arrangement of the positively charged residues in the pore region seems to help the negatively charged Neu5Ac (carboxylate group ($COO^-$) of Neu5Ac deprotonated at physiological pH, pKa ~ 2.6) to move through NanC. The positively charged tracks are likely to guide the movement of negatively charged solutes the way the steel tracks of a railroad guide trains, both passive passenger and freight cars, and active locomotives.

Interestingly, the earlier functional measurements of NanC carried out using patch-clamp experiments were unable to demonstrate any significant change in the function of NanC in the presence of even large concentrations of Neu5Ac, up to 50 mM (70). The high resolution structure suggests one explanation for this finding. The structure shows many basic (positive) side chains that would be screened (71) in the high salt concentrations used in these earlier measurements. Structural measurements also show HEPES binding to NanC. Two HEPES molecules were immobile enough under crystallizing conditions to produce diffraction in the crystal structure, one near each end of the channel. Binding strong enough to crystallize HEPES in place in a channel is likely to modify current flow through the channel.

The biophysical properties of NanC are thus characterized and the experimental conditions necessary for the measurement of transport of Neu5Ac in bilayer setups are determined. The biophysical conditions that produce the biological sialic acid transport are identified through single channel measurements of NanC in an artificial lipid bilayer determining the ion selectivity and conductance of the channel.

## 1.2. <u>Overview</u>

The work that is described in what follows is broadly focused on two topics: (i) Simulation studies characterizing the physical mechanisms underlying the micromolar calcium selectivity of voltage-gated calcium channels through a reduced model. The success of reduced model in reproducing the experimentally observed calcium binding under a range of ionic conditions is investigated. (ii). Single channel planar lipid bilayer experiments investigating the biophysical and biological properties of a bacterial outer-membrane channel NanC.

A novel aspect of this work is that it explores the fundamental principle determining the ion selectivity and permeation properties of two biologically dissimilar voltage-gated ion channels (one is eukaryotic the L-type calcium channel of cardiac muscle cell and the other one is bacterial NanC of *E.coli*) from two different approaches *i.e.* via simulations and bilayer experiments. A similar picture of a narrow crowded selectivity filter results where the charge/space competition principle seems to rule. The dominant interactions governing the ion selectivity and permeation are suggested to be electrostatics and entropic occurring between the mobile ions and fixed charges of the protein side chains in the selectivity filter of these ion channels.

In Chapter 2, a physical model accounting for the selectivity of calcium (and sodium) channels in many ionic solutions of different composition and concentration using just two parameters (radius of the pore R and protein dielectric constant $\varepsilon_p$) with unchanging values is introduced. The role of flexibility in the selectivity of calcium channels is studied using this simple model. The Grand Canonical Monte Carlo (GCMC) method applied in these simulations is described. The outputs of the GCMC simulations namely, the occupancies of cations $Ca^{2+}/ Na^+$, the concentration profiles (distribution of cations), and linearized conductance of $Na^+$ are compared.

In Chapter 3, the principle and practical issues related to the measurements of a single ion channel in a planar lipid bilayer are described. The main components of the planar lipid bilayer set-up used while conducting these measurements are discussed. The critical issue of voltage offsets for these sub- pico ampere current measurements in planar lipid bilayer experiments is addressed.

In Chapter 4, biophysical properties of a bacterial outer-membrane channel NanC are characterized using single channel planar lipid bilayer experiments. The ionic conditions favorable for studying its biological role are identified.

In Chapter 5, the biological role of a NanC in transport of sialic acid is studied in planar lipid bilayers under the ionic conditions identified in Chapter 4. The specificity and efficiency of NanC in transport of sialic acid is shown by comparison with OmpF.

**CHAPTER 2. SELF-ORGANIZED MODELS OF SELECTIVITY IN CALCIUM CHANNELS**

## 2.1. <u>Introduction</u>

Selectivity arises from the interactions of ions and side chains of proteins constrained by the geometry and properties of the rest of the protein, and the surrounding bathing solutions. For a long time—as long as the issue of selectivity has been considered by biologists—it has been believed that selectivity depends on the detailed structure of the binding sites for ions. Selectivity has been thought to depend on the exact atomic arrangements of side chains, ions, and other nearby molecules (72, 73). The selectivity of the calcium channel considered here is particularly important because of the enormous importance of the channel (74-92). The L-type calcium channel considered controls the contraction of the heart and signaling in skeletal muscle. Few channels are more important than the calcium channel because calcium concentration inside cells is used as a signal in almost every tissue of an animal.

It has come as a surprise that a model of selectivity that includes only a few features of the atomic structure has been able to describe the selectivity properties of calcium and sodium channels very well, in all solutions over a wide range of conditions, with only two adjustable parameters using crystal radii of ions (47, 93). This model is skeletal in its simplicity, representing side chains as charged spheres unrestrained by connections to the surrounding protein. This model represents the selectivity filter of calcium and sodium channels as cylinders containing spherical structural ions (modeling the terminal groups of side chains) free to move within the cylinder but unable to leave it (Figure 1). Note that water is present in this model only as a dielectric: the primitive implicit solvent (94-118) model of ionic solutions is used and extended into the channel. The *question* is how can this simple model possibly work, given that it includes no detail of the protein structure, and uses only a crude representation (to put it kindly) of the side chains and their interaction with ions.

The *answer* is that the model works because it captures the features of the binding sites of calcium and sodium channels that biology actually uses to produce selectivity. Many other physical phenomena could be used to produce selectivity, and probably are in other types of channels, transporters, enzymes, and proteins, let alone physical systems in general. It seems here however that biology has used only the simplest, the competition between electrostatic forces and volume exclusion in very concentrated systems of ions and side chains.

The two cations $Ca^{2+}$ and $Na^+$ that compete for the channel arrive from a bath of a given composition and mix with side chains of the channel protein in this model. The winner of the competition is the ionic species whose binding minimizes the free energy $F = U - TS$. Electrostatic attraction between the side chains and the cations decreases the energy $U$, while competition for space in the crowded selectivity filter chiefly influences the $-TS$ term. The entropic term is most directly influenced by the flexibility of side chains. Indirectly, of course, everything is influenced by everything.

The $U$ and $-TS$ terms are not uncoupled, of course (119), but separating $U$ and $-TS$ terms provides a useful framework to understand selectivity.

In this model a large mobility (flexibility) of the terminal groups (8 half-charged oxygen ions, in this paper) of side chains is assumed. In reality, the movement of side chains is restricted because they are tethered to the polypeptide backbone of a protein. Side chains of acidic and basic residues in real proteins are not fully rigid (though they are not entirely mobile either) and have considerable mobility during thermal motion.

In this chapter, the goal is to study the effect of the flexibility of the side chains on physical quantities computed by Monte Carlo simulation. The basic quantities computed by the simulation are the profiles of spatial distribution of the various ionic species. These distributions, as expected, are very sensitive to restrictions in the mobility of oxygen ions. Integration of the ionic profiles provides quantities comparable to experiments. These integral properties characterize selectivity from two profoundly different points of view.

The integral of the ionic profile (of a given species) itself over a given volume (of the selectivity filter, for example) describes the *binding affinity* of the given ionic species to this volume. This quantity—called as 'occupancy'—is proportional to the probability that a given ionic species binds to the selectivity filter in competition with other kind of ions. Occupancy is a result of minimized free energy, an equilibrium concept, but occupancy exists in non-equilibrium systems as well, of course. It just must be computed by a different theory in that case, one that includes non-equilibrium parameters (like conductance) and spatially non-uniform boundary conditions.(120, 121)

The resistance of the channel is the sum of the resistance of a series of (infinitesimally thin) discs, as described precisely later in Methods section. Resistance is a more dynamic variable than occupancy. Each disc has resistance determined by the amount and mobility of the ions in the disc. The amount is proportional to the reciprocal of the concentration of ions, times the volume of the disc. The sum is actually the integral of the reciprocal of the profile of ionic concentration over a length of the ionic pathway. The integral is proportional to the resistance of that length to the current of the ionic species in question. This resistance describes the selectivity of ion channels directly measurable from measurements of current or flux through channels under near equilibrium conditions. It is more sensitive to local variations in the ionic profiles than occupancy. It is expected, therefore, that conductance is more sensitive to variations in flexibility of side chains too.

The flexibility of the oxygen ions will be changed in different ways and the change in the physical quantities mentioned above will be analyzed as the flexibility is changed. A quite different behavior is obtained for the profiles, occupancies, and conductances as a function of varying flexibility.

An important result of analysis of this model is that the locations of ions and side chains have a 'self adjusting' nature as ion concentrations are changed in the bath, or the nature of the side chains is altered, for example, from EEEE to EEEA in a calcium channel or DEKA to DEEA in a sodium channel. Side chains are mobile in this model and so they change their distribution as parameters of the model change. In this picture, the structure of the binding site is a computed consequence of thermodynamic and geometrical constraints and all the forces present in the system. Thus, *'structure'* refers to the equilibrium profiles of spatial distribution of concentration of all the ions obtained as ensemble averages of converged simulations. This structure obviously changes as experimental conditions change.

The sensitivity of the structure to the location and flexibility of side chains helps answer the general paradox "How can a model without a detailed structure account for complicated properties of two different types of channels under so many conditions?" The answer to the question is that the model computes the structure. The channel structure is different in different conditions because of the sensitivity of spatial distributions. Thus, biological properties that are sensitive to structure will be sensitive to location and flexibility of side chains. Integrated properties like conductance are not so sensitive.

In other words, the model of selectivity is an induced fit model, in which the structure of the binding site is induced by all the forces in the system, depending on the ion concentrations in the bath, and the other parameters of the problem. The fit of the side chains to the ions is induced by the minimization of the free energies of the system. The model is a self organized model. In this model, the only energies that determine the structure are electrostatic and excluded volume. No energies with a more chemical flavor are included in the present version of the model.

Unfortunately, nothing is known yet about the actual locations of side chains in real calcium and sodium channels because X-ray structures are unavailable. It is possible to hypothesize the locations assuming various homologies with the known structure of the KcsA potassium channel (122, 123). However, a different route is followed as described in the Model section.

It is concluded that the ionic distribution profiles are sensitive functions of locations of immobilized side chains and their flexibility. Occupancy, on the other hand is less sensitive to flexibility because its first order determinant is the density of ions in the selectivity filter as we established earlier (124, 125). Conductance, the most interesting quantity of the three, lies in between because it depends on both occupancy (the number of available charge carriers) and fine details in the distribution profiles. The conductance of a single channel of fixed diameter depends on the sum of many differential 'resistors' because it measures the 'series resistance' of the single channel. The conductance depends on the reciprocal of the concentration for this reason and is very sensitive to high resistance obstructions in the permeation pathway even if they occur over a tiny length. These high resistances are produced by the low concentrations of mobile ions in the depletion zones in the ionic profiles (126). Depletion zones are responsible for many of the most important properties of transistors (127-129).

Transistors and channels involve quite similar physics and are described by quite similar equations, but the charge carriers of transistors are points with no size.

If three dimensional structures for calcium and sodium channels become available, the information gained must be built into the model. Until then, this approach—that allows the side chains to find their optimal distribution automatically in the simulation—seems to be a good way to proceed. It allows constructing a functional skeletal model that fits a wide range of data from two channel types under many conditions with only two parameters. It makes understanding of permeation and selectivity mechanisms of real calcium and sodium channels possible, if permeation and selectivity are viewed as outputs of that simplified model.

## 2.2. <u>Methods</u>

### 2.2.1. Model of Channel and Electrolyte

A reduced model is used to represent the L-type calcium channel because a crystallographic structure is not available. It is known from the experimental literature that $Ca^{2+}$ selectivity in this channel is determined and regulated by the four glutamates of the pore lining loops which form the selectivity filter (17, 20, 8, 9). The negatively charged carboxyl ($COO^-$) side chains of these glutamates extend into the selectivity filter region of the channel (130). As Sather and McCleskey put it in their review (20) "The calcium channel field is convinced that the EEEE carboxyl side chains project into the pore lumen" to form a mixture of ions and side chains that has been called an 'electrical stew'(131).

At room temperature of 300K, these side chains as well as the surrounding cations interact and exhibit thermal motions. The four glutamates (EEEE) contribute to a fixed charge of -4$e$ on the selectivity filter. This existing information about the L-type calcium channel is used to build a reduced model (132, 125).

In the reduced model the channel is represented as a doughnut-shaped object with a pore in the middle connecting the two baths (Figure 2.1). The protein which forms the pore is represented as a continuum solid with dielectric coefficient, $\varepsilon_p = 10$. The central, cylindrical part of the pore, the selectivity filter, is assigned a radius $R = 3.5$ Å and length $H = 10$ Å. The model is rotationally symmetric along the pore axis. The selectivity filter (–5 Å, 5 Å) contains the negatively charged side chains extending from the polypeptide backbone of the channel protein into the pathway for ionic movement. We represent the carboxyl (COO$^-$) side chains as a pair of negative half charged oxygen ions (O$^{1/2-}$) resulting in 8 oxygen ions (also called 'structural ions'). The net charge in the selectivity filter therefore sums to -4$e$. The side chains are free to rearrange inside the selectivity filter of the channel but cannot leave the selectivity filter. The mobile ions (Na$^+$, Ca$^{2+}$, Cl$^-$) as well as the oxygen ions of the carboxyl groups are hard spheres and are assigned Pauling crystal radii (see caption of Figure 2.1). The structural ions mix with the mobile ions producing a flexible but confined environment in the selectivity filter. The structural ions have different degrees of flexibility as described later. Water is represented implicitly as a dielectric ($\varepsilon_w = 80$).

The simulation cell is a cylindrical compartment that is kept small to save computation time. The simulation cell is checked to be sure that it is large enough that our final results do not depend on its size. The dimensions of the simulation cell are chosen depending on the ionic concentrations in the surrounding bath solutions. The baths are separated by a lipid membrane 20 Å thick except where the channel protein is found. Ions are excluded from the lipid membrane. The filter is assumed to have a dielectric coefficient of 80. If more realistic values were used, the amount of computation required for the electrostatics increased and no significant effects were noticed on the simulation results (49, 132).

### 2.2.2. Metropolis Monte Carlo Method

A Monte Carlo simulation generates configurations of a system by making random changes to the positions of the species present, together with their orientations and conformations wherever appropriate. Many computer algorithms are based upon 'Monte Carlo' method; a term coined by Metropolis in 1953 (133).In general, MC method refers to all those methods that use importance sampling which ensures that only those configurations be generated which make a large contribution to the average quantity of interest. The MC method based on Metropolis scheme (MMC) is essentially the importance sampling where selected configurations are generated that contribute significantly to the averages such that if the new configuration is accepted if it has lower energy than the preceding configuration. However, if the new configuration has a higher energy then we compare the difference in their corresponding Boltzmann factors *i.e.* $\exp\left[\dfrac{-\Delta U}{kT}\right]$ where $\Delta U$ is the difference in energies, to a random number generated uniformly between 0 and 1. The move is accepted if the difference is greater than 1. Therefore, the MMC method essentially does a statistical sampling of the configuration space efficiently by increasing the probability of the 'accepted' configurations for determining the thermodynamic properties of equilibrium systems.

The MMC method was initially applied in a canonical ensemble to simulate the calcium over sodium selectivity in the reduced model of the L-type calcium channel. This model was pretty good in describing the selective behavior but it was not able to simulate the micromolar selectivity for $Ca^{2+}$ as observed in calcium channels (134). This problem was overcome when the ion channels were modeled with Grand Canonical Monte Carlo (GCMC) method. The major difference between the MC simulation method in canonical and grand canonical ensemble is the way statistical sampling is done. In canonical ensemble the configuration space is sampled by accepting/rejecting random moves in the ensemble whereas in grand canonical ensemble the statistical sampling is ensured by making random insertions/deletions in the ensemble.

The grand canonical version of MMC method is useful while dealing with cases where the number of particles is varying and hence, insertions/deletions in the ensemble are natural to occur. Some earlier studies on GCMC simulations were done by Valleau et al., (135) to simulate primitive model electrolytes. This version of MC method is particularly useful while the concentration of particular species is very low in a mixture. Therefore, the use of MC method in grand canonical ensemble is quite useful while dealing with modeling systems in physiology that are sensitive to concentrations of order of micromolar. Following are the basic steps followed in GCMC simulation:

1. A particle is displaced using the usual Metropolis scheme (as in canonical version of MC method).

2. A particle is destroyed.

3. A particle is created at random position.

The probability of acceptance of creation is –

$$p_{creation} = \min\left[1, \frac{V}{\Lambda^3(N+1)}\exp\left(\beta\left(\mu - \Delta U\right)\right)\right], \tag{1}$$

and probability of acceptance of destroying particle is –

$$p_{deletion} = \min\left[1, \frac{\Lambda^3 N}{V}\exp\left(-\beta\left(\mu + \Delta U\right)\right)\right], \tag{2}$$

where, $N$ is the number of particles, $V$ is the volume for insertion/deletion of particles in region of interest, $\beta = \dfrac{1}{kT}$ , $\Lambda = \sqrt{\dfrac{h^2}{(2\pi m kT)}}$ is the thermal de-Broglie wavelength, $\mu$ is the chemical potential and $\Delta U$ is the corresponding energy change. The probability of creation is equal to the probability of destroying in order to ensure the symmetry.

### 2.2.3. Practical Implementation of Grand Canonical Monte Carlo Method

MC simulations performed using Metropolis sampling in the grand canonical ensemble (135) allows to efficiently simulate the very small ionic concentrations important for calcium channels. The details of the methods of sampling and their acceptance tests have been described in (136, 49, 48, 135) and earlier papers.

An equilibrium grand canonical ensemble is simulated at room temperature 300K. The chemical potentials of ions are the inputs chosen for the grand canonical ensemble and are determined separately using an iterative method (137, 138). The acceptance tests of new particle configurations involve the total electrostatic energy of a configuration. The net electrostatic energy of the system includes the Coulombic interactions between the ions and the side chains and the interactions resulting from the charges induced at dielectric boundaries which are computed using the induced charge computation method (45). The GCMC simulations presented in the results section are averages of many runs performed on multiple processors and beginning from different seed configurations. Each result is the average of $6 \times 10^8$ to $1.2 \times 10^9$ MC configurations.

### 2.2.4. Simulated Experimental Setup

The micromolar block of $Na^+$ current observed by McCleskey et al. (139, 140) is a characteristic behavior of the L-type calcium channel. In this experiment, $CaCl_2$ is gradually added to a fixed background of 30 mM NaCl. The experiments show that 1 μM of $Ca^{2+}$ reduces the current through the L-type calcium channel to the half its value in the absence of $Ca^{2+}$. This result implies that the selectivity filter of the calcium channel contains a high affinity binding site for cations, especially for $Ca^{2+}$. The strongly bound $Ca^{2+}$ obstructs the diffusion of $Na^+$ ions in the packed and narrow selectivity filter. This experiment was the main target of several theoretical and simulation studies of the L-type calcium channel (49, 132, 48, 93, 141, 124, 142).

As far as known this reduced model is the only one that is able to produce the strong $Ca^{2+}$ vs. $Na^+$ selectivity in a range of conditions. Indeed most computations of selectivity do not contain concentration as a variable at all (72, 73, 143). Moreover, classical kinetic models of selectivity (39, 4) assume the energy landscape or the barrier to be independent of ionic concentrations which is an implausible approximation given the ubiquity of shielding in ionic solutions and the reality of Gauss' law. These and other difficulties with classical models have been evident for a long time (144-154). Indeed, the 'law' of mass action itself has recently been shown to apply only to infinitely dilute solutions of non-interacting particles, if it is used with constant rate constants as in classical models of channels (39, 4).

The simulations (analyzed with the integrated Nernst-Planck equation, see later Eq. (5)) show that at 1 µM $Ca^{2+}$ the average number of $Na^+$ ions in the selectivity filter drops to half the value it has at zero $Ca^{2+}$. Under those conditions, $Ca^{2+}$ ions occupy the central binding site in the filter, but they do not contribute to the current because of the depletion zones formed at the entrances of the filter (93, 141). This mechanism is in agreement with an earlier intuitive description of the mechanism of this block (155).

Some simulations were carried out for the entire range of $Ca^{2+}$ concentrations as used by Almers and McCleskey in their experiment (from zero to $10^{-2}$ M). Others were just carried out for the special case when $[Ca^{2+}]$ is 1µM. In experiments, the replacement of $Na^+$ by $Ca^{2+}$ takes place in a narrow concentration range around this value.

### 2.2.5. Models of Flexibility

There are two limiting cases of the flexibility of side chains.

(1) The 'flexible' case is the usual model with maximum flexibility. In this case, oxygen ions are perfectly mobile inside the selectivity filter but they are confined within the filter by hard walls. In this model, oxygen ions automatically find their average distribution that minimizes free energy of the system. As cations ($Na^+$ or $Ca^{2+}$) enter the filter, oxygen ions rearrange and make it possible for the cations to pass the channel (126). The distribution of the oxygens is an output of the simulation.

There is agreement (156, 17, 20, 157, 9) that side chains in the filter of calcium channels are quite flexible. They are in constant thermal motion and the long (3-carbon) chains allow the $COO^-$ end groups considerable freedom to move within the channel. Nonetheless, the maximum mobility case obviously overestimates the flexibility of side chains.

(2) The '*fixed*' case places oxygen ions in fixed positions and allows zero mobility. This assumption roughly corresponds to the system at 0 K and underestimates the flexibility of side chains.

Unfortunately, the X-ray structure for the calcium channel is not yet known. Therefore, instead of assuming the initial positions of the oxygens, we choose certain fixed oxygen configurations from the billions of possible configurations that are computed during an MC simulation of the '*flexible*' case. The nature of MC simulations ensures that all these configurations occur in the sample in a Boltzmann distribution(158, 159). Each configuration chosen as a fixed configuration is one of the configurations of the equilibrated system. In other words, they can be called '*probable*' configurations.

The configurations chosen are even 'more probable' using a criteria based on electrostatic energy. These configurations are determined by carrying out the usual MC simulation for 50 blocks where each block consists of $5\times10^5$ trials. After a block is finished, the oxygen ions are displaced to ensure that the next block samples new configurations. After every MC trial in each block the electrostatic interaction energy of the oxygen ions with every other ion is calculated. A running average of this energy is updated. At the end of each block, the instantaneous locations of the oxygen ions and the corresponding running averages of the energy for that block are saved. Therefore, at the end of the simulation for each of the 50 blocks a set consisting of locations for the 8 oxygen ions is obtained which represent the instantaneous configurations at the end of the block and the corresponding average energy of that block. 10 configurations of the oxygens are chosen from these 50 configurations that have the lowest average energy and are called as the '*low energy*' configurations. Simulations were performed for these cases in the usual way except that MC movements were not performed for the oxygen ions. Simulations were performed for $[CaCl_2] = 10^{-6}$ M and $[CaCl_2] = 0$ M in order to relate the results for $10^{-6}$ M to results in the entire absence of $Ca^{2+}$.

In addition to these 10 '*fixed low energy*' configurations, configuration of oxygens is chosen that had the lowest energy in the sample checked. This configuration does not have the lowest energy in general; it just has the lowest energy in the 50 configurations of oxygens examined in the simulation. This case is called the '*lowest energy*' case. The lowest energy case is simulated for concentrations spanning the whole $Ca^{2+}$ concentration range used in experiments.

Because the exact structure of the selectivity filter is not known, the configurations just defined were not selected with the purpose to mimic any realistic structure. They were chosen to study the effect of fixing the oxygens somewhere. A large number—hundreds—of other configurations were chosen by *ad hoc* methods in the course of these calculations, as developed these procedures. From all this work, it is quite certain that the conclusions drawn from the results are quite general and independent of the actual positions of the oxygen ions. The qualitative conclusions of this work are not sensitive to the exact method of choosing the 'low energy' configurations. The exact configuration called as 'low energy' has little importance.

To study the effect of oxygen-flexibility in more detail models are constructed in which oxygen ions have partial flexibility. These models lie between the fixed and flexible cases defined above.

(1)　　The '*restricted*' case: It restricts the four pairs of oxygen ions to stay in four narrow regions. The glutamate residues assumed are not concentrated at one position of the selectivity filter, but rather they are distributed along the axis of the pore confined in cylinders defined by the filter wall and hard walls at fixed z-coordinates as shown in Figure 2.2. The regions in which the oxygen-pairs are confined overlap and they prevent a given pair to diffuse elsewhere in the filter. The distribution of oxygen ions is more uniform in this case than in the '*flexible*' case. Simulations were performed for this case in the whole $Ca^{2+}$ concentration range where the oxygen ions were perfectly mobile in the selectivity filter $|z| \leq 3.6$Å.

(2)    The 'confined' case: It starts with the '*lowest energy fixed*' position of the oxygen ions and gradually allows them more and more mobility in the following way. Spheres are defined of radii $R_{ox}$ around these fixed positions and allowed the oxygen ions to move freely inside these spheres. The oxygen ions cannot leave these spheres. Then, we gradually increase the size of these spheres thus increasing the mobility of the oxygen ions. Any radius that is larger than 7.2 Å effectively corresponds to the '*flexible*' case. Thus, we designed a scheme in which we smoothly change the model between the two limiting cases of flexible and fixed. We performed simulations in the confined case only for $[CaCl_2] = 10^{-6}$ M and $[CaCl_2] = 0$ M.

## 2.2.6. Pore Conductance

We use the integrated Nernst-Planck formulation of Gillespie and Boda (141, 160) to relate the equilibrium GCMC simulation to the linearized slope conductance estimated in experiments with equal concentrations of ions on both sides of the channel. The integrated Nernst-Planck formulation based on the resistors-in-series model is used to calculate the conductance of each ionic species in the pore region. The combination of the integrated Nernst Planck equation (Eq.(5)) and MC simulations has been successful in reproducing the anomalous mole fraction effect  in L-type calcium channel known from experiment (141). We reproduce here the key equations that we use in our calculations.

The ions are assumed to move diffusively and their current is described by the Nernst Planck equation

$$-\mathbf{J}_i(\mathbf{x}) = \frac{1}{k_B T} D_i(\mathbf{x}) \rho_i(\mathbf{x}) \nabla \mu_i(\mathbf{x}),$$

(3)

where, $\mathbf{J}_i(\mathbf{x})$, $D_i$, $\rho_i$, and $\mu_i$ are the local flux density, diffusion coefficient, density and the electrochemical potential respectively of ion species $i$. The value $k_B$ is the Boltzmann constant and $T$ is the temperature. We focus on the selectivity filter region of length $L$ and cross-sectional area $A(z)$ confining the structural ions. The chemical potential $\mu_i(z)$ and the diffusion coefficient $D_i(z)$ are assumed to be uniform over the cross-section of the selectivity filter. With these approximations, the total flux of ion species $i$ through the pore:

$$-J_i^T = \frac{D_i}{k_B T} \frac{d\mu_i(z)}{dz} n_i(z),$$

(4)

where, $n_i(z)$ is the axial number density of ions (number per unit pore length) at axial location $z$.

The total conductance $\gamma$ of the pore in the presence of symmetrical bath solutions at the end of the selectivity filter region containing several ion species of charge $z_i e_0$ and a very small voltage applied across the system is:

$$\gamma = e_0 \sum_i z_i \left. \frac{\partial J_i^T}{\partial V} \right|_{V_{L/2} - V_{-L/2}} = \sum_i D_i \frac{z_i^2 e_0^2}{k_B T} \left[ \int_{-L/2}^{L/2} \frac{dz}{n_i(z)} \right]^{-1} . \tag{5}$$

Note that the conductance depends on the square of the charge of the ions $z_i^2 e_0^2$. The axial densities of ion species $n_i$ are computed from the MC simulations. The diffusion coefficient of ions is an external parameter not determined by our simulation that must be provided as an input to our computation. We deal with normalized conductances and therefore, it is sufficient to specify the ratio of diffusion coefficients of the two cations $D_{Ca}/D_{Na}$. We use the value $D_{Ca}/D_{Na}^+ = 0.1$ in our computation. Dynamical Monte Carlo (DMC) simulation (126) recently verified an early suggestion of Nonner and Eisenberg (142) that the diffusion coefficient of $Ca^{2+}$ in the selectivity filter is much smaller than that of $Na^+$.

The approximations and equations used to calculate the pore conductance $\gamma$ quantifies the current through the channel. It is very important here to emphasize that the estimation of pore conductance requires only the spatial (i.e., cross sectional) uniformity of the chemical potential, which does not imply that the other variables, such as the spatial number density of the ions or the electrical potential are spatially uniform. The concentration profiles used in these calculations are obtained from simulations performed under equilibrium conditions and with self-consistent electrostatics. When the system is off equilibrium the concentration profiles will have to be recomputed to be self-consistent as done in Density Functional/Poisson Nernst Planck theory (DFT/PNP) (161, 52, 141, 160, 162, 163) or more recently with variational method (*EnVarA*) (120, 164, 121). Thus, our conductance $\gamma$ of Eq. (5) does not include the nonlinear effects that may occur when other voltages or concentration gradients are applied. The non-equilibrium effects on the potential profiles cannot be determined by the equations given here. Non-equilibrium effects are important in channels because channels are devices that use gradients of free energy to perform their function. Non-equilibrium effects are responsible, for example, for the properties of semiconductor devices and those effects are not present near equilibrium where our conductance equation applies. Such effects can be computed by Poisson Nernst Planck equations if charge carriers have zero diameter or by a modification of those equations if the charge carriers have finite diameter (165, 120, 164, 166, 167, 52, 168, 121). The slope conductance $\gamma$ is nonetheless quite informative, because experiments show that current-voltage curves are linear in wide voltage range around equilibrium.

## 2.3. <u>Results</u>

We analyze our results on various levels of abstraction. The primary output of our simulations is concentration profiles. These profiles describe the probability of various ions being found in a given position, or, looking at the same thing from a more dynamical point of view, they describe the relative amount of time that these ions spend in a given location. We will show profiles averaged over the cross-section of the channel. We can derive two kinds of integrated quantities from the concentration profiles.

(1) The integral of the concentration profile provides the average number of the given ionic species in a certain sub-volume of the system. We will show average number of ions integrated over the selectivity filter and we will call these numbers occupancies. The occupancies of $Na^+$ and $Ca^{2+}$ describe equilibrium binding affinity of these ions to the selectivity filter.

(2) The other integrated quantity of interest is the integral of the reciprocal of the concentration profile based on the Nernst-Planck equation (see Methods section, Eq. (5)). This integral is proportional to the resistance of the pore to a given ionic species in the region of validity of the equation. We will show conductance values $\gamma$ that are the reciprocals of the resistances. We will concentrate on conductances of $Na^+$, because $Ca^{2+}$ does not conduct significant current in the conditions of interest, at micromolar $[CaCl_2]$.

We work with normalized conductances. Usually, we normalize with respect to the $Na^+$ conductance computed in the absence of $Ca^{2+}$. Because in this paper we compare different systems, we have two choices regarding how we normalize the conductance.

(A) First, each model ('*flexible*', '*restricted*', '*fixed*' or '*constrained'* cases) can be normalized by its own zero-$Ca^{2+}$ conductance. Each model is normalized by itself. In this way, $g$ is always normalized to 1 in the limit $[CaCl_2] \rightarrow 0$. This kind of normalization is used in the experimental literature (139, 140, 156, 17, 20, 9).

(B) To compare the absolute values of conductances computed from different models, we can normalize with respect to one fixed value in all cases. This way, we can draw conclusions about how changing the flexibility of the oxygen ions influences the ability of the channel to conduct $Na^+$. Correlations between the integrated quantities are shown by plotting one integrated quantity on the ordinate, and the other integrated quantity on the abscissa.

Figure 2.3 shows concentration profiles for the (A) '*flexible*', (B) '*restricted*', (C) '*lowest energy fixed*', and (D) the average of the 10 '*low energy fixed*' cases. The oxygen concentration profiles (top panels of Figure 2.3 A-D) are similar in the '*flexible*' and '*restricted*' cases: oxygen ions tend to accumulate at the entrances of the filter because the negative oxygen ions repel each other to the confining walls of their compartment. A third peak appears in the center of the filter in the '*flexible*' case because of packing. In the '*restricted*' case, the total oxygen profile is a sum of the four profiles for the four oxygen pairs. The '*restricted*' case has two additional peaks. The distribution of oxygens is more flat than in the '*flexible*' case. The vertical lines in the '*lowest energy fixed*' case represent Dirac-delta functionals describing the positions of the fixed oxygens.

Each simulation performed with a given fixed oxygen configuration provides an adequate sampling of all possible configurations of the free ions ($Na^+$, $Ca^{2+}$, and $Cl^-$), because of the large number of configurations calculated (and examined) in each simulation. Many of the configurations we have chosen for the fixed oxygens produce configurations of the free ions that resemble results of simulations of the totally '*flexible*' case. The 10 selected '*low energy fixed*' simulations then correspond to a sample that resembles a sample of the '*flexible*' case. The resemblance is approximate because we have only 10 '*low energy fixed*' oxygen configurations compared to the millions of oxygen configurations found in a usual simulation for the entire '*flexible*' case. (Remember the result of a Monte Carlo simulation is a set of configurations, not a single configuration.) To test this idea, we averaged the concentration profiles obtained from the 10 simulations for the 10 '*low energy fixed*' cases. The oxygen profile shown in the top panel of Figure 2.3D was calculated using a wide (1 Å) bin. The overall behavior of the curve is very similar to the oxygen profile for the '*flexible*' case despite the small (10) sample for the oxygens in the '*low energy fixed*' case.

The bottom panels of Figure 2.3A-D show the results for the free ions ($Na^+$ and $Ca^{2+}$) for [$CaCl_2$] = $10^{-6}$ M and [$CaCl_2$] = 0 M. A micromolar amount of $Ca^{2+}$ in the bath is sufficient to decrease the $Na^+$ concentration in the pore to the half of its value in the absence of $Ca^{2+}$. (Compare the thick solid and thin dashed lines.) From this point of view, the four models behave similarly. The similarity is especially striking in the case of the '*lowest energy fixed*' oxygen configuration: although the details of the distribution of $Na^+$ ions are very different from those in the '*flexible*' and '*restricted*' cases, the relative behavior with respect to the zero-$Ca^{2+}$ curve is similar. This is also true for the average of the 10 '*low energy fixed*' cases.

This is the first important conclusion of this work: adding $Ca^{2+}$ to the systems simply scales the $Na^+$ spatial profiles but it does not change the shape of the spatial distribution. Furthermore, adding 1 μM $Ca^{2+}$ scales the $Na^+$ profiles similarly in the three different cases, thus producing similar selectivity behavior. This scaling is shown by Figure 2.4, where we plot the ratio of the $Na^+$ concentration spatial profiles for $[CaCl_2]$ = $10^{-6}$ M and $[CaCl_2]$ = 0 M. This ratio is similar for the three cases plotted showing that these three models have similar selectivity behavior. The 10 '*low energy fixed*' cases have the same behavior on average. The shape of the curves is very similar in the individual cases too (data not shown).

$Ca^{2+}$ profiles also behave similarly in the various cases in this respect. As $Ca^{2+}$ is added, it appears in the selectivity filter wherever space is available (at the minima of the oxygen profiles).The common feature is that $Ca^{2+}$ is absent at the filter entrances (3.0 $\leq$ |z| $\leq$5Å).

Depletion zones—here at the filter entrances—have the property that the reciprocal of the $Ca^{2+}$ concentration is large at these locations, so the integral (of the reciprocal of the concentration profile, which is the resistance) is also large. Various slices of the channel along the pore axis behave as resistors connected in series. One high-resistance element makes the resistance of the whole circuit high. Therefore, $Ca^{2+}$ does not carry any current at this concentration; it only blocks (reduces) the current of $Na^+$.

Here we see the origin of the most noted experimental property of the calcium channel, calcium block. Calcium block of sodium current is produced by the selective binding of $Ca^{2+}$ in the selectivity filter. The absence of calcium current (an important part of the block) is produced by depletion zones. The depletion zones are computed outputs of our model, not assumed inputs.

To describe further the selectivity behavior of the channel models using the integrated quantities, we show titration curves. We plot these integrated quantities as a function of the concentration of added $Ca^{2+}$.

Figure 2.5A shows the occupancy curves for both $Na^+$ and $Ca^{2+}$. $Ca^{2+}$ gradually replaces $Na^+$ in the filter as $[CaCl_2]$ increases. At about 1 μM, the two ions have equal amount in the filter. Also, at this $Ca^{2+}$ concentration the amount of $Na^+$ drops to half in the filter. The three different cases behave similarly in spite of the differences in fine details in the concentration profiles (see Figure 3).

Figure 2.5B shows the normalized $Na^+$ conductances (normalized by their own zero-$Ca^{2+}$ values) for the three different cases. The current carried by $Na^+$ is gradually decreased as $Ca^{2+}$ is added in accordance with the experiment of Almers and McCleskey (139, 140). As we demonstrated before, we reproduce the micromolar block of the current by $Ca^{2+}$, our curves agree well with the experimental curve in the low $[CaCl_2]$ range. $Ca^{2+}$ starts to conduct at high $[CaCl_2]$. In this regime, our agreement with experiments is only qualitative due to differences between the theoretical and experimental situations as described previously (141). (In experiments, $Ca^{2+}$ is added only to the extracellular side, while our setup is symmetrical.) Note that the occupancy

and conductance curves for $Na^+$ behave similarly indicating a strong correlation between these two quantities (see later discussion).

Figure 2.6 shows the correlation between $Na^+$ conductance and $Na^+$ occupancy. The correlation was evident from Figure 5, but this figure shows it clearly: the more $Na^+$ we have in the filter, the larger the pore's conductance for $Na^+$. Again, the various '*fixed*' oxygen points scatter over a relatively wide range, but this figure clearly shows that the large drop in both occupancy and conductance occurs in a relatively narrow concentration range around 1 μM. In the range of lower and higher [$CaCl_2$] (below and above $10^{-6}$ M), the values do not change very much. Also, the average of the 10 '*low energy fixed*' cases is quite close the '*flexible*' and '*restricted*' cases that allow (some) movement of the oxygens.

In the previous figures, conductances were normalized by the conductances obtained at zero $Ca^{2+}$ for a given case. Figure 2.7 was designed to show correlation between the conductances normalized in the two different ways described previously, if correlations existed. No correlations were found. Normalization of results from a model by the zero $Ca^{2+}$ conductance of that model shows the selectivity of that model. The lower this normalized value (shown on the abscissa), the more selective the model is for $Ca^{2+}$. On the other hand, if we normalize by one specific value—the zero $Ca^{2+}$ conductance for the '*flexible*' case—we can tell how the conductance of the model for $Na^+$ changes as we change the flexibility of the oxygen ions. The lack of correlation indicates that a given *fixed* oxygen configuration can favor selectivity and conductance independently. Conductance is sensitive to the presence of depletion zones, while selectivity is more sensitive to the degree of competition between $Ca^{2+}$ and $Na^+$.

The relative conductance in comparisons between different models depends primarily on the oxygen configuration. A configuration can be 'fortunate' (large conductance) so that $Na^+$ ions find enough space between the crowding oxygen ions. A configuration could also be 'unfortunate', meaning that the oxygen ions are in positions that act as obstacles for the passing $Na^+$ ions and produce deep depletion zones for them. Indeed, sometimes the location of oxygens create depletion zones where $Na^+$ concentration is zero, so the conductance is also zero. The '*flexible*' and '*restricted*' cases are statistical averages of the many '*fixed*' configurations in some sense or other. Therefore, these averaged cases do not suffer from the unfortunate configurations found in individual unaveraged cases: there are always enough configurations in the sample in the averaged cases in which oxygen ions move away and give way to the $Na^+$ ions. The average of the 10 '*low energy fixed*' cases is another example where the fortunate cases balance the unfortunate cases. Another interesting result shown in the figure is that the conductance of the '*restricted*' model is larger than that of the '*flexible*' model. The conductance is larger because the distribution of $Na^+$ ions is more spatially uniform for this model (see Figure 2.2). Therefore, the depletion zones of $Na^+$ are less deep in this case.

Figure 2.8 shows the concentration profiles for $O^{-1/2}$, $Na^+$, and $Ca^{2+}$ for our other model with different oxygen flexibility. When the oxygen ions are fixed ($R_{ox} = 0$ Å), the $Na^+$ and $Ca^{2+}$ ions have high peaks in the regions where oxygen ions are absent. As the $R_{ox}$ is increased, the peaks become lower, the valleys become less deep, and the curves eventually converge to those of the '*flexible*' model.

Figure 2.9 shows the conductance and occupancy of $Na^+$ as a function of $R_{ox}$. The conductances are normalized by the value at $R_{ox} = 0$ Å. There are two maxima in the conductance. It seems that conductance is larger in a model that is not perfectly '*flexible*', but not perfectly '*fixed*' either. In the '*fixed*' model, the oxygen ions act as obstacles as described above, especially if their configuration is 'unlucky'. In the '*flexible*' case, the oxygen ions act as obstacles because they pile up at the filter entrances thus producing depletion zones for the $Na^+$ ions.

There is a clear correlation between conductance and occupancy: less $Na^+$ means more $Na^+$-conductance. Intuitively, we might expect the opposite. The explanation is that although we have less $Na^+$ (smaller peaks), as $R_{ox}$ is increased, we also—in the same profile—have shallower depletion zones (see Figure 8).

## 2.4. <u>Discussion and Conclusions</u>

### 2.4.1. Some Flexibility of Side Chains is Required for Calcium Selectivity

We investigated the role of flexibility by comparing differential quantities and integral quantities in the selectivity filter of the 'flexible', 'fixed' and 'restricted' models. We compared the differential quantities, the distribution of the side chains (oxygens) and the cations ($Na^+$ and $Ca^{2+}$) and the integrated quantities, the occupancies of cations $Ca^{2+}$/ $Na^+$ and linearized conductance of $Na^+$.

The differential quantities that we obtain from our model show a strong dependence on the locations of fixed side chains and the flexibility of the side chains. We show that holding the side chains fixed at certain predetermined locations in the selectivity filter distorts the distribution of $Ca^{2+}$ and $Na^+$ in the selectivity filter. This distortion can result in a loss in selectivity because the distribution of the cations is determined by the distribution of the side chains (Figure 2.3C). This reasoning is further supported (1) by calculations (Figure 2.8) which vary the mobility of the side chains in the radial direction from frozen to perfectly mobile and (2) by calculations (Figure 2.3, A and B) of the 'restricted model' in which the side chains are allowed to have a limited flexibility along the channel axis. The results are similar to the usual 'flexible' case.

However, the behavior described by the integrated quantities (occupancy and normalized conductance) is much less sensitive, as might be expected from averaged, i.e., integrated quantities which are, as a rule, much less sensitive than differential quantities. Behavior of integrated quantities is similar in the 'restricted', 'fixed', and 'flexible' models. The occupancy of $Ca^{2+}$ increases with the increasing bath concentration of $Ca^{2+}$ (while occupancy of $Na^+$ decreasing) and the normalized conductance of $Na^+$ is reduced with increasing amount of $Ca^{2+}$ in bath. Thus, the integrated quantities obtained from our model are much less sensitive to details of structure. They do not show sensitivity to the flexibility and the locations of the side chains.

Our results show that some flexibility of side chains is necessary to avoid obstruction of the ionic pathway by oxygen ions in 'unlucky' fixed positions. When oxygen ions are mobile, they adjust 'automatically' to accommodate to the permeable cations in the selectivity filter.

Beyond the rigid (fixed oxygen) case, however, our results seem quite insensitive to how and what degree do we make the side chains flexible. Density profiles and selectivity (expressed in term of either occupancy or conductance) behave similarly in the different models of flexibility ("flexible", "restricted", and "confined").

The exact location and mobility of oxygen ions (let alone even finer details of structure) have little effect on the selectivity behavior of calcium channels. Seemingly, nature has chosen a robust mechanism to control selectivity in calcium channels: the first order determinant of selectivity is the volume of the selectivity filter with the charge and excluded volume of structural ions confined within it. Flexibility of side chains seem to belong to the group of second order determinants. These conclusions of course apply to what we study here. Flexibility and fine structural details may have important role in other properties of calcium channels that are not studied in this paper.

These results justify our early assumption—suggested by Nonner et al. (125) using theory and further studied with MC simulations (169, 132, 124)—that the important factor in Ca$^{2+}$ vs Na$^+$ selectivity is the density of oxygen ions in the selectivity filter. Later studies (41, 42) showed the importance of (charge) polarization (i.e., dielectric properties). The assumption of maximum mobility of oxygens ('flexible' case), seems to be an excellent approximate working hypothesis in the absence of exact structural information. We look forward to seeing how well the real structure fits within this hypothesis, when it becomes available.

### 2.4.2. Self-organized Induced Fit Model of Selectivity

Our results from the simulations of the zero flexibility model and the restricted flexibility model suggest that the reduced model is actually an induced fit model of selectivity, a specific version of the induced fit model of enzymes (170) in which biological function is controlled by the flexibility of the side chains that allows the side chains to self-organize into structures that change with changing ionic conditions.

The variation of binding with concentration and type of ions arise from different structures that self-organize under different ionic conditions. The fit of the protein side chains to the ions, and the fit of the ions to the protein, change with conditions. The different fit in different types of ions produces selectivity. The structures—in the sense defined in this paper—vary with concentration as well as type of ion. The energy of the structures varies with concentration and type of ion and we know of no simple theory to calculate this change in energy. Simulations are needed in a range of concentrations and types of solutions.

Calculations that characterize selectivity by a single free energy of binding do not address these issues. They also do not address the issue of how selectivity occurs in life or in experiments in which ions appear in mixed solutions of varying concentration.

Induced fit and self-organized models have traditionally been focused on the average structure of the protein. Here we view the locations of ions as part of the structure and we find that the distribution of locations ('flexibility', 'entropy') is also important. Monte Carlo methods used by Boda et al. (171, 169, 136, 172, 48, 141) seem ideally suited to make the qualitative idea of self-organized systems and induced fit of enzymes into a quantitatively specific (and testable) hypothesis of protein function. The self-organized/induced fit theory says that all relevant atoms are in an equilibrium distribution of positions and (perhaps) velocities. Monte Carlo methods are used to estimate such distributions in many areas of physics. These methods seem to be less sensitive to sampling errors than traditional forms of molecular dynamics for many reasons discussed at length in the literature (173, 174).

Selectivity in these channels arises from the interaction of the flexible side chains with the ions. The balance of the two main competing forces—electrostatic and excluded volume—in the crowded selectivity filter of the reduced model determines the binding site for $Ca^{2+}$. The structure of the binding site is an output of the calculations which rearranges according to the surrounding ionic conditions. The word 'structure' is somewhat inadequate to describe what is happening here. The thermal motions of the structure are as important as the average location. The biologically important properties depend on the entire ensemble of trajectories of ions and side chains. The distributions of location and velocities are involved. Traditional self-organized/induced fit models need to be generalized to include the self-organized/induced entropy (i.e., flexibility) as well as the self-organized/induced energy (i.e., location).

The success of the self-organized induced fit model of selectivity arises because it calculates structures instead of assuming them. Assuming a preformed structure, independent of conditions, distorts the model significantly. Evidently, assuming a constant structure involves applying an artificial constraint not present in real channels. We suspect, but have not proven, that the difficulty is fundamentally similar to that which arises when a protein is described by a potential surface independent of conditions instead of as a distribution of permanent and dielectric charge (149, 150).

Monte Carlo methods developed by Boda et al. seem ideally suited to compute the equilibrium properties of these self-organized systems. Other methods are needed to extend to the general non-equilibrium conditions in which most channels and proteins function, for example, variational methods like *EnVarA* (165, 168) or PNP-DFT (166, 167, 52).

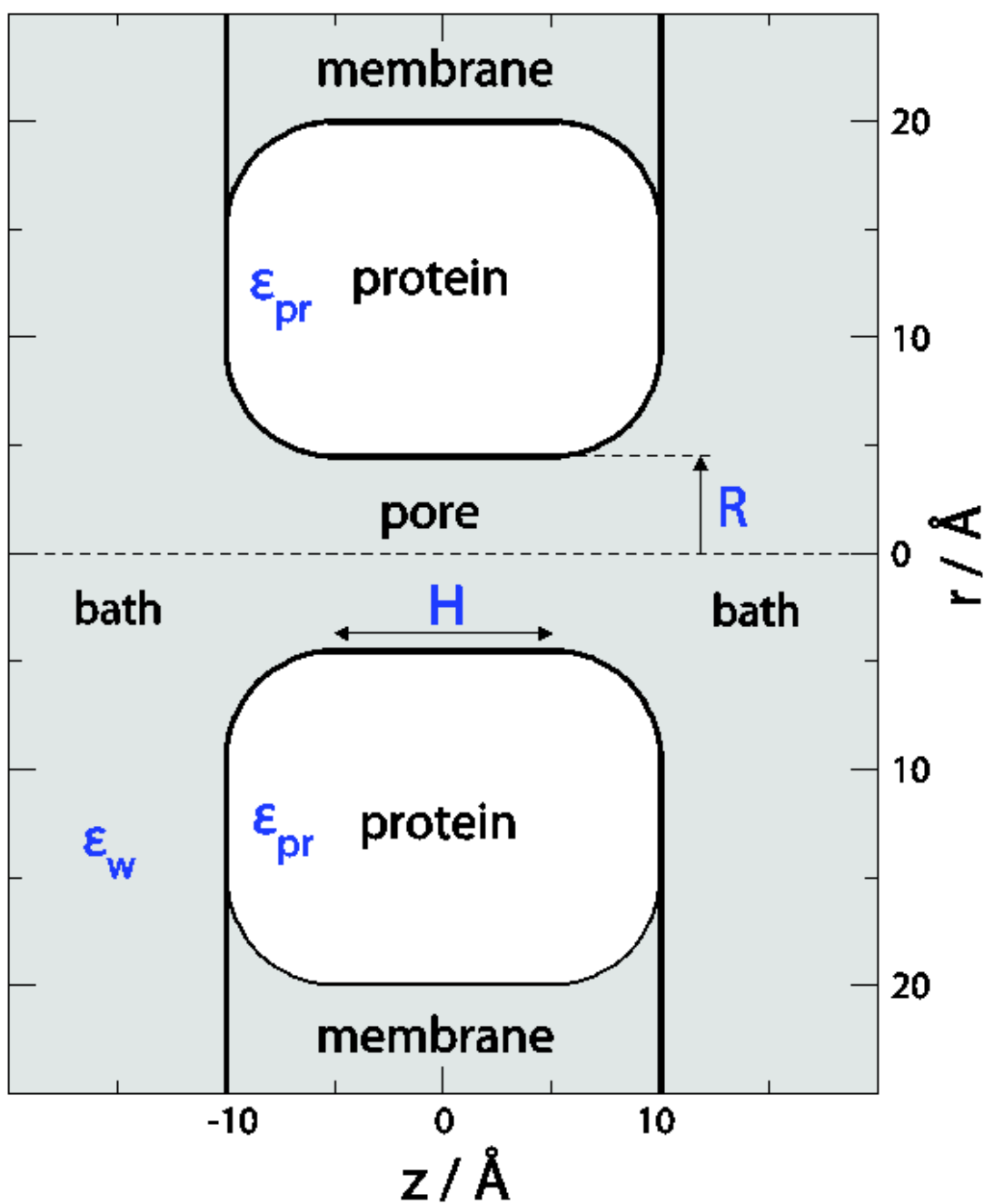

Figure 2.1: Geometry of the model of the ion channel. The parameters R = 3.5 Å, H = 10 Å, and $\varepsilon_{pr}$ = 10 are used in the simulations in this paper.

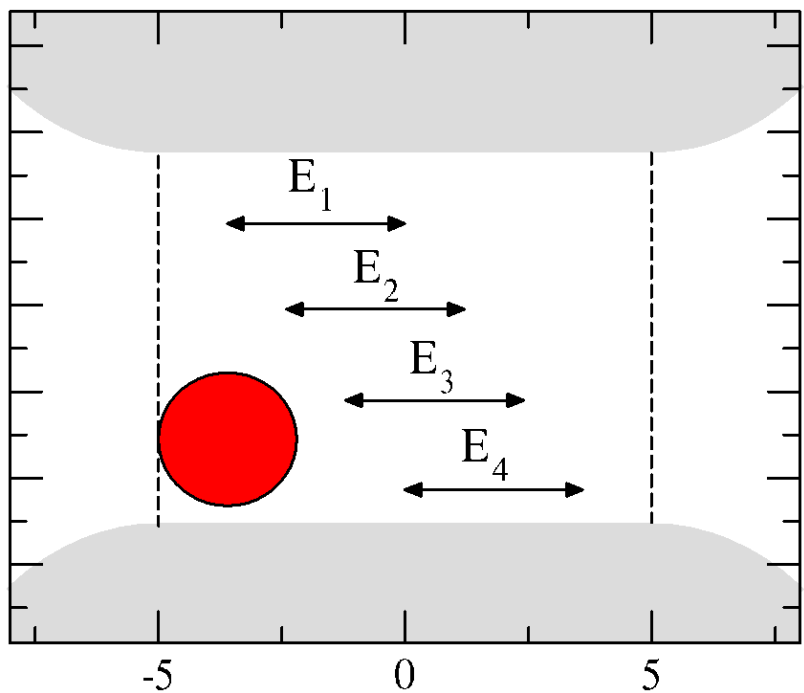

Figure 2.2: In the 'restricted' case, four pairs of oxygen ions are restricted to four overlapping regions in the selectivity filter along the z-axis of the pore. The ion centers are restricted to the following intervals: $E_1$: [-3.6, 0], $E_2$: [-2.4, 1.2], $E_3$: [-1.2, 2.4], and $E_4$: [0, 3.6].

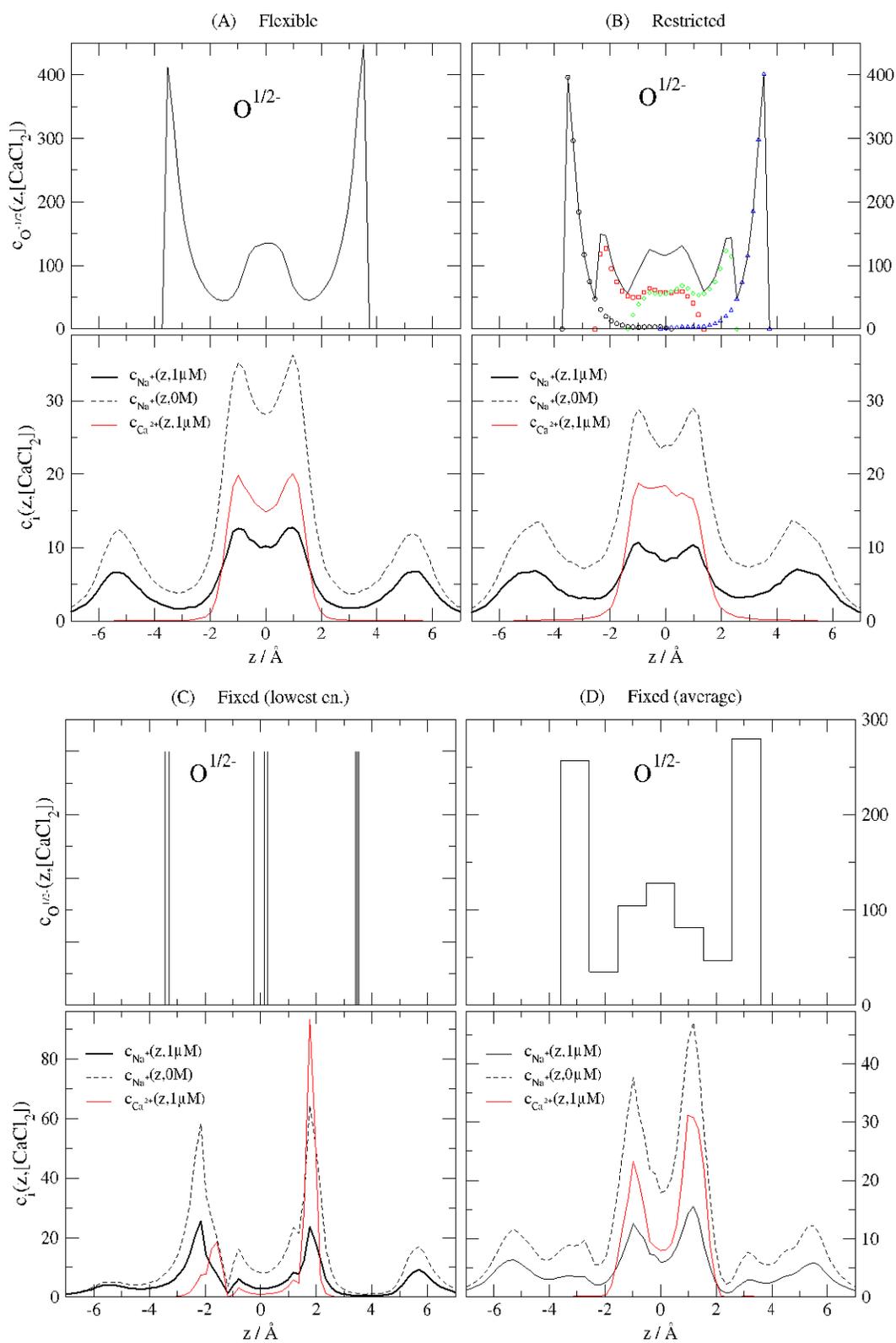

Figure 2.3: Concentration profiles for oxygen (top row) and free ions (bottom row) for three different restrictions of the oxygens. (A) 'Flexible' case: the oxygen ions are fully mobile inside the selectivity filter, but cannot leave it. (B) 'Restricted' case: four pairs of oxygens are restricted to four overlapping regions inside the selectivity filter as shown in Figure 2.2. Profiles with different symbols and colors in the top panel of Figure 2.3B refer to these four oxygen pairs. The solid black line is the sum of these four profiles. (C) 'Fixed' case: the eight oxygens are fixed in positions that correspond to the lowest-energy configuration of a finite sample. The vertical lines in the top panel of Figure 2.3C represent Dirac deltas corresponding to these fixed positions. In the bottom row, we show the profiles for $Na^+$ at $[CaCl_2] = 0$ M (dashed black line) and $[CaCl_2] = 10^{-6}$ M (thick solid black line). The thin solid red line represent the $Ca^{2+}$ profiles for $[CaCl_2] = 10^{-6}$ M. (D) 'Average' case: is similar to Figure 2.3C, but represents the averages of the ten low energy configurations.

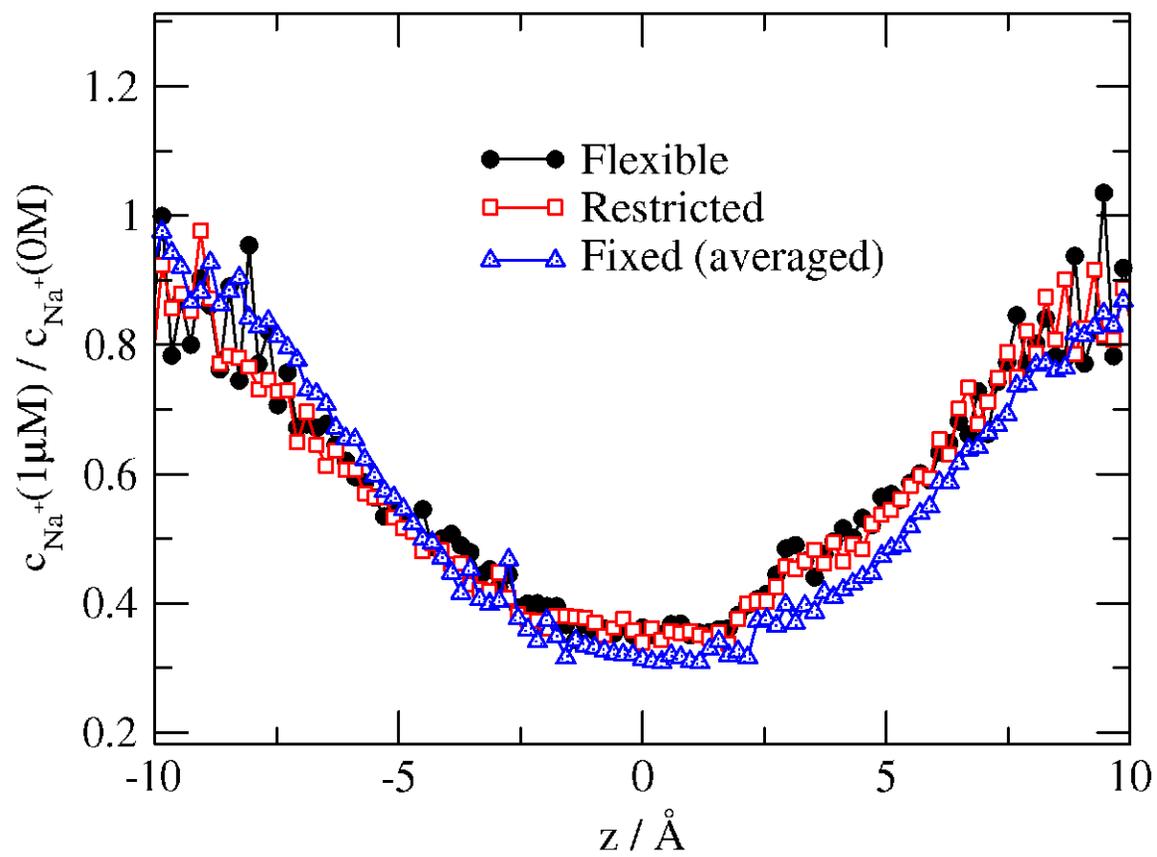

Figure 2.4: The ratio of the $Na^+$ concentration profiles for $[CaCl_2] = 10^{-6}$ M and 0 M for the three cases depicted in Figure 2.3.

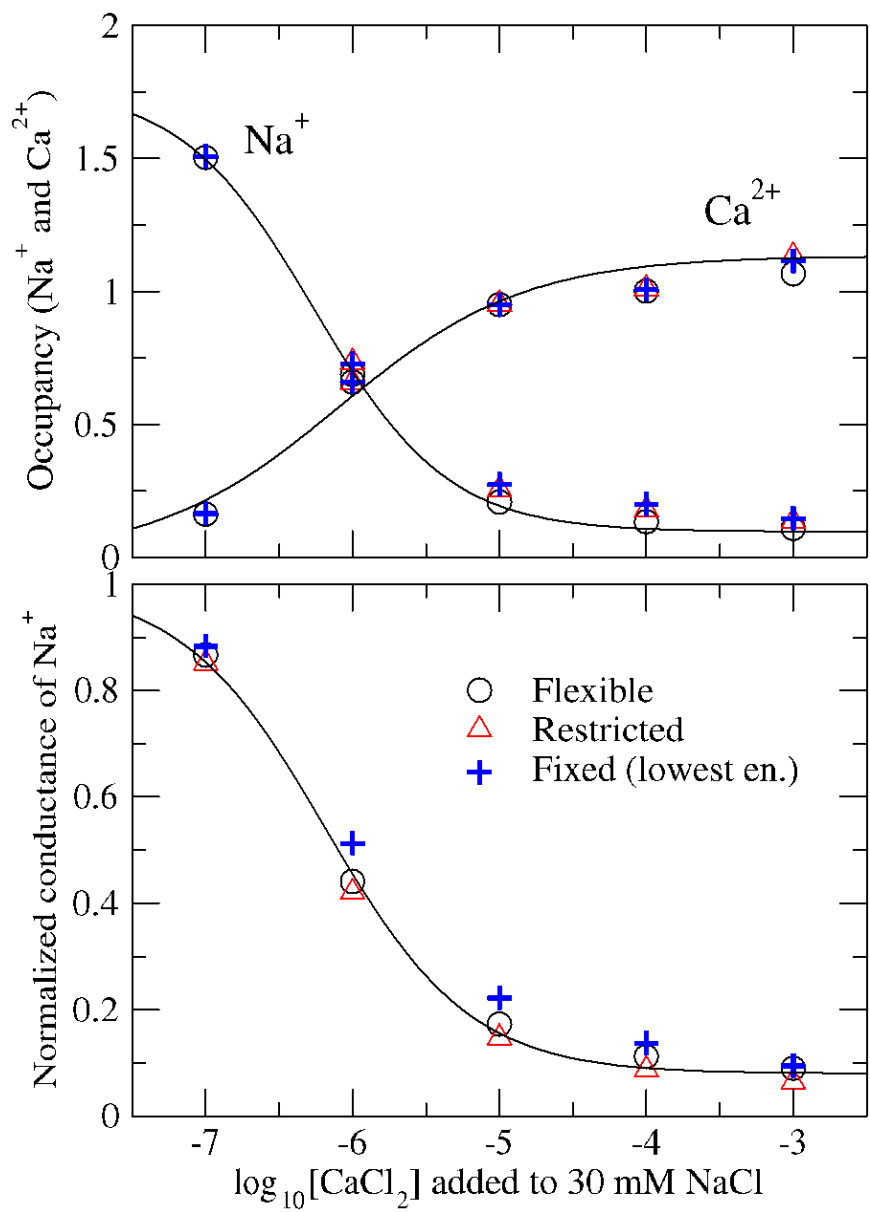

Figure 2.5: (A) The average number (occupancy) of $Na^+$ (decreasing curves) and $Ca^{2+}$ (increasing curves) ions as functions of $\log_{10}[CaCl_2]$ for the three different cases depicted in Figure 2.3. (B) The conductance of Na+ ions normalized by the conductance at $[CaCl_2] = 0$ M as a function of $\log_{10}[CaCl_2]$ for the three different cases.

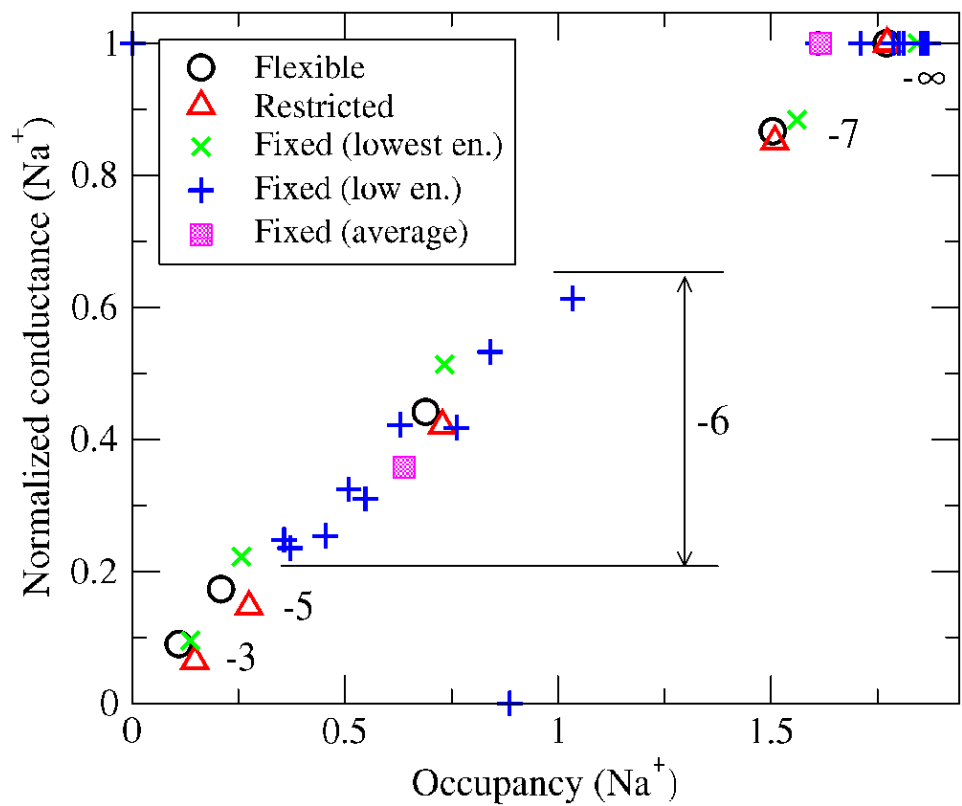

Figure 2.6: Normalized conductance of Na$^+$ versus occupancy of Na$^+$ in the selectivity filter at different CaCl$_2$ concentrations for the three different cases depicted in Figure 2.3. The green $\times$ symbols represent results for the lowest-energy 'fixed' oxygen configuration used in Figure 2.3. The blue + symbols represent results for 10 selected 'fixed' oxygen configurations with low energies. The pink $\square$ symbols represent averaged results for 10 selected 'fixed' oxygen configurations with low energies. The numbers near the symbol denote the values of log$_{10}$ [CaCl$_2$]. Na$^+$-conductance for a given case is normalized by the Na$^+$-conductance for that particular case in the absence of Ca$^{2+}$.

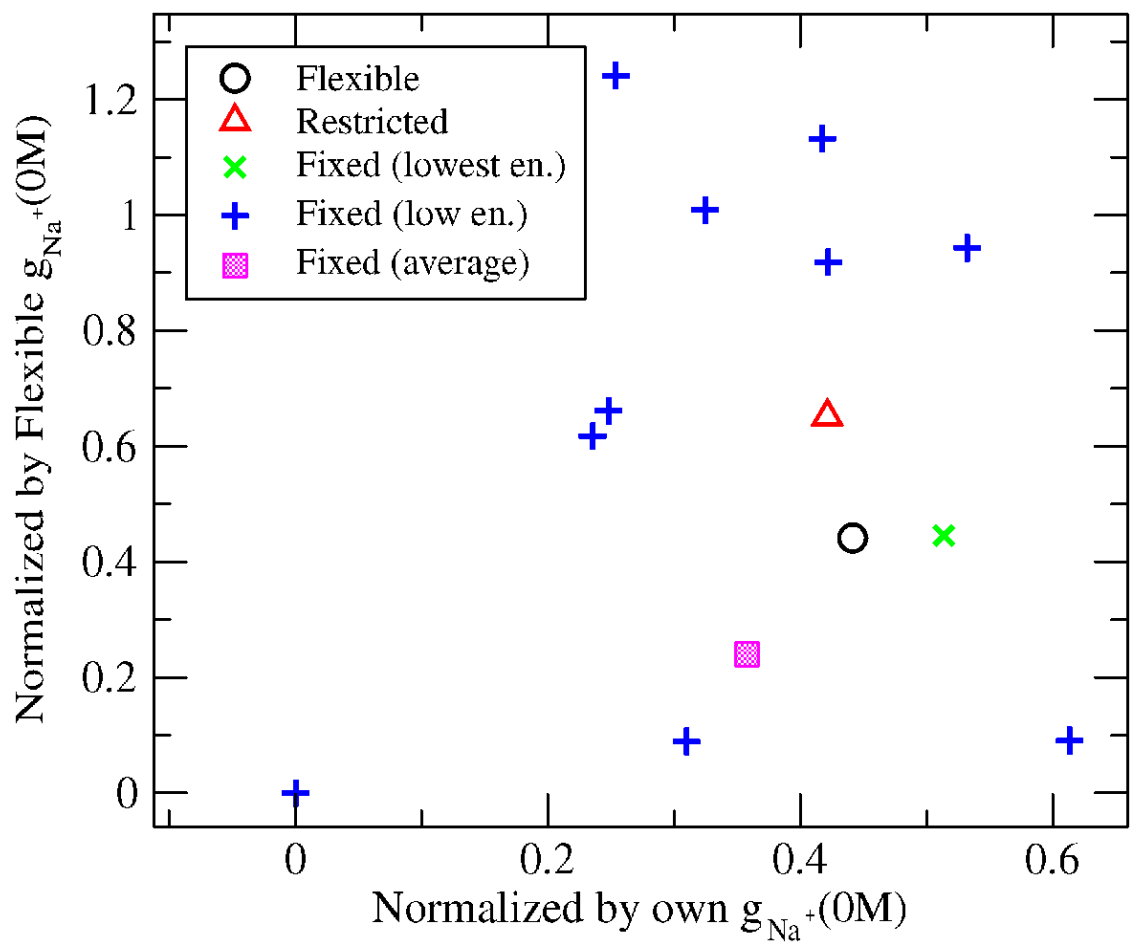

Figure 2.7: Correlation between $Na^+$ conductances normalized by two different ways for $[CaCl_2] = 10^{-6}$ M. The ordinate shows conductance of $Na^+$ normalized by the $Na^+$ conductance in the absence of $Ca^{2+}$ for the 'flexible' case. The abscissa shows conductance of $Na^+$ for a given case normalized by the $Na^+$ conductance in the absence of $Ca^{2+}$ for that particular case. Symbols have the same meaning as in Figure 2.6.

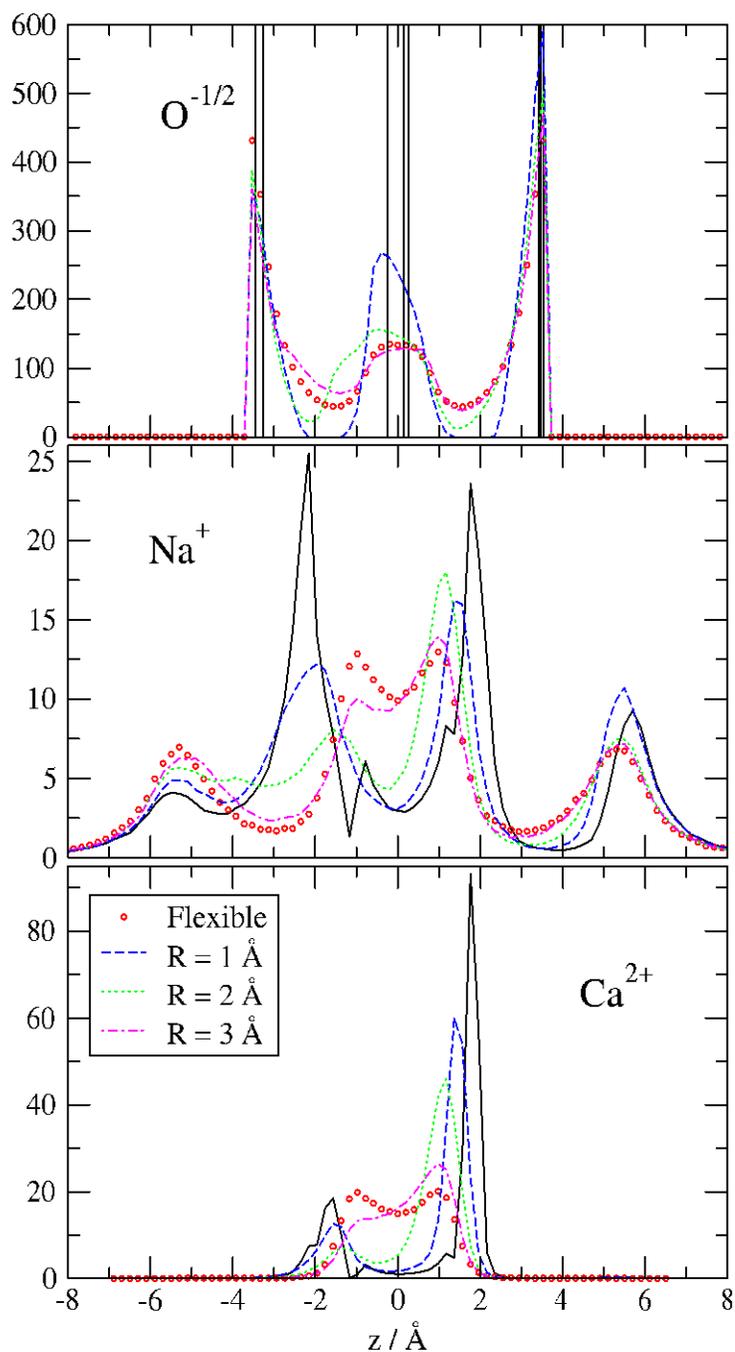

Figure 2.8: Concentration profiles of $O^{-1/2}$, $Na^+$, and $Ca^{2+}$ for the 'flexible' (red circles) and the lowest-energy 'fixed' (solid black line) cases, as well as cases when the oxygens are restricted in spheres of radii $R_{ox}$ centered around the 'fixed' positions (various colored non-solid curves) at $[CaCl_2] = 10^{-6}$ M.

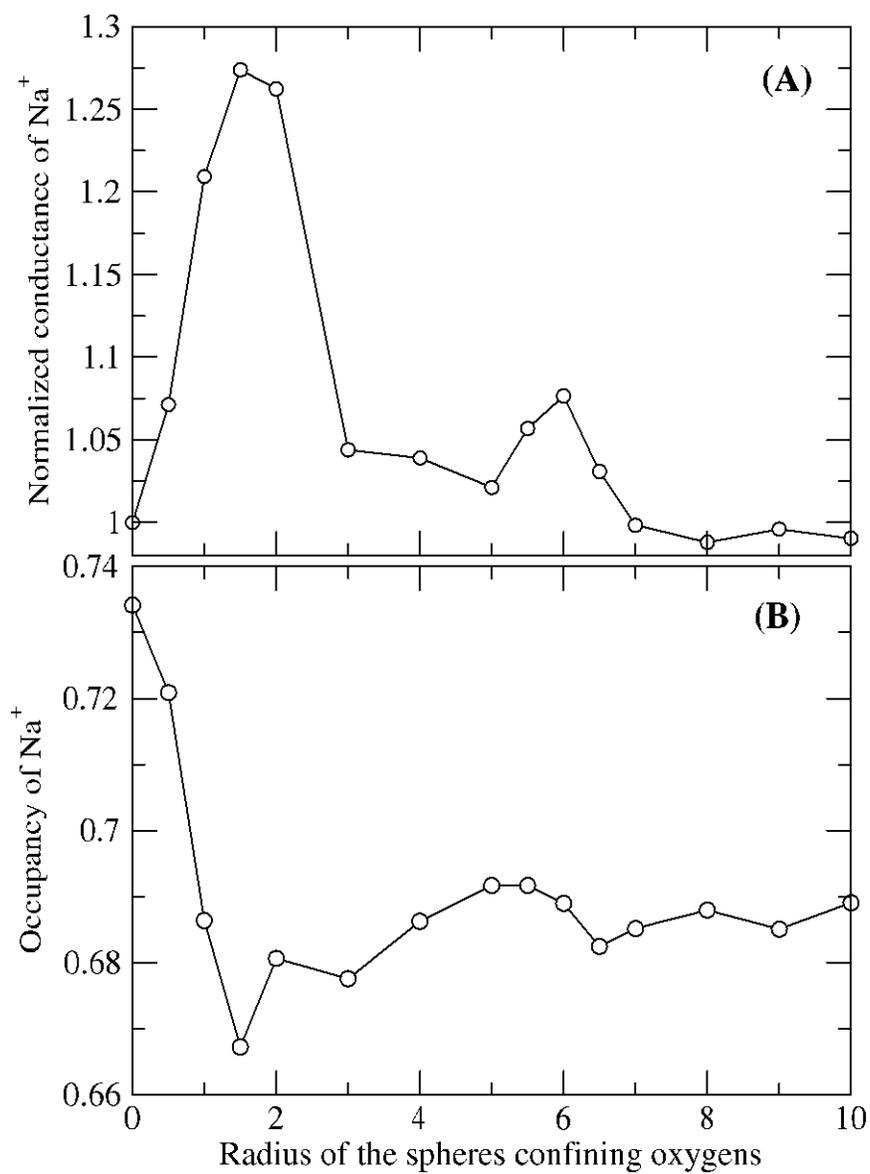

Figure 2.9: (A) Normalized conductance and (B) occupancy of Na+ ions as functions of the spheres confining the oxygens $R_{ox}$ for $[CaCl_2] = 10^{-6}$ M. The conductance is normalized by the 'fixed' ($R_{ox} = 0$ Å) case.

# CHAPTER 3. PRINCIPLES AND PRACTICE OF SINGLE CHANNEL RECORDING

## 3.1. <u>Introduction</u>

The mechanisms underlying the biophysical properties of ion channel function can be understood by measuring their function at the molecular level. Patch-clamp and planar lipid bilayer techniques of electrophysiology are used extensively for measuring the function of ion channels at the single protein-level (175-179). The patch-clamp technique measures current flowing through single (or multiple) ion channel in intact cell membranes (180, 181) whereas the planar lipid bilayer method measures the current flowing through a single ion channel in an artificial lipid membrane (182). These methods are widely used for single channel measurements (183, 184) and have their own advantages as well as disadvantages.

In this chapter, the advantages/disadvantages of the patch-clamp and the planar lipid bilayer method are discussed. The main components of the bilayer set-up used for recording electrical measurements of sub-picoamp range with lowest noise are described. The underlying principles of bilayer experiments that have made possible the measurement of the function of a single channel protein in a controlled environment are explained. The critical issue of voltage offsets in bilayer experiments is addressed. The sources of voltage offsets and the practical approaches followed for handling these offsets are described.

## 3.2.  Patch Clamp vs. Planar Lipid Bilayer Method

The major advantage of the patch-clamp method is its higher electrical resolution. The signal to noise ratio is much larger in the patch-clamp method where the area of the cell membrane—i.e., the 'patch' containing the proteins—is much smaller than the (macroscopic) area of a lipid bilayer. Noise varies (approximately) as the capacitance in parallel with the channel and capacitance is proportional to area (185-191). The noise advantage of the patch clamp comes at a price, however lipids of the cell membrane patches may flow into the pipette and exhibit voltage dependence/independence thereby modifying the functional properties of many channels (192, 193) and membrane composition. The density of channels and the associated specific accessory proteins varies from patch to patch creating unbiological variability. Another practical issue with the patch-clamp method is the difficulty in keeping the ionic conditions of the cell interior comfortable for the cells so channels can survive. Only specialized apparatus allows patch solutions to be changed while maintaining low noise (194, 195).

In terms of these issues, the planar lipid bilayer method appears to be more useful compared to the patch-clamp method as it provides a better experimental control. Experiments can be performed with the planar lipid bilayer method in ionic conditions that are not possible to achieve while performing patch-clamp recordings on intact cells. In the planar lipid bilayer method, the ionic environments present on both sides of the bilayer can be controlled simultaneously. If required, it is also possible to modify the lipid environment in the planar lipid bilayer method. Note that in patch-clamp experiments, protein-protein interactions in the membrane may often remain intact affecting the measurements. Therefore, in such situations, the planar lipid bilayer method seems to be an appropriate choice. The planar lipid bilayer method seems to work best for the functional investigation of channel proteins that can be purified from their native cell membranes or from cells that heterologously express the channel proteins.

### 3.3. Planar Lipid Bilayer Set-Up for Single Channel Measurements

The planar lipid bilayer set-up shown in Figure 3.1 is uniquely designed to carry out single channel measurements in a planar lipid bilayer with great ease, high efficiency and quality. It is a smart integrated engineered device making possible the measurement of single channel currents in pA range with lowest noise. The main components of a complete bilayer set-up are categorized as mechanical and electrical.

### 3.4. Mechanical Components

The mechanical components include the air table, Faraday cage, experimental chambers, perfusion system and stirring devices.

### 3.4.1. Vibration Isolation Table

The lipid bilayer is very fragile and is extremely sensitive to mechanical disturbances and vibrations. It is therefore very important to isolate and shield the lipid bilayer from the mechanical interference by placing the components of the set-up on a vibration isolation table. In the commercially available anti-vibration table or 'air table' air cushions are used in the table legs supporting a very heavy table top. These air cushions act as shock absorbers damping the mechanical vibrations. The table top is kept afloat by a gas source (nitrogen tank or air pump). These air tables are therefore portable, sturdy, and stable. The air tables and the table tops are available commercially in a variety of sizes.

### 3.4.2. Faraday Cage

Faraday cage —a hollow metal box— is used to enclose the bilayer chamber/cup, the electrodes, and the headstage of the Axopatch $^{©}$ amplifier (Molecular Devices, Inc.) in the bilayer set-up. The purpose of Faraday cage is to provide a shielding to the enclosed components carrying out pA current measurements from the influence of external electromagnetic field. When an external electric field acts on the walls of the Faraday cage, the charge carriers within the conductor experience an electric force thus, generating a current that rearranges the charges in the metal cage. Once the charges are rearranged, it cancels the effect of the external fields inside the metal cage and the current stops. Shielding by a Faraday cage requires connection to ground. If the cage is 'floating' (*i.e.*, infinite impedance to ground) it does not shield.

### 3.4.3. Experimental chamber, perfusion system and stirring devices

The experimental chamber is a crucial component of the bilayer set-up. It consists of two sections or compartments that are filled with aqueous solutions separated by an artificial membrane. The bilayer chambers are commercially available in various designs. The chamber is chosen on the basis of the technique (painted (182) or folded(196)) used to form the bilayers. In our set-up we use the painted bilayer technique. The chamber and the stir bars in our set-up are purchased from Warner Instruments (197). The design of the chamber is such that when equal volumes of aqueous solutions are added in the front and the rear side, the height of the solutions is balanced which further, minimizes the mechanical gradients across the bilayer.

The chamber consists of two parts: the basic block and a cup or cuvette. The basic block is made from black Delrin (197). There are two wells drilled into the block. The well in the rear fits the cup. There is a slight depression (~ 0.5 mm) in the front well or the non-cup side in order to confine the motion of magnetic stir bar (coated with Teflon) and to reduce mechanical noise artifacts. The volume of the front well is ~3-4 ml. Ports are drilled into the side walls of the block for holding the pipette tips filled with Agar and 3 M KCl acting as salt bridges for the electrodes.

The cup or the cuvette is made from Delrin consisting of a 150 μm diameter hole in the front wall and can hold ~ 3-4 ml of aqueous solution. A lipid bilayer is painted across the hole. The base of the cup consists of a 0.5 mm well that confines the motion of the Teflon coated magnetic stir bar and reduces the mechanical noise artifacts. A viewing window made from quartz or glass is fitted in the front well for visualizing the cuvette hole under a lamp placed outside the Faraday cage while painting the bilayer.

The Teflon coated magnetic stir bars that we use in our set-up for stirring solutions in the chamber are driven by batteries in order to avoid alternating current interference that is likely to be present even with the most shielded power supply. The stirring is turned off while recording the measurements to prevent mechanical artifacts.

One of the major advantages of the planar lipid bilayer technique is the ease with the solutions can be exchanged while the experiment is carried out. However, the efficiency of the solution exchange technique highly depends on the choice of the perfusion system. An efficient perfusion system allows changing solutions present on both sides of the bilayer with minimum noise. A perfusion system especially designed to maintain low noise while the solution exchange is carried out without breaking the bilayer (194, 195) is integrated in our set-up.

## 3.5.  Electrical Components

The primary electrical components of the bilayer set-up includes the silver-silver chloride (Ag/AgCl) electrodes with Agar-salt bridges and the headstage of the Axopatch© 200 B amplifier (Molecular Devices, Inc.).

### 3.5.1. Ag/AgCl electrodes: Design and Operation Principle

The Ag/AgCl electrode is made of a silver (Ag) wire with a coating of silver chloride (AgCl). These electrodes are the most commonly used reference electrodes in biomedical applications. The usage of Ag/AgCl electrodes is popular because they are simple to design, inexpensive, exhibit relatively low electrode potential and are non-polarizable *i.e.* their electrode potential is quite stable and it does not drift upon passage of current.

When electric current passes through the electrode a redox reaction occurs between the Ag and AgCl described as follows:

$$Ag \Leftrightarrow Ag^+ + e^-$$
$$Ag^+ + Cl^- \Leftrightarrow AgCl \downarrow \text{ (precipitate).}$$

(6)

Under equilibrium conditions, the solubility product $K_s$ determining the rate of precipitation of silver chloride is the product of $Cl^-$ and $Ag^+$ activities such that,

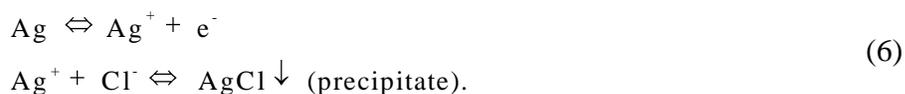

$$K_s = a_{Ag^+} \times a_{Cl^-} .$$

(7)

The electrode potential arising because of this fast occurring redox reaction in Eq. (6) is described approximately by the classical Nernst equation of electrochemistry (198) as:

$$V_{Ag/AgCl} = V_0 + \left(\frac{RT}{F}\right)\ln(a_{Ag^+}). \tag{8}$$

From Eq. (7),

$$V_{Ag/AgCl} = V_0 + \frac{RT}{F}\ln\left(\frac{K_s}{a_{Cl^-}}\right), \tag{9}$$

or

$$V_{Ag/AgCl} = V_0 + \frac{RT}{F}\ln(K_s) - \frac{RT}{F}\ln(a_{Cl^-}), \tag{10}$$

where, $V_{Ag/AgCl}$ is the electrode potential with respect to the electrolyte solution, $V_0$ refers to the standard electrode potential for the chemical reaction (a constant), $K_s$ is the solubility product (a constant), $R$ is the gas constant, $T$ is the absolute temperature, $F$ is the Faraday constant and $a_{Cl^-}$ is the activity of chloride ions in the electrolyte solution (varies with the concentration of $Cl^-$). The first and second terms in Eq. (10) are constants and only the third term is dependent on ionic activity. The electrode potential of the Ag/AgCl electrode is in fact a chemical term that does not fluctuate due to the passing current because Ag oxidation does not contribute considerably to the electrode potential. The activity of $Cl^-$ (see Eq. 10) is the major contributor since the activity of $Ag^+$ is relatively lower and of the same order of magnitude as the solubility product for AgCl, Ks ~ $10^{-10}$. Therefore, the Ag/AgCl electrode potential is fairly stable when current is passed provided the activity of $Cl^-$ remains stable.

In order to minimize and stabilize the electrode potential (*i.e.* activity of $Cl^-$), the Ag/AgCl electrodes are preferably used with salt bridges that contain electrolyte solution with higher concentration of $Cl^-$ ions. For example, in our bilayer set-up, the electrical connectivity between the contents (salt solutions) of the experimental chamber and the amplifier is attained through a pair of Ag/AgCl electrodes that are connected via Agar-3M KCl salt bridges where one of the electrodes is connected to the amplifier (the *voltage* side) and the other one to the ground (the *ground* side) as shown in Figure 3.1.

### 3.5.2. Axopatch 200 B Capacitor Feedback Patch Clamp Amplifier

The Axopatch 200 B capacitor feedback patch clamp amplifier (199) is an integral component of a bilayer set-up making possible the measurement of sub-picoamp current signals with high efficiency, speed, and ultra low noise. The Axopatch 200 B amplifier is the most widely used patch-clamp amplifier around the world for conducting single channel measurements. The Axopatch 200 B amplifier is based on the novel capacitor-feedback technology that enables us to carry out quiet single channel measurements. The Axopatch 200 B amplifier unit includes a integrating headstage that operates in both capacitive as well as resistive feedback mode and a module or a 'box' with a built in circuitry for conducting various pre/post signal processing operations for example, gain adjustment, offset control, modes of operation (patch or whole cell etc.). The details of the various controls and related settings can be found in the Axopatch manual (199). In the bilayer set-up, the bilayer chamber is electrically connected to the headstage of the amplifier via the Ag/AgCl electrodes. The headstage has two ends, the front end is for the *ground* and the rear end referred as the *voltage* is connected to the Axopatch 200 B amplifier module. The principle of operation and calibration of the integrating headstage will be described in more detail.

### 3.5.2.1. Principle and Calibration of Axopatch 200 B Amplifier Integrating Headstage

The Axopatch 200B patch clamp amplifier integrating headstage is essentially a current (I) to voltage (V) converter; the output voltage is proportional to the current input. The integrating capacitive feedback mode of the headstage enables measurements of sub-picoamp current signals at lowest noise because a capacitor can be nearly an ideal element thus, suitable for single-channel current measurements.

In the integrating headstage, the current is measured as the rate of change of the voltage across a capacitor, $C_f$ shown in Figure 3.2. The headstage measures the integral of the current which is subsequently differentiated to allow measurements of the current itself. The capacitive feedback allows a substantial reduction of noise and exhibits better linearity compared to the resistive feedback headstage (not shown here). This noise reduction is particularly significant in the frequency band of interest for single- channel recordings (10 Hz-10 kHz).

However, it is worth taking into account that the voltage across the capacitor cannot ramp in one direction always and at some point the capacitor voltage will approach the supply limits. In such a situation the integrator needs to be reset to start again near zero volts. The reset process takes place for the integrator and the differentiator as well interrupting the ongoing current measurements. The frequency of these resets depends on the current amplitude passed through the headstage. Therefore, larger current will require more frequency resets. However, this issue is not encountered while conducting single-channel measurements because the average current amplitude is not more than a few pico-amperes and therefore, the frequency resets are rare.

### 3.5.2.2. Transfer Function of Axopatch 200 B Amplifier Integrating Headstage

The transfer function is a mathematical representation of the relation between the input and output of a system. It describes how the input is transferred to the output as,

$$\text{Transfer Function}(\omega) = \frac{\text{Output }(\omega)}{\text{Input }(\omega)},$$

where, $\omega$ is the angular frequency.

(a) Transfer function of Integrator

$$\frac{V_{out}}{V_{in}} = \frac{-I_f Z_f}{I_p R_p} = \frac{-Z_f}{R_p} \ (\because \ I_f = I_p) \tag{11}$$

where, $V_{out}$ is the output voltage of the integrator circuit at **A**, $V_{in}$ is the input voltage, $I_f$ is the current through the feedback capacitor $C_f$ with impudence $Z_f$, $I_p$ is the current through the pipette with resistance $R_p$. For lipid bilayer experiments, $I_p$ gets replaced by $I_m$ (membrane current) and $R_p$ by $R_m$ (membrane resistance).

(b) Transfer function of Differentiator

$$\frac{V_{out}^{'}}{V_{in}^{'}} = \frac{-I_R R_f}{I_o Z_o} = \frac{-R_f}{Z_o} \ (\because \ I_o = I_R), \tag{12}$$

where, $V_{out}^{'}$ is the output voltage of the differentiator circuit at **B**, $V_{in}^{'}$ is the input voltage, $I_R$ is the current through the feedback resistor $R_f$, $I_o$ is the current through the capacitor $C_o$ with impedance $Z_o$.

(c) Transfer function of I-V converter

Multiply Eq. (11) by (12)

$$\frac{V_{out}^{I-V}}{V_{in}^{I-V}} = \frac{-Z_f}{R_p} \times \frac{-R_f}{Z_o} = \frac{j\omega C_o R_f}{j\omega C_f R_p} = \frac{R_f}{R_p} \ (\because \ C_o = C_f) \ , \tag{13}$$

where, $V_{out}^{I-V}$ is the output voltage of the I-V converter and $V_{in}^{I-V}$ is the input voltage.

$$V_{out}^{I-V} = V_{in}^{I-V} \times \left( \frac{R_f}{R_p} \right) = V_{cmd} \times \left( \frac{R_f}{R_p} \right) \tag{14}$$

Therefore, after scaling and offset corrections of $V_{out}^{I-V}$ from Eq. (14), the current $I$ is measured in response to the applied voltage at the input of the I-V converter $V_{cmd}$.

### 3.6. <u>Voltage Offsets in Single Channel Planar Lipid Bilayer Experiments</u>

The measured membrane voltage ($V_m$) in planar lipid bilayer experiments is corrected for offset voltages ($\Delta V^{offset}$) such that,

$$V_m^{corrected} = V_m^{measured} + \Delta V^{offset} = V_m^{measured} \pm V_{amplifier}^{offset} \pm V_{electrodes}^{offset} \pm V_{solutions}^{offset}.$$

The primary sources of offset voltages include the amplifier, the silver-silver chloride electrodes, and the salt solutions in the experimental chamber. The typical magnitudes for amplifier offsets is $\pm$ 30 mV, up to 100 mV for electrode offsets (depending on Cl⁻ concentrations), and up to $\pm$ 15 mV for liquid junction potentials at interfaces between different solutions.

### 3.6.1. Source 1: Axopatch 200 B Amplifier

The current-voltage converter based amplifier in the bilayer set-up could contribute a finite offset to the measured voltage when it draws a finite current at its input in response to the applied voltage such that at 0 mV the current is not zero. The amplifier is tested by calibrating it against a dummy circuit with known components.

### 3.6.1.1. Calibration of Axopatch 200 B Amplifier Integrating Headstage

The bilayer set-up is calibrated with an *electric circuit module* before beginning the single channel experiments. The goal is to test and verify the functioning of the integrating headstage of the Axopatch 200 B amplifier in the set-up. The electric circuit module also referred as the dummy circuit is designed such that it represents the electrical equivalent of a lipid bilayer surrounded by salt solutions before a channel protein is added for reconstitutition.

The dummy circuit consists of a parallel combination of a capacitor ($C_m=80 \ pF$) and resistor ($R_m \sim 10 \ G\Omega$) representing the electrical equivalent of a lipid bilayer membrane connected in series with a resistor ($R_e= 100 \ \Omega$) representing the resistance of the electrolytes in baths. The values of resistors and capacitor for the electrical equivalent representation were based on experimental observations. The dummy circuit is connected to the headstage of the Axopatch 200 B amplifier set-up such that one end of the circuit module is connected to the input terminal of the headstage and its other end is connected to the ground terminal of the headstage as shown in Figure 3.3.

For a given input (step/ramp voltage), the output current generated from the module is recorded. The measured current-voltage data is further analyzed for estimating the membrane resistance ($R_m$) and the time constant ($\tau$) of the bilayer's capacitive response. The value of $R_m$ is determined from the inverse of the slope of the measured I-V curve and compared to its known value in the dummy circuit. The time constant $\tau$ calculated from the dummy circuit's response to voltage (step/ramp) is compared to the experimental current-voltage response for verification of the capacitive response of the bilayer membrane.

If all the components of the set-up are configured, the measured values of the current, membrane resistance, and capacitive time constant should match with their known values. This is how the Axopatch 200 B amplifier headstage is calibrated. In case the output current, the passive component values (resistor (s) and capacitor) do not match with their expected values then the set-up needs to be investigated systematically for finding out the faulty source.

For calibration of the Axopatch 200 B amplifier headstage the dummy circuit is connected to the headstage and the output current is measured for an input step voltage. The accessory settings on the Axopatch 200 B amplifier box (pipette offset adjustment, gain, filter frequency and sampling frequency) are kept same as for a real experiment and are not changed during the measurements.

The current flowing in the dummy circuit on application of a step voltage is determined. The capacitive time constant is also estimated.

The dummy circuit in Figure 3.4 is connected and a step voltage is applied representing a typical voltage-clamp experiment where the current is measured in response to a fixed applied voltage *i.e.* step voltage $V_s$, such that at time $t \geq 0$, $V_{in} = V_s$.

For the circuit shown in Figure 3.4, let $I$ denote the current, $V_{in}$ is the applied input voltage, and $Z$ the impedance. The current $I$ in the circuit is,

$$I = \frac{V_{in}}{Z} = \frac{V_{in}}{R_b + \dfrac{R_m}{1 + j\omega C_m R_m}}.$$
(15)

The current $I$ determined in Eq. (15) contains a complex impedance term $Z$. The further analysis is therefore, carried out in the Laplace domain for unit step voltage $V_{in}(s) = \dfrac{1}{s}$, the current $I(s)$ in the circuit is given as,

$$I(s) = \frac{1}{s} \frac{1}{Z(s)} = \frac{1}{s} \frac{1}{R_b + \dfrac{R_m}{1 + s C_m R_m}},$$
(16)

Inverse Laplace transform of $I(s)$ in Eq. (16) gives $I(t)$ as,

$$I(t) = \frac{1}{R_b + R_m} + \frac{R_m}{R_b(R_m + R_b)} e^{-\frac{t}{R_{eq}.C_m}},$$
(17)

such that the time constant $\tau = R_{eq} C_m$ where, $R_{eq} = \dfrac{R_b R_m}{R_m + R_b}$.

Therefore, for unit step voltage $V(t) = 1\,\mathrm{mV}$, at $t = 0$ the current is

$$I(t = 0) = \frac{1}{R_b} = \frac{1\,\mathrm{mV}}{100\,\Omega} = 0.01\,\mathrm{mA}, \tag{18}$$

and at $t = \infty$, the current will be

$$I(t = \infty) = \frac{1}{R_m + R_b} \sim \frac{1}{R_m} \because R_m >> R_b, \tag{19}$$

*i.e.*

$$I(t = \infty) = \frac{1\,\mathrm{mV}}{10\,\mathrm{G}\Omega} = 0.01\,\mathrm{pA}, \tag{20}$$

and the time constant therefore should be $\tau = R_{eq} C_m \sim 100\,\Omega \times 80\,\mathrm{pF} = 0.8\,\mathrm{nsec}$ ( $\because R_m >> R_b$ ).

### 3.6.1.2. Experiment – Calibration of Axopatch 200 B amplifier headstage using dummy circuit module

The dummy circuit is connected to the headstage of the Axopatch 200 B amplifier as shown in Figure 3.3. The control settings on the amplifier panel are adjusted similar to the usual single channel bilayer experiment. Output current from the amplifier is measured at 0 mV as well as in response to ramp voltage (from -100 mV to + 100 mV), Figure 3.5 and Figure 3.6 and step voltages (+/-100 mV) (Figure 3.8, Figure 3.9, Figure 3.10, and Figure 3.11). The slope conductance is measured from the ramp current (Figure 3.6) recorded in response to the ramp voltage protocol as shown in Figure 3.5. The headstage is calibrated when the measured current is 0 pA at 0 mV (Figure 3.7, Figure 3.9, and Figure 3.11) and the slope conductance estimated from the I-V curve *i.e.* the membrane conductance ($G_m$) as shown in Figure 3.7 is equal to the resistance $R_m$ = 10 G$\Omega$ in the dummy circuit.

### 3.6.2. Source 2: Solid-Liquid Junction Potential at Electrode-Electrolyte Interface

In planar lipid bilayer experiments, an electrical connection between the salt solutions in the experimental chamber and the amplifier is established through a pair of Ag/AgCl electrodes in presence of Agar-salt bridges. In the bilayer set-up, one of the electrodes connects the front of the experimental chamber filled with salt solution to the headstage of the Axopatch amplifier representing the *voltage* side and the other one connects the rear of the experimental chamber to the ground representing the *ground* side via Agar-salt bridges as shown in Figure 3.1.

It is absolutely essential while using the Ag/AgCl electrodes that they are never in direct contact with the salt solutions in the experimental chamber. The foremost reason being the sensitivity of the electrode potential to the $Cl^-$ concentration in the surrounding solution. The electrodes respond to the total electrochemical potential of $Cl^-$ that depends on the $[Cl^-]$ further dependent on the electric potential (see Eq. (10)). The electrode potentials drift with changing $Cl^-$ concentration. The drifting electrode potentials produce a measurable potential difference between the two electrodes contributing a finite current to the output current measured.  It is therefore, recommended that the electrodes face a constant ionic environment preferably saturated with $Cl^-$. Secondly, silver ions are known to be toxic to biological preparations and to channel proteins producing irreversible changes.

In planar lipid bilayer experiments, the voltage offset issue dealing with the stability of Ag/AgCl electrode potential and the toxicity of silver ions to the biological preparations are handled by using Agar-KCl salt bridges. The Agar-KCl salt bridges are designed using pipette tips filled with Agar (2% v/v) and 3 M KCl. Agar is used to keep the silver ions away from the bilayer. In salt bridges, the use of KCl is preferred because the mobility of $K^+$ and $Cl^-$ are nearly identical resulting in a significantly smaller electrode potential at the solid-liquid junction of the Ag/AgCl electrode and 3 M KCl. Further, in presence of the Agar-KCl salt bridges the $Cl^-$ concentrations match and are constant resulting in electrode potentials at the electrode-electrolyte interfaces that are equal in magnitude and opposite in polarity thus, canceling out each other. The Agar bridge moves the variable $Cl^-$ interface away from the silver wire to the end of the salt bridge.

A small voltage offset still appears due to the imperfect behavior of the electrodes/Agar-salt bridges, dirt as well as some unknown reasons. This offset can be corrected in the beginning or at the end of every experiment. Before starting the experiment *i.e.* even before painting the bilayer, the pipette offset control on the Axopatch 200 B amplifier is tuned until the resulting offset current is nullified. The pipette offset control is a built-in potentiometer injecting a finite current of opposite polarity for cancelling the offset current contributed by the electrodes/Agar-KCl bridges. The measurements made can also be corrected for offset voltage resulting from the electrode asymmetry at the end of the experiments.

### 3.6.3. Source 3:  Liquid-Liquid Junction Potential at Electrolytic Interface

The liquid junction potentials usually of the order of tens of millivolts create a considerable offset in the applied voltage at the lipid bilayer during the planar lipid bilayer experiments and in the command voltage under voltage clamp or measured voltage in current clamp at the cell membrane patch-pipette interface in patch-clamp experiments. Liquid junction potential (LJP) is the potential difference that exists at the interface of two ionic salt solutions of different concentrations in contact with each other containing ions of different mobilities and valence. In a bilayer set-up, LJP appears at the interface of the 3 M KCl in salt bridge and the bath solution. Besides the magnitude, the polarity (negative or positive) *i.e.* the direction in which the liquid junction potential correction has to be applied is also important. The liquid junction potential correction is a crucial step of data analysis and should be carefully executed as it may lead to misinterpretation of important biological ion channel functions such as ion selectivity, ion permeability ratios etc. (200-203).

Depending on the experiment protocol, the liquid junction potential correction can be performed online (during the experiment) or offline (at the end of the experiment). For example, in planar lipid bilayer experiments examining the ion selectivity of an engineered bacterial outer-membrane channel OmpF, the liquid junction potential correction was applied to the measured reversal (zero current) potential at the end of the experiment (55, 204, 57).

A software package JPCalc used extensively by electrophysiologist (205, 206) is commercially available with the Pclamp software for calculating these potentials under various experimental settings. The JPCalc software is used widely by electrophysiologist all over the word for calculating the liquid junction potentials is based on this classical treatment of liquid junction potentials as described by the generalized Planck-Henderson equation. Even though this formulation of calculating liquid junction potential is practiced in the electrophysiology field yet there are serious issues associated with this approach that are difficult to be ignored (assumptions of electroneutrality and steady state) as it may contribute to erroneous estimations of the liquid junction potential for voltage offset corrections to the measured values of reversal potential determining ionic selectivity of the ion channel.

### 3.6.4. Theory of Liquid Junction Potential: Planck-Henderson Equation

The liquid junction potential can be measured directly. However, these experimental values have to be corrected as well. In such a situation it is essential to be able to calculate these potentials theoretically and validate the calculated values with the experimental measurements. The significance, practical application, experimental measurement, and theoretical calculation of liquid junction potential for voltage offset correction in patch-clamp experiments has been addressed in great detail (201, 200, 207, 203, 184, 208, 209).

The liquid junction potential is developed due to the diffusing ions at the interface of the two ionic solutions. In electrochemistry, the diffusive ionic flux under the influence of the electrochemical gradient is customarily determined by the Nernst-Planck equation of electrodiffusion. The general expression of the diffusion potential or the liquid junction potential is derived by assuming a steady state condition where the net electrodiffusive flux of ions becomes zero. The ionic fluxes are determined from the Nernst-Planck equation including both the independent and the interacting ionic drifts as described by the Onsager phenomenological equations (210). The forces governing the electrodiffusive flux of ions at the interface of the two solutions correspond to the change in electrochemical free energy $d\overline{G}$ given as:

$$d\overline{G} = \sum_i \frac{t_i}{z_i} d\overline{\mu}_i \qquad (21)$$

where, $\overline{\mu}_i$ is the electrochemical potential of ion $i$, $t_i$ is the transference number or transport number for the ion $i$ ( defined as the contribution of conductivity made by that ionic species divided by the total conductivity *i.e.* $t_i = \dfrac{\left|z_i\right|u_i C_i}{\sum_j \left|z_j\right|u_j C_j}$ , $C_i$ being the ion molar concentration, and $u_i$ the ion mobility). The liquid junction potential is therefore, the net electrochemical

free energy change obtained by integrating $d\overline{G}$ from the α phase (ionic solution 1) to the β phase (ionic solution 2) given as:

$$V_{LJP} = \int_{\alpha}^{\beta} d\overline{G} = \sum_i \int_{\alpha}^{\beta} \frac{t_i}{z_i} d\overline{\mu_i} \tag{22}$$

A general expression for the liquid junction potential is obtained by assuming a steady state where the net transport of ions at the boundary ceases as the diffusion flux gets balanced by the electric flux which implies that the concentrations and the electrostatic potential throughout the interface (also static/unchanging) do not vary with time. On the basis of this assumption, the net electrochemical free energy change obtained by integrating $d\overline{G}$ from the α phase (ionic solution 1) to the β phase (ionic solution 2) is equated to zero,

$$\int_{\alpha}^{\beta} d\overline{G} = 0 = \sum_i \int_{\alpha}^{\beta} \frac{t_i}{z_i} d\overline{\mu_i}. \tag{23}$$

As both phases are aqueous solutions therefore, the standard chemical potentials in phase α and β are equal *i.e.* $\mu_i^0(\alpha) = \mu_i^0(\beta)$,

$$\sum_i \int_{\alpha}^{\beta} \frac{t_i}{z_i} RT \ d\ln a_i + \left(\sum_i t_i\right) F \int_{\alpha}^{\beta} d\phi = 0 \tag{24}$$

since, $\sum_i t_i = 1$,

$$V_{LJP} = \phi^{\beta} - \phi^{\alpha} = \frac{-RT}{F} \sum_i \int_{\alpha}^{\beta} \frac{t_i}{z_i} \, d \ln a_i \qquad (25)$$

where, in the general expression for the liquid junction potential $V_{LJP}$, $\phi^{\beta} - \phi^{\alpha}$ is the electric potential difference at the interface of the two ionic solutions in contact and the negative sign in front of the chemical potential gradient term implies that the resulting electric field opposes the diffusion of ions due to the chemical gradient. The integration of Eq. (25) is not trivial even after a steady state is assumed (net flux of each ionic species is zero) because the variation of ionic concentrations $C_i$ at the interface is not known. It is also not clear how the activity coefficients $a_i$ and transport number $t_i$ vary with $C_i$.

The widely used approach for calculating the liquid junction potential is the generalized Planck-Henderson equation (198, 210) that essentially integrates Eq.(25) based on the assumptions that (a) the concentrations of ions everywhere in the junction are equivalent to the activities and (b) the concentration of each ion follows a linear transition between the two phases. Under these assumptions, the liquid junction potential is estimated as:

$$V_{LJP} = \phi^{\beta} - \phi^{\alpha} = \frac{\sum_i \frac{|z_i|u_i}{z_i}[C_i(\beta) - C_i(\alpha)]}{\sum_i |z_i|u_i[C_i(\beta) - C_i(\alpha)]} \frac{RT}{F} \ln \frac{\sum_i |z_i|u_i C_i(\alpha)}{\sum_i |z_i|u_i C_i(\beta)}, \qquad (26)$$

where, α and β represents the two electrolytes in contact, $u_i$ is the mobility of ionic species $i$, $z_i$ is the valence, $C_i$ is the molar concentration, $R$ is the gas constant, and $T$ is the absolute temperature.

It is worth noting that most experiments using electrochemical cells in the last ~120 years have used salt bridges to minimize irrelevant effects of changes in bath solutions on the potential at the electrodes. The electrodes are bathed in one unchanging solution, typically 3M KCl. That solution is connected to the test chamber through U shaped tubes filled with concentrated agar (with the consistency of solid Jello) filled itself with 3M KCl. The solution on the far side of the U shaped tube (away from the electrode side) is varied. Ion selectivity determination experiments of ion channels, measuring reversal potential thus, depend for voltage offset corrections on the potential difference between 3M KCl in agar and the solution of variable composition. This potential *i.e.* the liquid junction potential is almost always estimated by the Planck-Henderson Liquid Junction equation (Eq.26). We follow that same practice in the experimental part of this work, and use the most sophisticated treatment available for the Planck-Henderson liquid junction equation (206). However, as physical scientist working the 21st century, with powerful tools of analysis and computation available, that were not available to Planck and Henderson, and perhaps with the benefit of some accumulated wisdom in the professions, we have further perspectives.

The Planck-Henderson equation is supposed to be a solution of the Poisson-Nernst-Planck (PNP) equations, but unfortunately, it is not. The correct solution of PNP equations can only be evaluated numerically by computationally methods that were not available to Planck and Henderson. However, the PNP equations are known furthermore, not to be an adequate representation of the properties of most salt solutions, certainly those with divalents, those that are mixtures, or even pure monovalent solutions of concentration larger than some 0.1 M as quoted by W. Kunz "It is still a fact that over the last decades, it was easier to fly to the moon than to describe the free energy of even the simplest salt solutions beyond a concentration of 0.1M or so (211)".

Excess chemical potentials (or ion activities) are large and cannot be ignored in such solutions. The finite size of ions is an important cause of excess chemical potentials, and other effects are involved as well. Excess chemical potential means by definition any component of chemical potential beyond the ideal. Some of the papers that present numerical solutions of PNP point out other assumptions of PNP that may be in error. For example, PNP ignores all convective flow and it is well known that tiny amounts of convection can have larger effects on concentration gradients produced by electrodiffusion (212). Indeed, it is not obvious that the solution of the full problem of the liquid junction potential including convection set up by the connection of 3M KCl (in agar) to the varying solution has a steady state solution at all. Until recently, there was no way (known to us) to deal with the full convective flow of electrolyte solutions, including convection, migration, and diffusion. The variational approach of Chun Liu has been applied to ionic solutions to create such a theory. The theory can in fact be solved numerically for cases like that of the liquid junction potential (213), but the utility of the solutions and theory is not yet known.

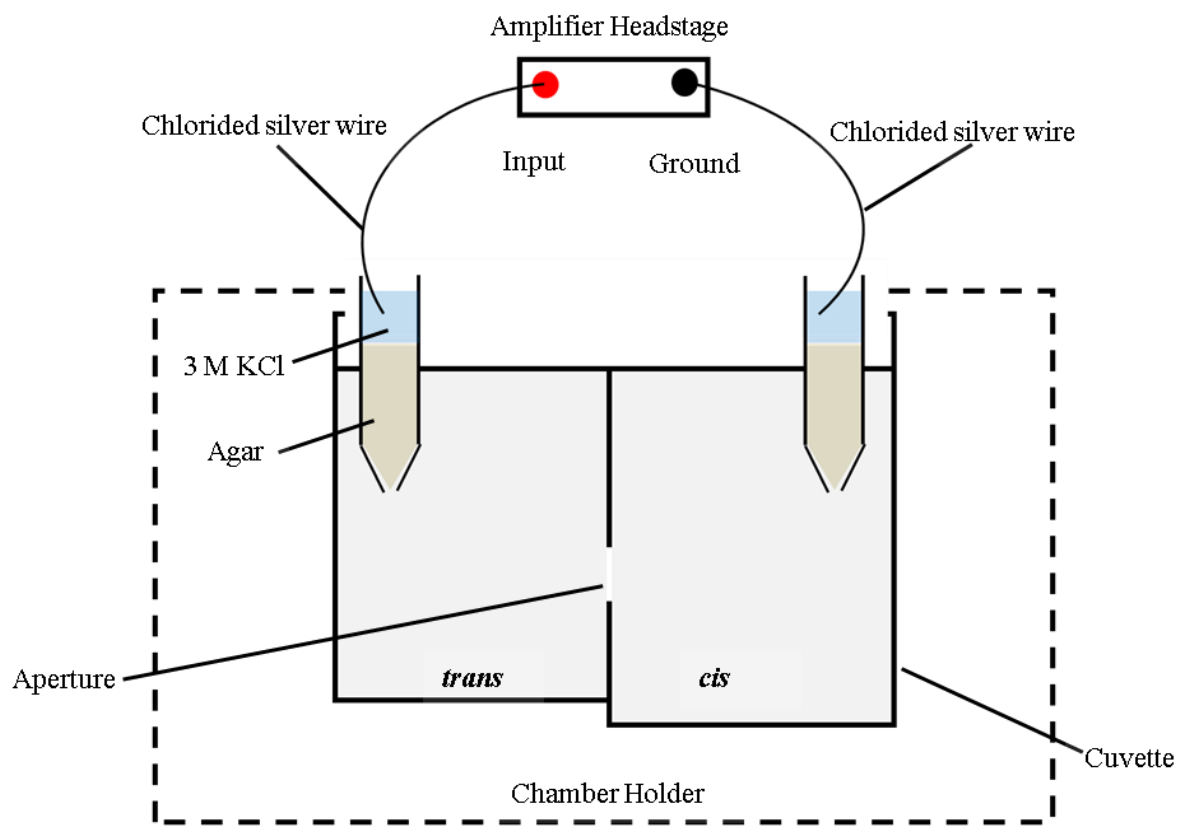

Figure 3.1: Bilayer Setup for single channel measurements: The *cis* (ground) and *trans* (voltage) compartments are filled with aqueous salt solutions. The chlorided silver (Ag/AgCl) wires are the electrodes connecting the baths via Agar-KCl salt bridges. An artificial lipid bilayer is painted over the aperture. The voltage applied across the lipid membrane is $V_m = V_{trans}- V_{cis}$. Protein solution (~0.1-0.2 μl) is added to the *cis* compartment and stirred for initiating the incorporation of a single channel into the bilayer.

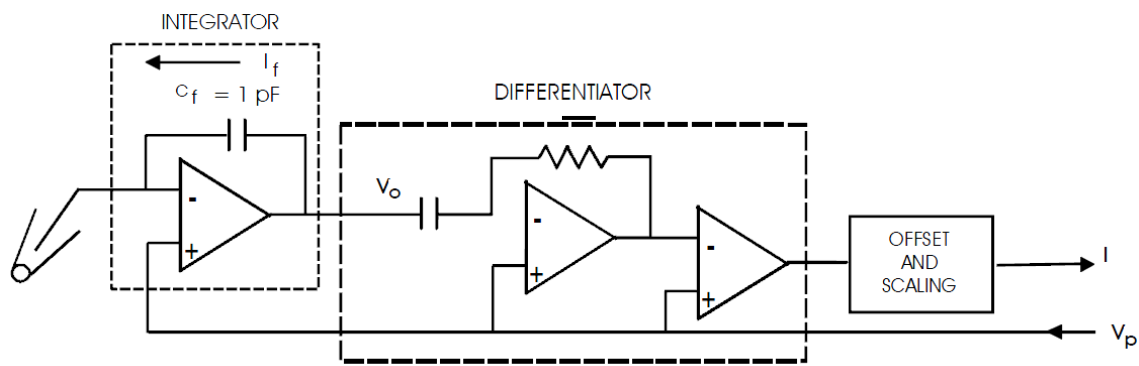

Figure 3.2: Capacitive headstage of the Axopatch 200 B amplifier. Circuit diagram representing the principle components of the headstage *i.e.* the integrator (output A) and the differentiator (output B). The command potential is $V_p$ and the corresponding current measured after offset and scaling is $I$.

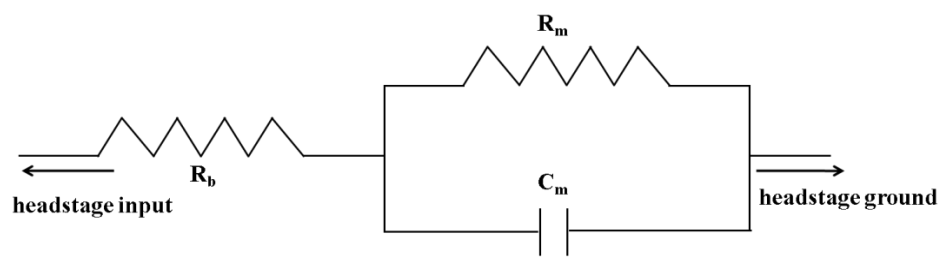

Figure 3.3: Dummy circuit for calibrating the headstage, $R_m = 10\ G\Omega$, $C_m = 80\ pF$ (electrical equivalent of the lipid bilayer), and $R_b = 100\ \Omega$ (electrical equivalent of the bath solutions).

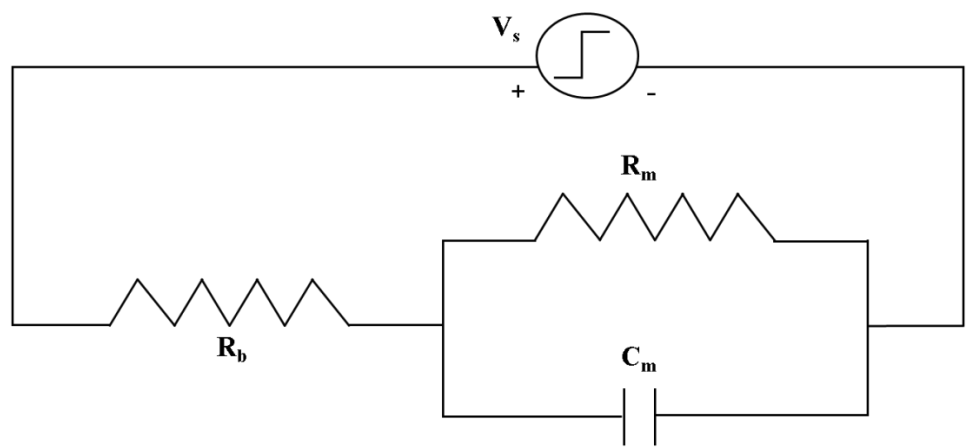

Figure 3.4: Dummy circuit connected to a voltage source generating step voltage $V_s$ (voltage clamp), $R_m = 10\ G\Omega$, $C_m = 80\ pF$ (electrical equivalent of the lipid bilayer), and $R_b = 100\ \Omega$ (electrical equivalent of the bath solutions).

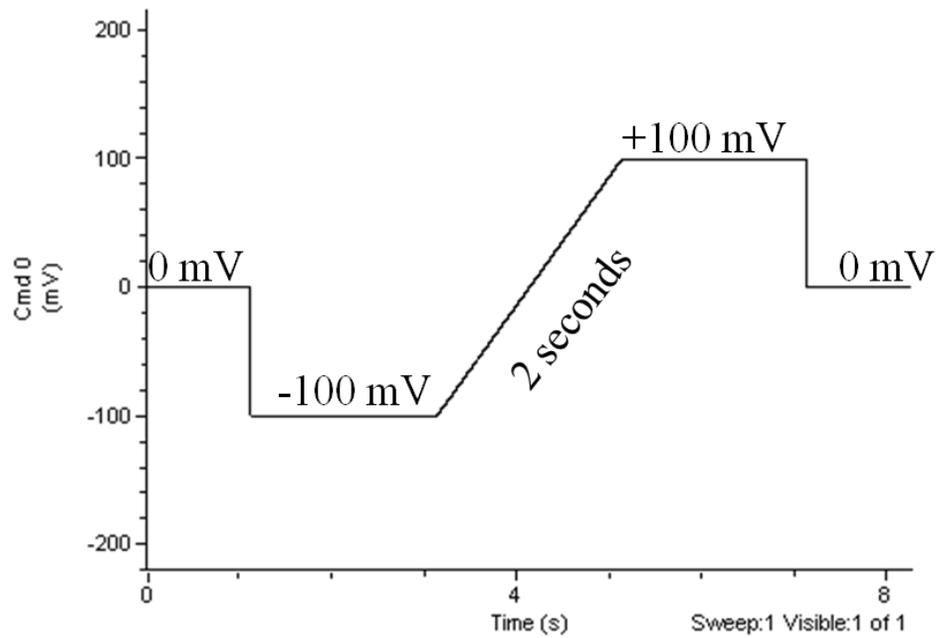

Figure 3.5: Ramp voltage protocol. Voltage is stepped down from 0 mV to -100 mV after ~ 1 second. Voltage stays at -100 mV for 2 seconds and then ramped to + 100 mV for ~ 2 seconds staying at + 100 mV for 2 seconds before being set back to 0 mV.

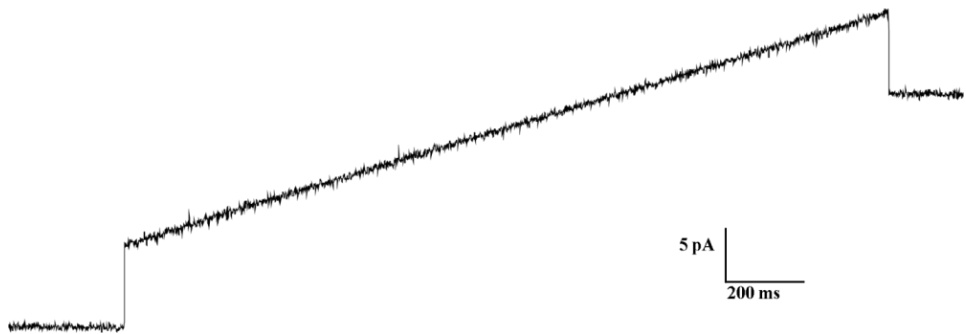

Figure 3.6: Calibration of the headstage via dummy circuit shown in Figure 3.3. Output current measured in response to the ramp voltage protocol shown in Figure 3.5. The current trace shown is filtered digitally at 300 Hz.

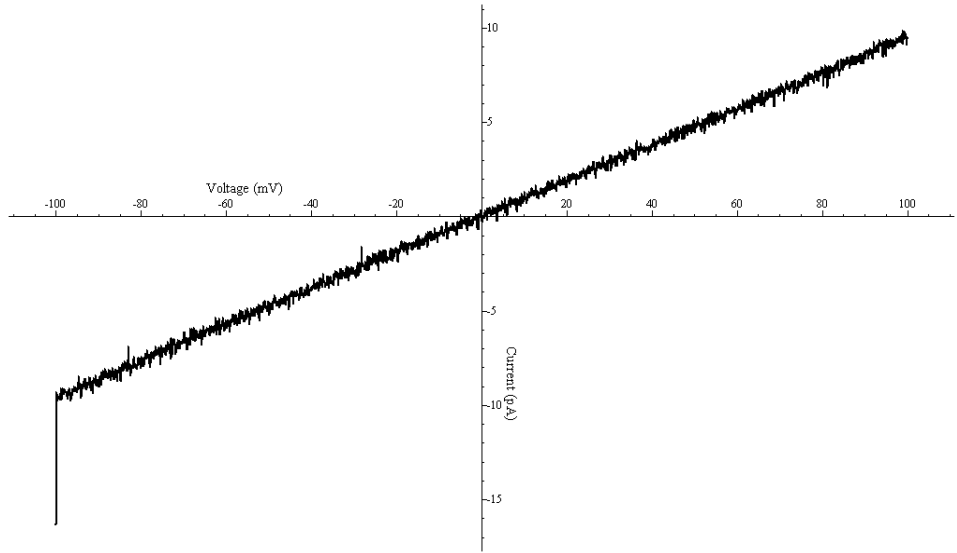

Figure 3.7: Current-voltage trace from Figure 3.6, for determination of the slope conductance given as $G_m \text{ (nS)} = \dfrac{\Delta \text{ I (pA)}}{\Delta \text{ V (mV)}}$, over the $-100$ mV to $+100$ mV range is shown. The value of membrane resistance or seal resistance is measured as $R_m = \dfrac{1}{G_m} \sim 10 \text{ G} \Omega$. At 0 mV, the output current is 0 pA.

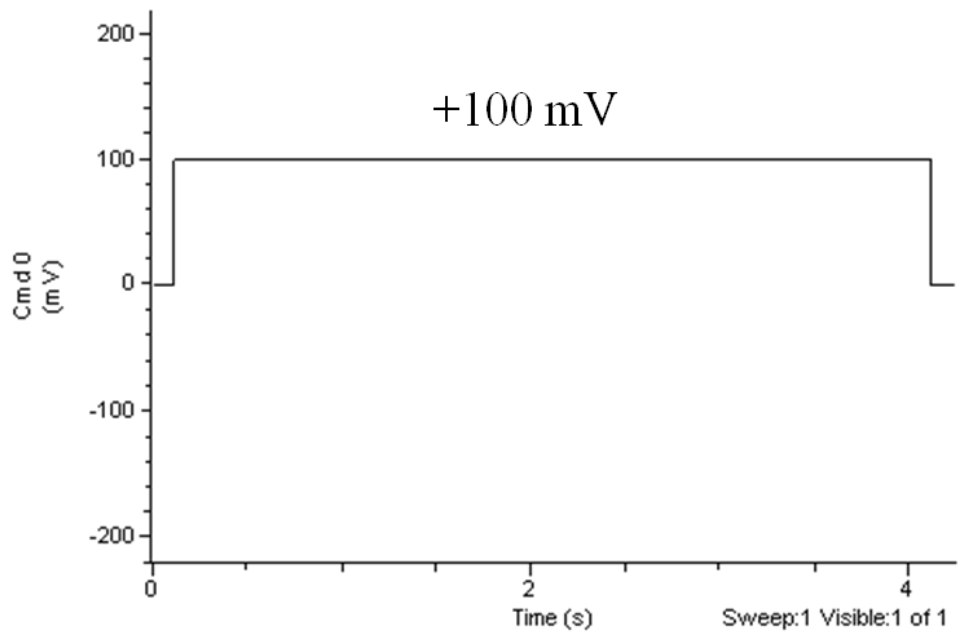

Figure 3.8: Positive step voltage protocol where voltage is stepped up from 0 mV to + 100 mV in less than 1 second and stepped back to 0 mV after 4 seconds.

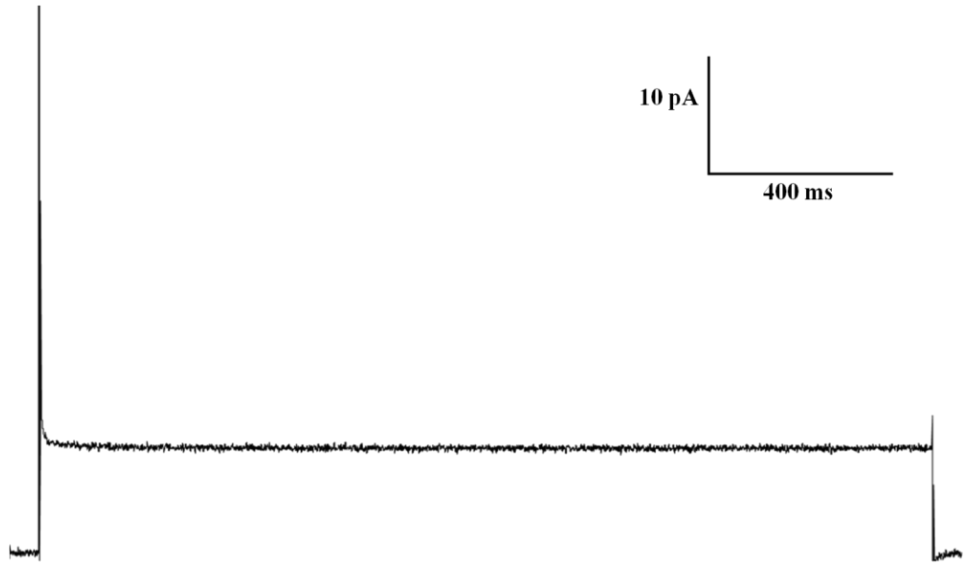

Figure 3.9: Step current response at +100 mV (step voltage shown in Figure 3.8) recorded for the dummy circuit shown in Figure 3 connected to the headstage. The "spikes" seen at the beginning and at the end of the current trace represent the instant charging and discharging of the capacitor $C_m$. The measured current value at a constant voltage of + 100 mV is + 10 pA in agreement with the value as estimated from Eq. (20), where $V_{in}$ = + 100 mV. The current trace is filtered digitally at 300 Hz.

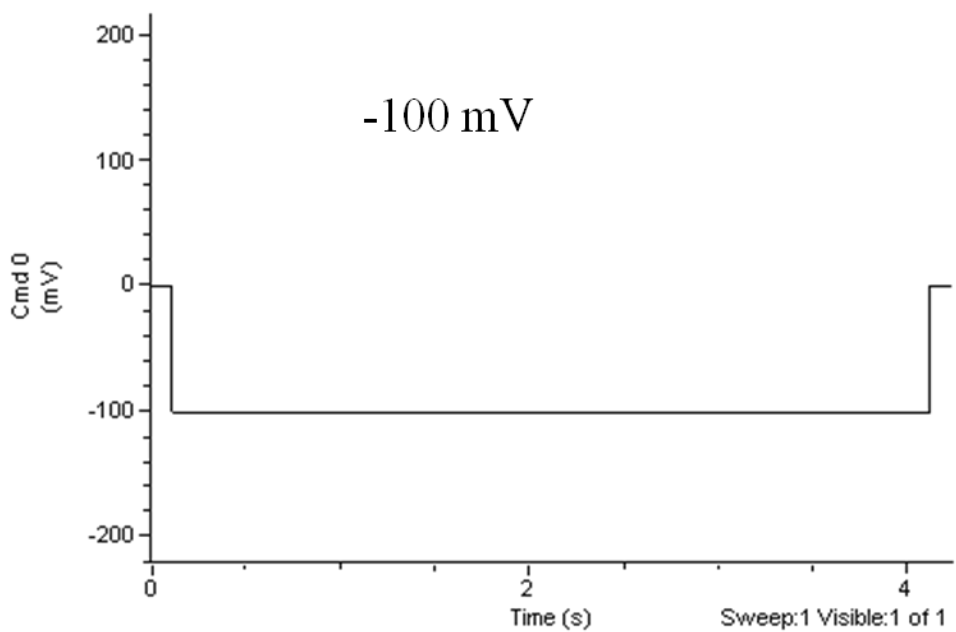

Figure 3.10: Negative step voltage protocol where voltage is stepped down from 0 mV to - 100 mV in less than 1 second and set back to 0 mV after 4 seconds.

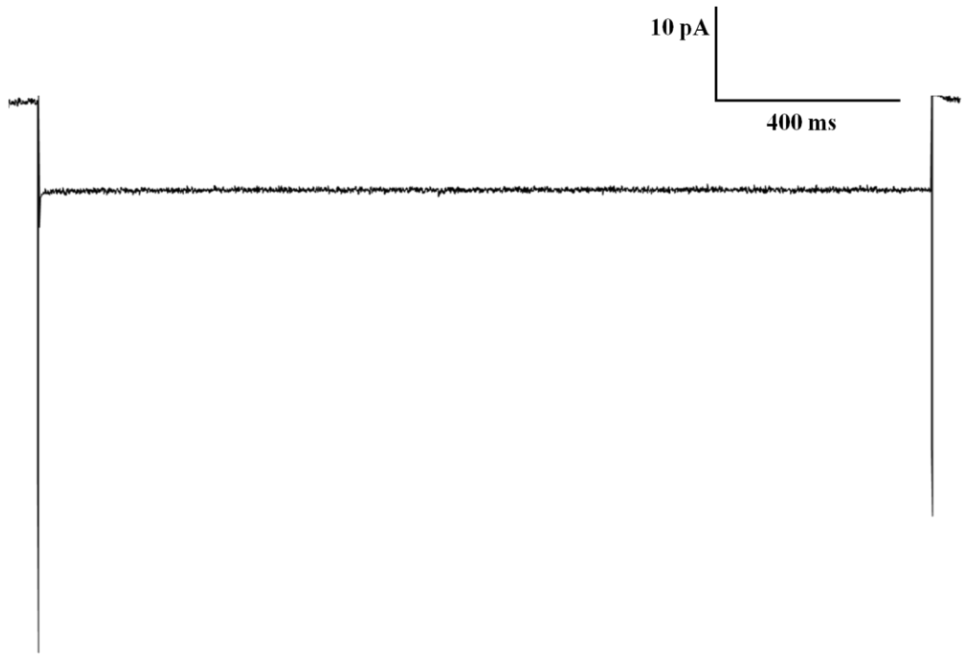

Figure 3.11: Step current response - 100 mV (step voltage shown in Figure 3.10) recorded for the dummy circuit in Figure 3 connected to the headstage. The "spikes" seen at the beginning and at the end of the current trace represent the instant charging and discharging of the capacitor $C_m$. The measured current value at a constant voltage of - 100 mV is -10 pA in agreement with the value as estimated from Eq. (20) where $V_{in}$ = - 100 mV. The current trace is filtered digitally                     at                     300                     Hz.

# CHAPTER 4. SINGLE CHANNEL MEASUREMENTS OF N-ACETYLNEURAMINIC ACID-INDUCIBLE OUTER MEMBRANE CHANNEL IN *ESCHERICHIA COLI*

## 4.1. <u>Introduction</u>

Purified porins from the outer-membrane of bacteria are easily available in large amounts. The intrinsically strong structure of the β-barrels enables the porins to resist chemical and mechanical stress and probably make the protein easier to crystallize. The structures of many porins are now known, including the classical porins OmpF, PhoE (214) and OmpC (215), as well as the specific porins LamB (216) and TolC (217). The nonspecific general porins were the first membrane proteins crystallized for X-ray diffraction (218). Indeed, more porin structures are known than of any other class of membrane passive transporters. The biophysical and biological functions of bacterial outer-membrane porins have been characterized through state of the art electrophysiological techniques such as patch-clamp and planar lipid bilayer reconstitution (219). The molecular mechanisms for the diverse roles and survival of porins thus, can be studied using structural and functional measurements, and molecular modeling. The detailed characterization of porins reveals the fundamental mechanisms that determine the functional properties of physiological ion channels at the molecular level and helps in the design and construction (i.e., 'engineering') of porins with desired functions (220, 55, 221, 57).

In this chapter, the functional properties of a sialic acid specific outer membrane porin NanC of *E. coli* are characterized using electrophysiological methods. The goal is to characterize the biophysical properties of NanC and determine the experimental conditions necessary for the measurement of transport of sialic acid in bilayer setups. The biophysical conditions that produce biological transport are investigated. The ionic conditions suitable for single channel measurements of NanC in an artificial lipid bilayer are determined and then the ion selectivity and conductance of the channel are measured.

NanC is named because it is a N-acetylneuraminic acid (Neu5Ac) inducible channel. N-acetylneuraminic acid is one of the family of naturally occurring sialic acids that include more than 40 nine carbon negatively charged sugar moieties (67). The sugars are found primarily at the terminal positions of many eukaryotic surface-exposed glycoconjugates. Many commensal and pathogenic bacteria are known to use the sialic acids of the host as sources of carbon, nitrogen, and amino sugars. The ability of the bacteria to colonize, persist and cause disease (68, 69) depends on their ability to use sialic acids in many cases.

The most abundant and well studied sialic acid is N-acetylneuraminic acid, Neu5Ac. If general porins OmpF/OmpC are not expressed by the *E. coli* bacteria (as is often the case (70)), NanC must be present (i.e., it must be induced) to promote efficient uptake of Neu5Ac across the outer membrane. Induction of NanC is important to the bacteria in its real life. Induction is not an artificially contrived situation that only occurs in the laboratory. Most of the animal tissues contain free sialic acid (69). When the general porins OmpF/OmpC are not expressed by the bacteria, a channel like NanC allows efficient uptake of sialic acid.

Multiple approaches have been followed for demonstrating the functional necessity of induction of these specific uptake systems under particular environmental conditions (63, 222). The classical approaches monitoring the expression and role of specific porins include: 1) growth experiments and 2) liposome swelling experiments.  Under the growth experiment approach the role of a specific membrane protein (or a transport system)  in the uptake of a primary solute into bacteria is demonstrated by monitoring the growth of the wild type and 'knock out' mutant bacterial strains (missing specific proteins). The growth is studied under controlled conditions in external solutions ('growth media') enriched or deprived of a particular solute for example, the maltodextrin specific transport system in *E. coli* (223). In the liposome swelling experiment the specificity of a membrane protein for a particular solute is demonstrated by the monitoring the swelling rate of protein containing liposomes in various solutions for example, the LamB protein of *E.coli*  (224).

These classical techniques are responsible for most of our knowledge of transport in bacteria and set the stage for biophysical analysis of specific uptake systems. Biophysical analysis is needed to reveal the mechanistic details of the uptake system at molecular and atomic resolution. For example, membrane proteins must be characterized using the patch-clamp or the planar lipid bilayer methods of channel biology to unravel the underlying mechanisms transport through single protein molecules (176, 184).

Recently, a high resolution structure of NanC has been reported (58). The structure of NanC shares many of the characteristic features of other outer membrane channel proteins.

NanC is a monomer, a 12-stranded β-barrel with a relatively narrow pore of average radius 3.3Å. However, NanC is also different in many ways from other porin structures. In most of the porins, 'loop' partially occludes the pore region and is the most obvious location where specific interactions and gating might occur. Interestingly, NanC has no 'loop' occluding the pore. The pore region of NanC is predominately decorated by positively charged residues that are arranged to form two positively charged tracks facing each other across the pore. This particular arrangement of the positively charged residues in the pore region seems to help the negatively charged Neu5Ac (carboxylate group ($COO^-$) of Neu5Ac deprotonated at physiological pH, pKa ~ 2.6) to move through NanC. The positively charged tracks are likely to guide the movement of negatively charged solutes.

Interestingly, the earlier functional measurements of NanC carried out using patch-clamp experiments were unable to demonstrate any significant change in the function of NanC in the presence of even large concentrations of Neu5Ac, up to 50 mM (70). The high resolution structure suggests one explanation for this finding. The structure shows many basic (positive) side chains that would be screened (71) in the high salt concentrations used in these earlier measurements. Structural measurements also show HEPES binding to NanC. Two HEPES molecules were immobile enough under crystallizing conditions to produce diffraction in the crystal structure, one near each end of the channel. Binding strong enough to crystallize HEPES in place in a channel is likely to modify current flow through the channel.

## 4.2. <u>Methods</u>

### 4.2.1. NanC expression and purification

NanC was expressed and purified as reported previously (58). In short, NanC was expressed in BL21/Omp8 strains. After cell disruption, the whole membranes were pelleted by ultracentrifugation (1h at 100000g). The inner membrane was solubilized using 1% Lauroyl sarcosine and the remaining outer-membrane was collected by ultracentrifugation. NanC protein was solubilized from the membrane by iterative extractions using increasing concentrations of Octylpolyoxyethylene (OPOE). NanC was then purified using anion-followed by cation-exchange chromatography. A final size exclusion chromatography (Superdex75, GE healthcare) allowed exchanging the buffer to 10mM Tris pH 8.0, 150mM NaCl and 1% OPOE in which the protein was stored.

### 4.2.2. Electrophysiology: Planar Lipid Bilayer Experiments

Planar lipid bilayer method is used to measure the functional properties of purified channel protein NanC. We reconstitute a single NanC into a preformed lipid bilayer in a set-up shown in Figure 3.1. Planar lipid bilayers can be formed in three ways. The first is the painted or the Mueller-Rudin technique (182) , the second is the folded or the Montal-Mueller (196)  technique and third is a variation of the Montal-Mueller technique known as the Schindler technique (225). We use the painted technique to form the planar lipid bilayer. In this method lipid bilayer is formed by painting a lipid solution across a $150\mu$ m diameter aperture in the cup. The stability of the lipid bilayer is increased if the aperture is pretreated ('primed') with the lipid solution before the bilayer is formed. The lipid solution is made of phospholipids DOPE: DOPC in a 4:1 ratio (v/v) dissolved in solvent n-decane (10 mg/ml). The solvent orients the monolayers of lipids to form a bilayer. Initially, a thick film of lipid solution is formed that eventually thins out as the solvent evaporates and a lipid bilayer is obtained. A 'healthy' (presumably thin) lipid bilayer is essential to ensure the incorporation of the protein. Membrane thickness is measured by membrane capacitance. In our set-up, the electrical capacitance of a 'healthy' lipid bilayer is 60pF-80pF with a specific capacity of ~ 0.4-0.6 µF per cm$^2$.

Membrane proteins are reconstituted into the bilayer in different ways depending on their water solubility. Three commonly used methods are: 1) direct fusion of 'pure' proteins, 2) fusion of protein-containing liposomes with preformed bilayers (226, 227) and 3) spreading of protein containing liposomes at air-water interface (228, 229). We used direct fusion to reconstitute a single NanC into the lipid bilayer. Note that membrane proteins are often not water soluble and therefore, detergent is added to its stock solution to make a soluble mixture of protein and detergent. The effect of the detergent on protein function must be checked, in every case. We use a stock solution of ~ $1.1\mu$ g/ml of the purified NanC in 150 mM NaCl, 10 mM Tris, pH 8.0 containing 1 %( v/v) n-OPOE detergent. We add ~0.1-0.2 $\mu$ l of the stock solution to the *cis* or *ground* side of a stirred solution to reconstitute NanC. The ground side is the side of the bilayer connected through a bath electrode to zero (ground or earth potential).

In our set-up Ag/AgCl electrodes are used to connect (1) the solutions in the *cis* compartment to ground and (2) the *trans* compartment (the voltage side of the bilayer) to the patch clamp amplifier. Ag/AgCl electrodes are stable, robust and the most commonly used electrodes in electrophysiology but they respond to the free energy (per mole) of $Cl^-$, i.e., to the activity of $Cl^-$. They respond to $Cl^-$ concentration as well as electrical potential, approximately as described by the classical Nernst equation of electrochemistry (198). The Ag/AgCl electrodes are isolated from bathing solutions by Agar bridges (2%) containing (typically) 3M KCl. The 3M KCl provides a fixed stable $Cl^-$ concentration to the Ag/AgCl wire so the electrode potential is stable and minimum. The changes in electrical potential in the wire are produced by changes in the electrical potential not by changes in the chemical potential (i.e., concentration) of $Cl^-$ because the chloride concentration is constant in this setup. A (so called liquid junction) potential appears of course at the interface of the 3 M KCl Agar with the bath solutions. This potential is small because the mobility of $K^+$ and $Cl^-$ are nearly equal. The small offset potential is calculated by custom from reduced models of liquid junction potentials that use the generalized Henderson liquid junction potential equation (230, 198, 200-202, 231). This equation is not customarily derived by mathematics from the appropriate description of nonideal solutions (necessary when salt concentrations are high as they are in and near the salt bridge). We look forward to such work in the future. The Agar bridge has the further advantage that it keeps $Ag^+$ ions away from the bilayer and reconstituted channel. $Ag^+$ ions can be toxic (i.e., produce irreversible changes) to channels in very low concentrations.

### 4.2.3. Pulse Protocol and Data Analysis

Conductance is determined from measurements of current through a fully open single channel. The voltage across the channel follows a ramp time course: the voltage is a linear function of time as it ranges in amplitude from -100mV to +100mV in ~2 seconds, see Figure 3.5. The amplitude of the single channel current and the probabilities that the channel is open or closed are also measured in another set of experiments in which the potential is changed from one value to another. Step voltage pulses are of varying amplitudes/durations for example, 3 and 10 seconds long +/-100 mV, +/-150 mV and +/-200mV step voltages as shown in Figure 3.8 and Figure 3.10**.** The data being recorded is low-pass filtered at 2 kHz using the built-in analog low-pass Bessel filter in the Axopatch 200B amplifier (Molecular Devices, Inc.) and is sampled or digitized at the 5 kHz rate that the Nyquist sampling theorem implies is needed to avoid aliasing. After the recording, the recorded data is filtered digitally (at 300/500 Hz) for further analysis using the low pass 8-pole digital Bessel filter in the PClamp software, version 10 Molecular Devices, Inc.

### 4.2.4. Leakage subtraction and offset current correction

There are two main corrections performed on the recorded single channel current-voltage traces after the digital filtering. This post-filtering correction of the single channel recordings includes the leakage subtraction and offset current correction.

Before the addition of protein sample, the *control* or the *baseline* current Figure 4.1B is measured in response to the same voltage (ramp and step) pulse protocol in the same ionic conditions as the experimental recordings.

Leakage is defined as the current that 'leaks' or flows through the lipid bilayer (without a channel) and is measured as the conductance of the *control* or the *baseline* current trace. The leakage conductance is determined by estimating the slope of the *baseline* current trace recorded for the same voltage pulse protocol that is applied while measuring the corresponding single channel current recordings, for example, a definite voltage in case of step pulse protocol or over a range of voltages for the ramp protocol.

The single channel current-voltage recordings are thus corrected for leakage by subtracting the corresponding measured leakage conductance from the filtered single channel current traces. This procedure is necessary in case the leakage conductance has nonlinear, ion dependent, and/or time dependent behavior. One must not assume the leakage conductance to be a fixed constant value resistor.

Under symmetric ionic conditions *i.e.* when equal concentrations of ions are present on both sides of the bilayer, the current flowing through the single channel in a bilayer should be equal to zero at 0 mV because the system is passive. However, under experimental situations various asymmetries in apparatus usually drive a finite current even when the baths are identical. Current offset is this 'residual' current that flows through the single channel in bilayer at 0 mV under symmetric ionic conditions.

The single channel current-voltage recordings under symmetric ionic conditions are corrected for current offset (that arises from external artifactual sources) by subtracting the corresponding measured residual current from the filtered single channel current traces.

However, under asymmetric ionic conditions i.e. when unequal amount of ions are present on both sides of the bilayer the offset current is determined from the control or the baseline current that is measured before the channel protein is incorporated. The offset current under asymmetric ionic conditions contains the artifactual offset current already described. It also contains a diffusive current that flows through the leakage conductance due to the ionic gradient across the lipid bilayer and leakage conductance. The single channel current-voltage recordings under asymmetric ionic conditions are therefore, corrected for current offset by subtracting the corresponding measured diffusive current from the filtered single channel current traces.

The potential difference ($V$) is defined as, $V = V_{trans} - V_{cis}$ where *cis* represents the *ground* side of the chamber and *trans* is where we apply the *voltage*. A positive (outward) current ($I$) is defined as a flux of positive charge from *trans to cis*.

The measurement of reversal potential $V_{rev}$ determines the ion selectivity of the channel in a predefined ionic gradient. The measured $V_{rev}$ is corrected for measured liquid junction potentials (LJP, mentioned in the figure legends where applicable). The LJP is determined from the Junction Potential Calculator (JPCalcW) made available by Molecular Devices, Inc. with their PClamp package (205, 206) based on publications(200, 201, 205).

*Conductance* is defined as the slope conductance of the fully open single channel in presence of equal concentration of ions on both sides of the bilayer measured over a 60 mV interval ranging from -30 mV to +30 mV. We have carried out the single channel measurements of slope conductance of NanC in symmetric ionic conditions i.e. in presence of equal concentrations of ions on both (*cis* and *trans*) sides of the bilayer because these conditions simplify the experimental behavior that any theory or simulation of non-equilibrium behavior will show. Moreover, using equal ionic concentrations (i.e. symmetrical solutions) provides dimensional reduction of the complexity of the system (theory or simulation or experiment for that matter) to thermodynamic equilibrium. Note that when flows are zero the behaviors are much easier to simulate or analyze.

## 4.3. <u>Results</u>

We have investigated the function of a single channel protein NanC in an artificial lipid bilayer using the planar lipid bilayer technique in order to identify appropriate ionic conditions that will permit the study the proposed biological role of NanC in the transport of Neu5Ac across the outer-membrane of *E. coli* (70, 58).

While searching for experimental conditions for single channel studies of NanC, we found that NanC is an anion selective channel that exhibits voltage-dependent gating and has a large conductance. Moreover, NanC operates in two-modes. In the first mode, a single channel of NanC has two states, i.e., one open state and one closed state. In the second mode, NanC exhibits two distinct open states that we call sub-conductance states and one closed state. The second mode of NanC occurs in only a small fraction experiments but we have seen the mode many times nonetheless. Although we do not yet understand the 'second' mode of NanC, it seems that both modes of operation are native to NanC's behavior: all the gating mechanisms of NanC are still unknown. It seems that NanC is an excellent preparation to study the molecular and atomic basis of gating.

We also find that NanC interacts strongly with the HEPES buffer. NanC interacts so strongly that HEPES actually decreases the ionic current carried by NanC. These experiments are easy to understand given our structural knowledge of NanC. The NanC structure contains two HEPES molecules bound to the channel. Since we do not wish to study a blocked channel, we explored using alternative pH buffers instead of HEPES. We found that the (classical inorganic) phosphate buffer (a commonly used buffer in electrophysiology experiments of anion channels) interfered with the channel's ionic conductance and TRIS (another commonly used buffer) damaged the Ag/AgCl electrodes in our set-up, even though the electrodes were separated from the buffer by an Agar/KCl bridge.

Therefore, we decided to study NanC in ionic solutions of concentration ~ 250 mM that are adjusted to pH 7.0 but without using a pH buffer. This is the only way we could study a channel that was not blocked, assuming of course that the $\sim 10^{-7}$M hydroxyl ion present in water at pH 7.0 does not itself block.

In Figure 3.1 we have shown the main components of the bilayer set-up used to perform the single channel measurements of NanC in an artificial lipid bilayer (see Methods). The recording voltage pulse protocols used throughout the study for the single channel measurements of NanC in lipid bilayer are the ones that were shown in Figure 3.5, Figure 3.8, and Figure 3.10. There are two main kinds: the step pulse protocol (Figure 3.8 and Figure 3.10) and the ramp protocol (Figure 3.5). In step pulse protocols, the recordings are made at a fixed voltage i.e. +/-100 mV, +/- 150 mV etc. for a definite time interval for example, 3 seconds or 10 seconds. In ramp pulse protocol, the recordings are made in order to capture the single fully open channel current –voltage behavior by measuring recordings for a range of voltages where the channel is expected to be mostly open, for example, from + 100 mV to – 100 mV. The duration of the voltage ramp is ~ 2 seconds.

We show the sequence of steps used to measure the slope conductance of single channels of NanC in lipid bilayer in Figure 4.1. First, the 'baseline' current as shown in B is measured i.e. the current flowing through the lipid bilayer before a channel is reconstituted in response to the voltage ramp pulse protocol shown in A. The current measured through the reconstituted single channel NanC in bilayer is shown in C. The offset and the leakage current correction as described in the Methods Section is carried out and the resulting the single channel current-voltage trace is shown in D. The slope is measured in the voltage-range denoted by the dashed lines and gives the unit conductance of NanC. The ionic conditions for this representative experiment were symmetric i.e. equal ion concentration on both sides of the bilayer 500 mM KCl, 20 mM HEPES, pH 7.4.

Figure 4.2 and Figure 4.3 show the voltage dependent gating of the NanC channel. The single channel measurements of NanC at +/- 100 mV and +/- 200 mV are shown in symmetric 500 mM KCl, 20 mM HEPES, pH 7.4. NanC is mostly open at voltages ≤ 100 mV and closes at voltages ≥ 200 mV in bilayer. The corresponding single channel current amplitudes determined from the amplitude histogram analysis of the current traces are shown.

The effect of the buffer HEPES on the single channel conductance and the ionic current carried by NanC in bilayer is shown in Figure 4.4. HEPES reduces the single channel conductance of NanC significantly. The "blocking" action of HEPES is not surprising and has been observed earlier in other anion channels as well (232, 233). HEPES alters the functioning of NanC and therefore, the native functional properties of NanC need to be investigated without HEPES.

Figure 4.5 shows the single channel current-voltage behavior of NanC measured in a range of KCl concentration solutions (100 mM – 3 M) without any buffer. We observe that the single channel current carried by NanC increases with increase in KCl concentration. From these measurements, we chose 250 mM for further experiments. At larger salt concentrations (≥ 500 mM), the electrostatic charge screening dominate, altering the net charge of the 'pore' region and the function of NanC.

Figure 4.6 shows the reversal potential used to determine the ion selectivity of a single NanC in the bilayer. We anticipate charge selectivity from the crystal structure of NanC that shows two tracks of positively charged residues in the pore region. The Nernst potential is a measure of the gradient of chemical potential. If NanC were a cation channel, then $V_{rev} = E_{K^+}$. If NanC were an anion channel, then $V_{rev} = E_{Cl^-}$. The negative direction of current at 0 mV and the measured value of the $V_{rev}$ (close to $E_{Cl^-}$) indicate that NanC is is an anion channel. It is more selective for the anion $Cl^-$ than the cation $K^+$. Table 4.1 shows reversal potentials for single NanC in bilayer under different ionic gradients.

The second mode of function of single NanC in bilayer is shown in Figure 4.7. The two distinct sub-conductance levels are seen here and the corresponding amplitudes at +/-100 mV are determined from the amplitude histograms. The distinct sub-conductance levels are not observed in the first mode, see Figure 4.2.

### 4.4. <u>Discussion and Conclusions</u>

The single channel bilayer experiments of NanC reveal the biophysical properties of ion selectivity, conductance and gating. NanC functions as a monomer, exhibits large conductance, anion selectivity and has two distinct modes of function (with or without sub-states). These experiments have been successful in identifying the native behavior of a single NanC in bilayer and suggested the ionic conditions (salt concentration and pH buffer) that seem to be appropriate for characterizing sialic acid transport through NanC.

The interactions of the commonly used pH buffers in electrophysiology experiments (HEPES and phosphate) with NanC are interesting. HEPES binds to the pore region (as seen in the crystal structure) and actually decreases the ionic current carried by a single NanC in bilayer. The measurements that we report here are thus carried out in ionic solutions adjusted to neutral pH without any buffer. We have been careful to measure the pH of the ionic solutions before and at the end of the experiment to be sure that the pH has not drifted. We found no drift and so we can carry out the measurements of biological function of NanC without buffer. We can measure Neu5Ac transport in salt solutions at concentration around 250 mM, pH 7.0.

However, for future functional studies of NanC we may still need to explore a variety of pH buffers in order to identify a buffering agent that has the least interaction and does not modify NanC's function in lipid bilayer.

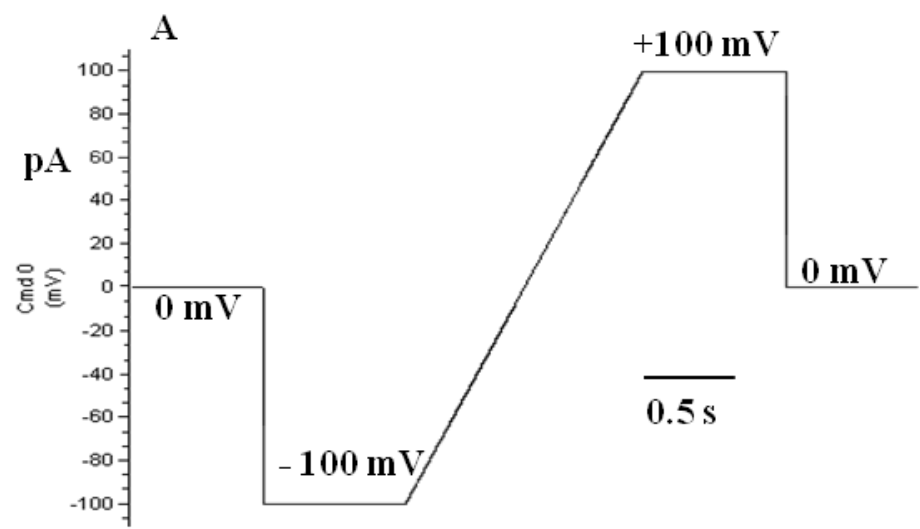

A

0 mV

+100 mV

0 mV

- 100 mV

pA

Cmd 0 (mV)

0.5 s

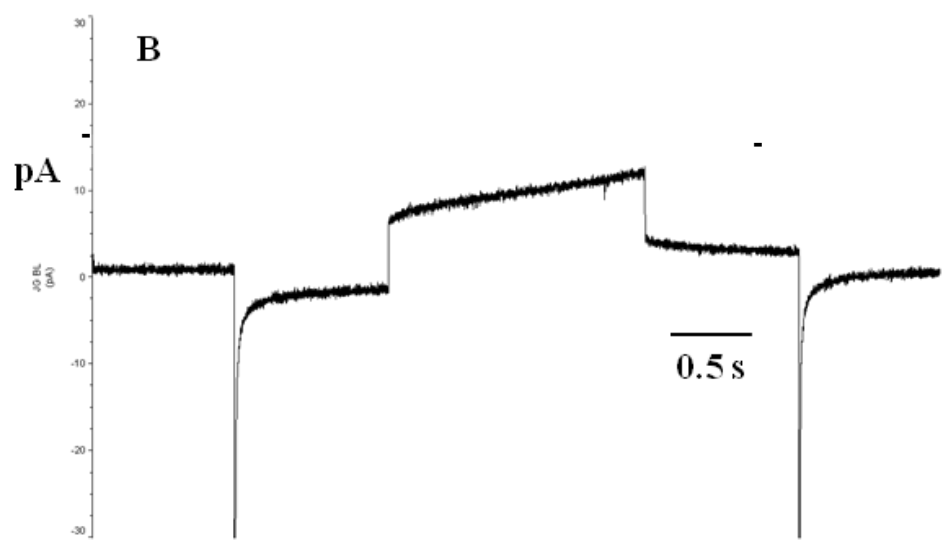

B

pA

i/O BL (pA)

0.5 s

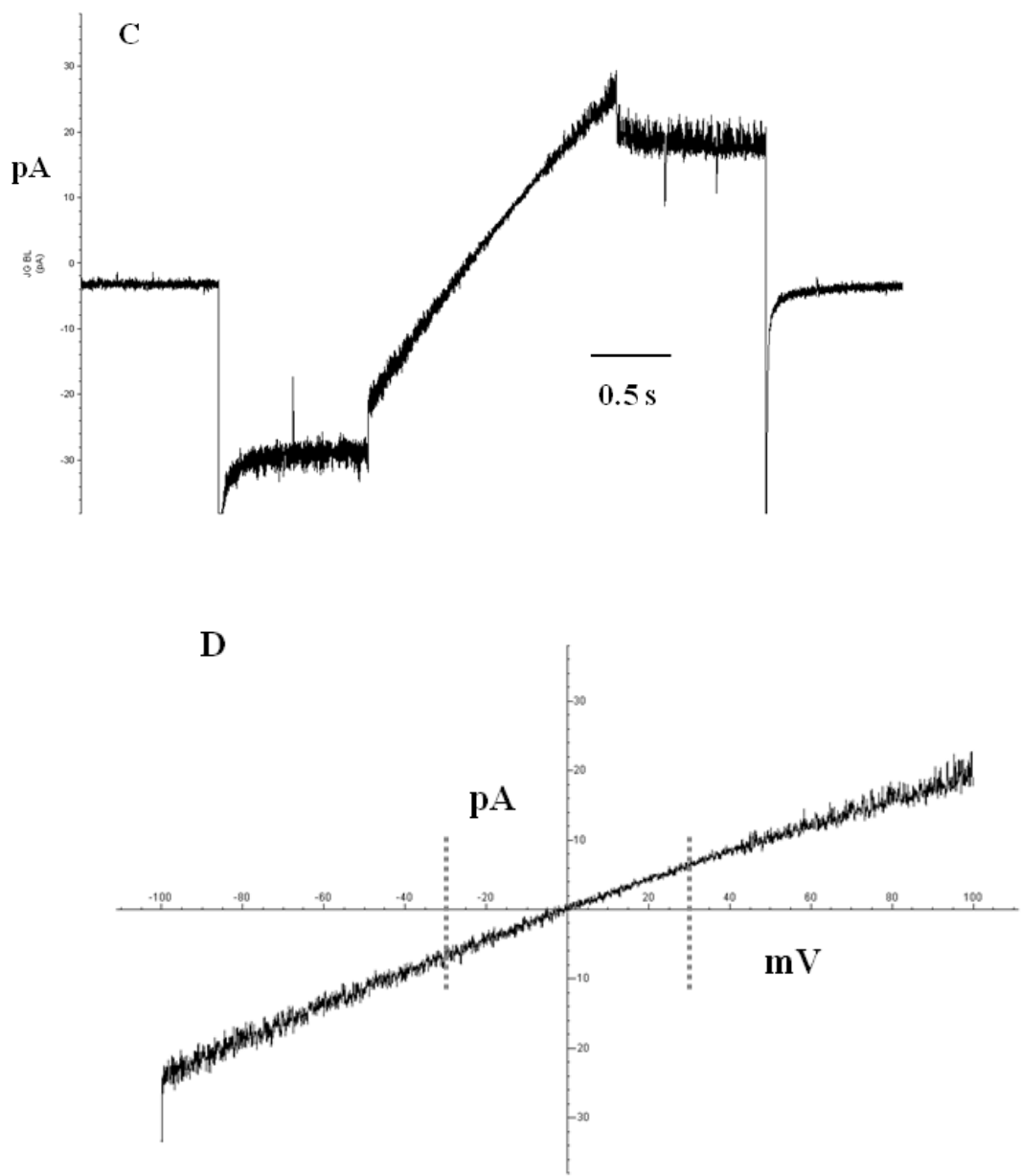

Figure 4.1: Measurement of single channel slope conductance of NanC in lipid bilayer. **A**. Ramp voltage pulse protocol. Voltage is stepped down from 0 mV to - 100 mV after ~ 1 second. Voltage stays at -100 mV for 1 second and then ramped to + 100 mV for ~ 2 seconds staying at + 100 mV for 1 second before being set back to 0 mV. **B**. Baseline current recorded in response to the ramp voltage pulse protocol as shown in **A**, through the lipid bilayer before the Figure 4.1(contd.) : channel reconstitutes in symmetric (i.e. same ionic conditions in *cis* and *trans*) 500 mM KCl, 20 mM, pH 7.4. Baseline current trace filtered digitally at 300 Hz is shown. **C**. Single channel current recorded as NanC reconstitutes in the lipid bilayer. Single channel current trace filtered digitally at 300 Hz is shown. **D**. Single channel current (I) –voltage (V) trace shown after correction for leakage and offset currents following the procedure as mentioned in the Methods section. Dashed lines represent the -30 mV to + 30 mV range chosen for determining the slope conductance given as $g \ (nS) = \dfrac{\Delta \ I \ (pA)}{\Delta \ V \ (mV)}$ , where g is the single channel conductance obtained as the slope of I-V trace over the 60 mV range. The measured single channel slope conductance of NanC in symmetric 500 mM KCl, 20 mM HEPES, pH 7.4 is $215.80 \pm 0.96$ pS, (N = 15). N is the number of measurements.

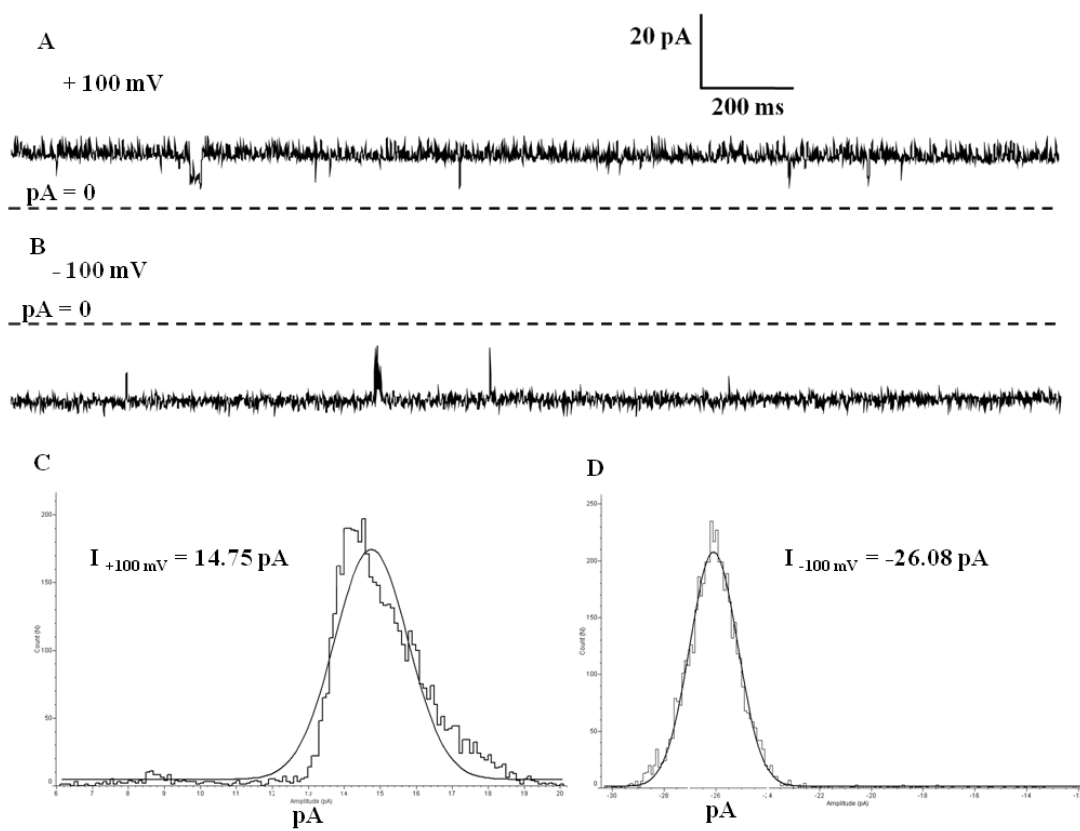

Figure 4.2: Single channel current of NanC in bilayer in the presence of symmetric 500 mM KCl, 20 mM HEPES, pH 7.4 shown at step voltages of + 100 mV shown in **A** and -100 mV in **B,** respectively. Dashed lines represent the zero current level. The corresponding single channel amplitudes are 14.75 pA at + 100 mV and -26.08 pA at -100 mV determined from amplitude histogram analysis shown in **C** and **D** respectively. In bilayer at +/-100 mV step voltages single channel NanC is mostly open and exhibits asymmetric (non-equal) conductances.

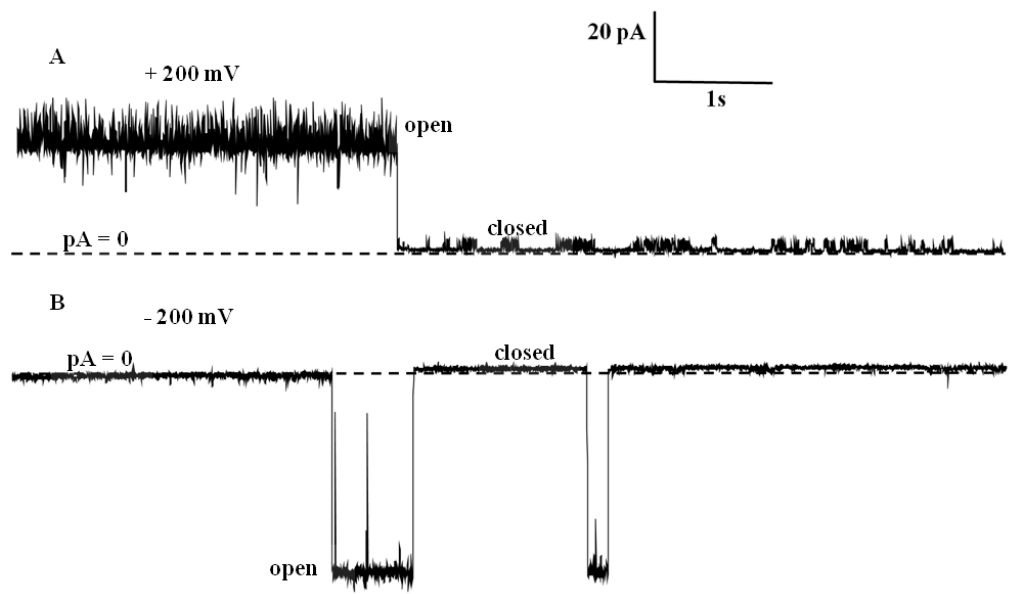

Figure 4.3: Single channel current of NanC in the bilayer in the presence of symmetric 500 mM KCl, 20 mM HEPES, pH 7.4 shown at step voltages of + 200 mV shown in **A** and -200 mV in **B** respectively. Dashed lines represent the zero current level. The corresponding single channel amplitudes are 27.71 pA at + 200 mV and -53.78 pA at -200 mV determined from amplitude histogram analysis shown in **C** and **D** of Figure 4.2 respectively. In the bilayer step voltages of +/-200 mV produce closing of NanC.

**A**

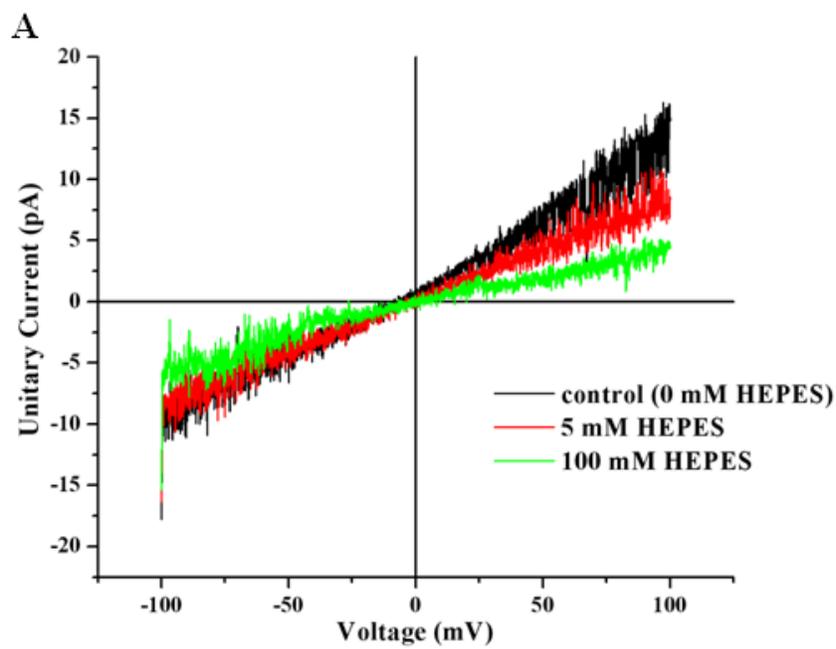

**B**

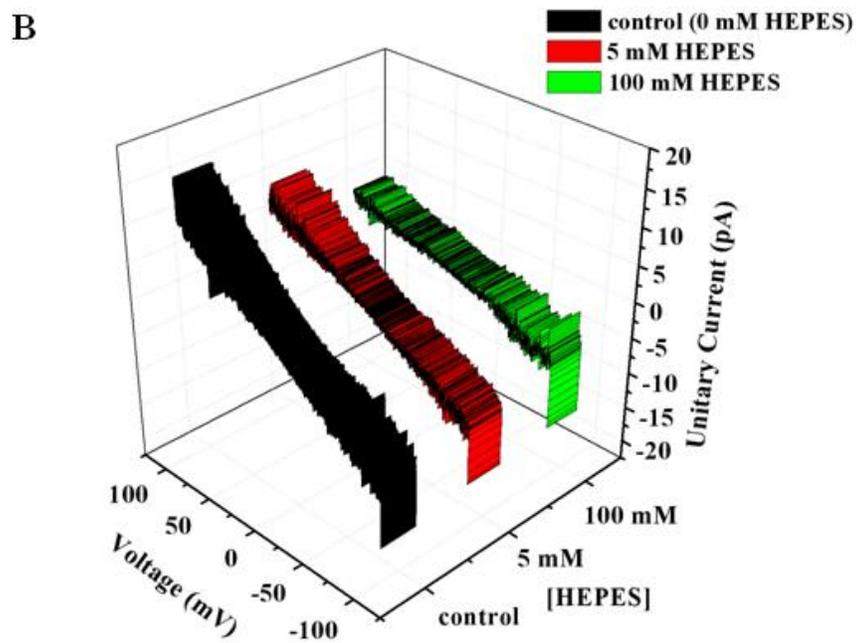

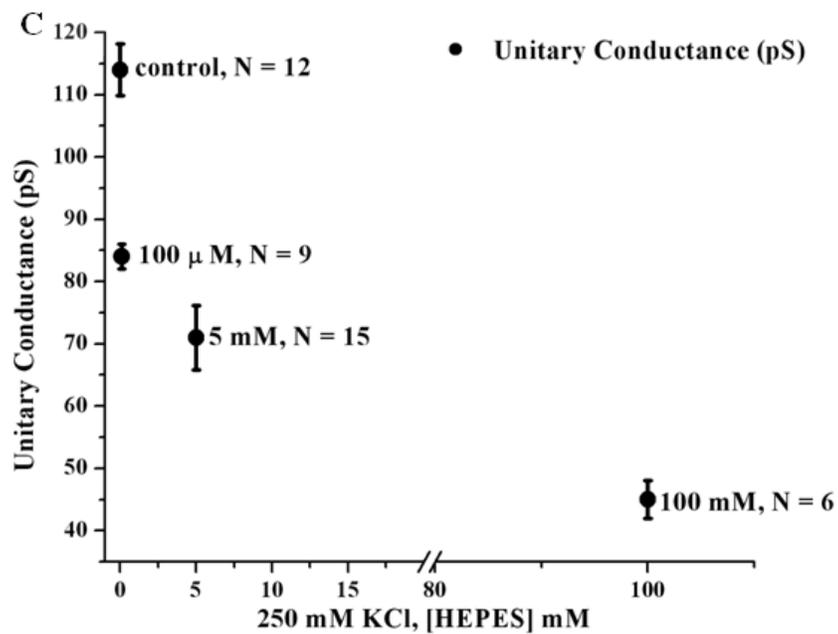

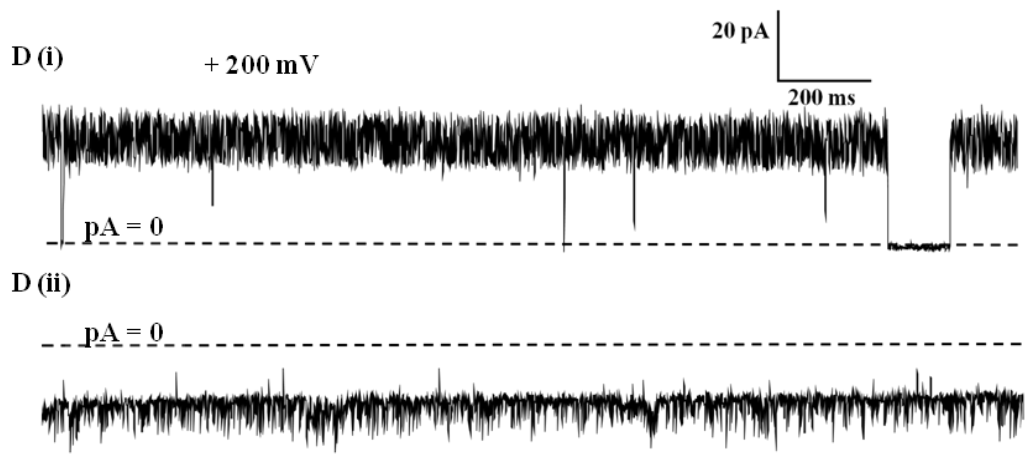

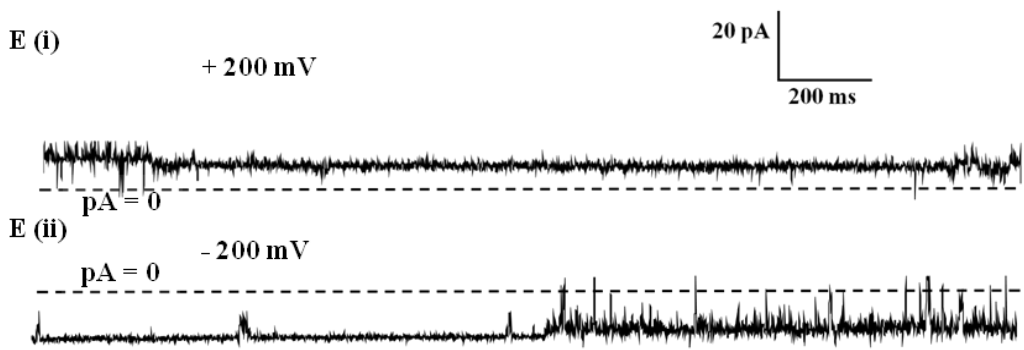

Figure 4.4: Effect of the buffer HEPES on single channel current carried by NanC in bilayer. **A.** Single channel current (I)-voltage (V) traces measured in presence of varying concentration of the buffer HEPES (0 mM (control in black), 5 mM (red) and 100 mM (green)) in 250 mM KCl on both sides of the bilayer shown. The current traces are filtered digitally at 300 Hz. **B.** Single channel I-V-C traces vs. the concentration of HEPES in 250 mM KCl on both sides of the bilayer. The single channel current decreases as the concentration of HEPES is increased with 250 mM KCl on both sides of the bilayer. HEPES binds to the channel's pore and reduces the unitary current as shown in **A** and **B**. The current traces are filtered digitally at 300 Hz. **C.** Single channel slope conductance measured as the concentration of HEPES is varied from 0 mM to 100 mM in 250 mM KCl on both sides of the bilayer. The unitary slope conductance decreases significantly as the concentration of HEPES increases. N is the number of measurements and the error bars represent the standard error of the mean. **D (i)** and **D(ii)**. Control- Single channel current carried at +200 and -200 mV in presence of symmetric 250 mM KCl, pH 7.0 without HEPES. **E(i)** and **E(ii)**. Single channel current carried at + and − 200 mV respectively in presence of symmetric 250 mM KCl, 100 mM HEPES, pH 7.0. The unitary current decreases by ~50 % compared to the control (0 mM) because of HEPES. Long closures are seen at +/- 200 mV. The current traces are digitally filtered at 500 Hz.

**A**

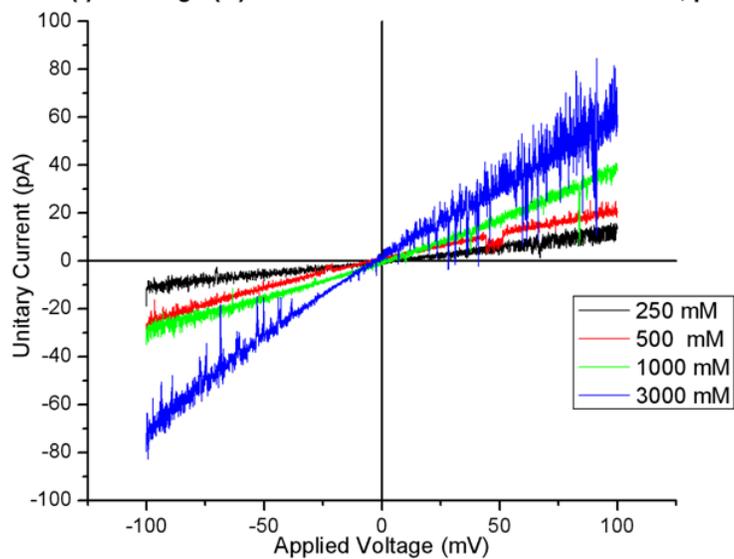

**Current (I) - Voltage (V) curves for unbuffered KCl solutions, pH 7.0**

**B**

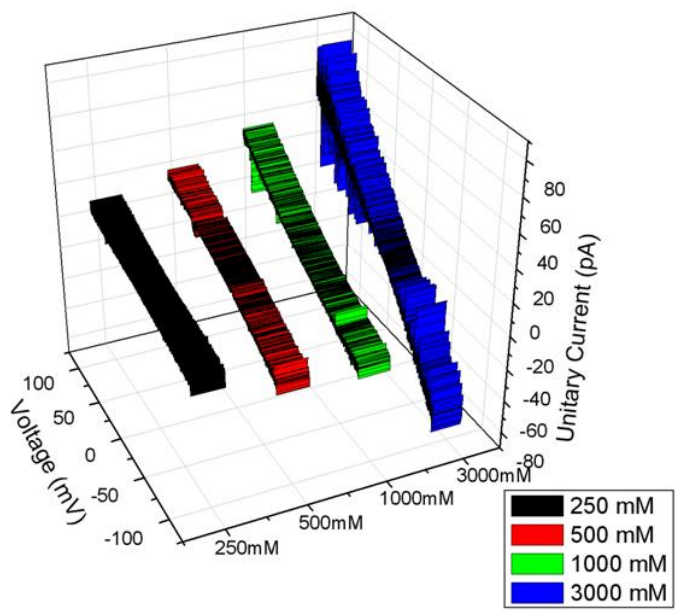

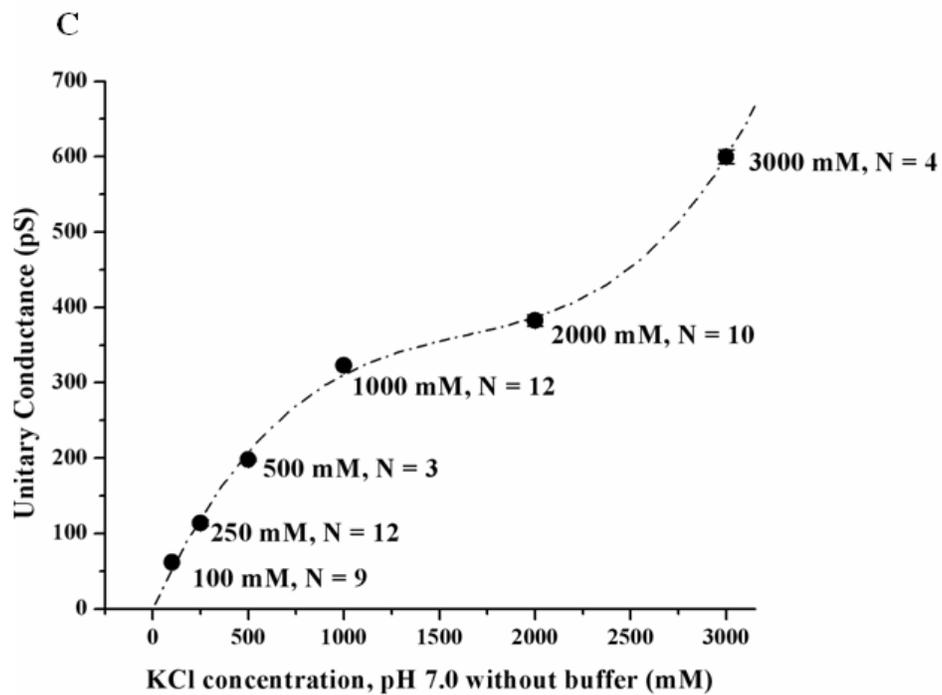

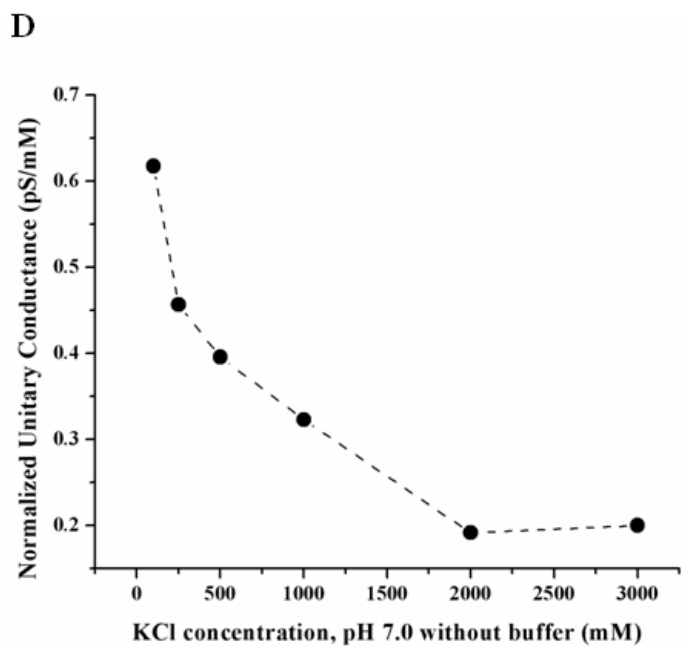

Figure 4.5: Single channel current (I) – voltage (V) traces measured in presence of symmetric KCl (from 250 mM to 3 M), pH 7.0 (without buffer). **B.** Single channel I-V-C traces shown as KCl concentration is varied from 250 mM – 3M, pH 7.0. The unitary current carried by NanC in bilayer increases as the concentration of KCl is increased. The current traces shown in **A** and **B** are filtered digitally at 300 Hz. **C.** Unit slope conductance of NanC determined in KCl salt solutions from 100 mM to 3M, pH 7.0 (without buffer). N denotes the number of measurements and the error bars (too small to be seen) is the standard error mean. The unit slope conductance is determined over the 60 mV range as described in the methods section. The curve in solid line is obtained through polynomial fitting to the measured unit slope conductance data. The single channel slope conductance of NanC increases as the KCl concentration is increased. From these measurements, we chose 250 mM KCl, pH 7.0 (without buffer) as the working solution for further experiments. **D.** Normalized conductance decreases as salt concentration increases.

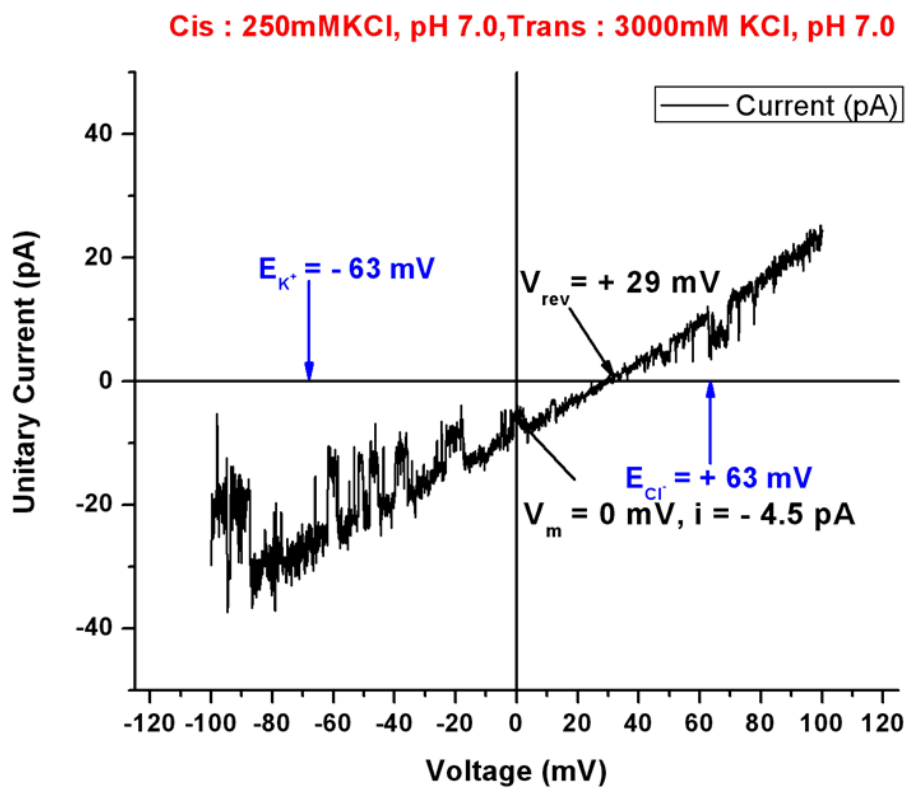

Figure 4.6: Measurement of reversal potential of single NanC in bilayer in presence of 250 mM KCl (*cis*) and 3 M KCl (*trans*) shown. The calculated Nernst potentials $E_K^+$ and $E_{Cl}^-$ are compared to the measured $V_{rev}$. At 0 mV, the current is negative and $V^*_{rev}$ (corrected for Liquid Junction Potential (LJP)) is close to $E_{Cl^-}$ indicating the anion selectivity of NanC. The current trace shown is filtered digitally at 300 Hz.

Table 4.1: Measurement of ion selectivity of single NanC in bilayer. Reversal potential Vrev measured in presence of ionic gradient in cis (250 mM KCl, pH 7.0) and trans (1 M/ 3 M KCl, pH 7.0) compartments compared to the calculated Nernst potentials $E_K^+$ and $E_{Cl}^-$ under those conditions. The measured Vrev indicates the anion selectivity of NanC. $V_{rev}^*$ is the reversal potential obtained after correction for the liquid junction potentials along with the standard error of the mean. N denotes the number of measurements.

| Expt. | *cis* (ground) KCl (mM) | *trans* (voltage) KCl (mM) | LJP (mV) | $E_{Cl^-}$ (mV) | $E_{K^+}$ (mV) | $V_{rev}^*$ (mV) |
|---|---|---|---|---|---|---|
| 1 | 250 | 1000 | -0.7 | + 35 | -35 | +15.89±1.01 (N = 2) |
| 2 | 250 | 3000 | -1.2 | + 63 | -63 | +28.31±0.37 (N = 9) |

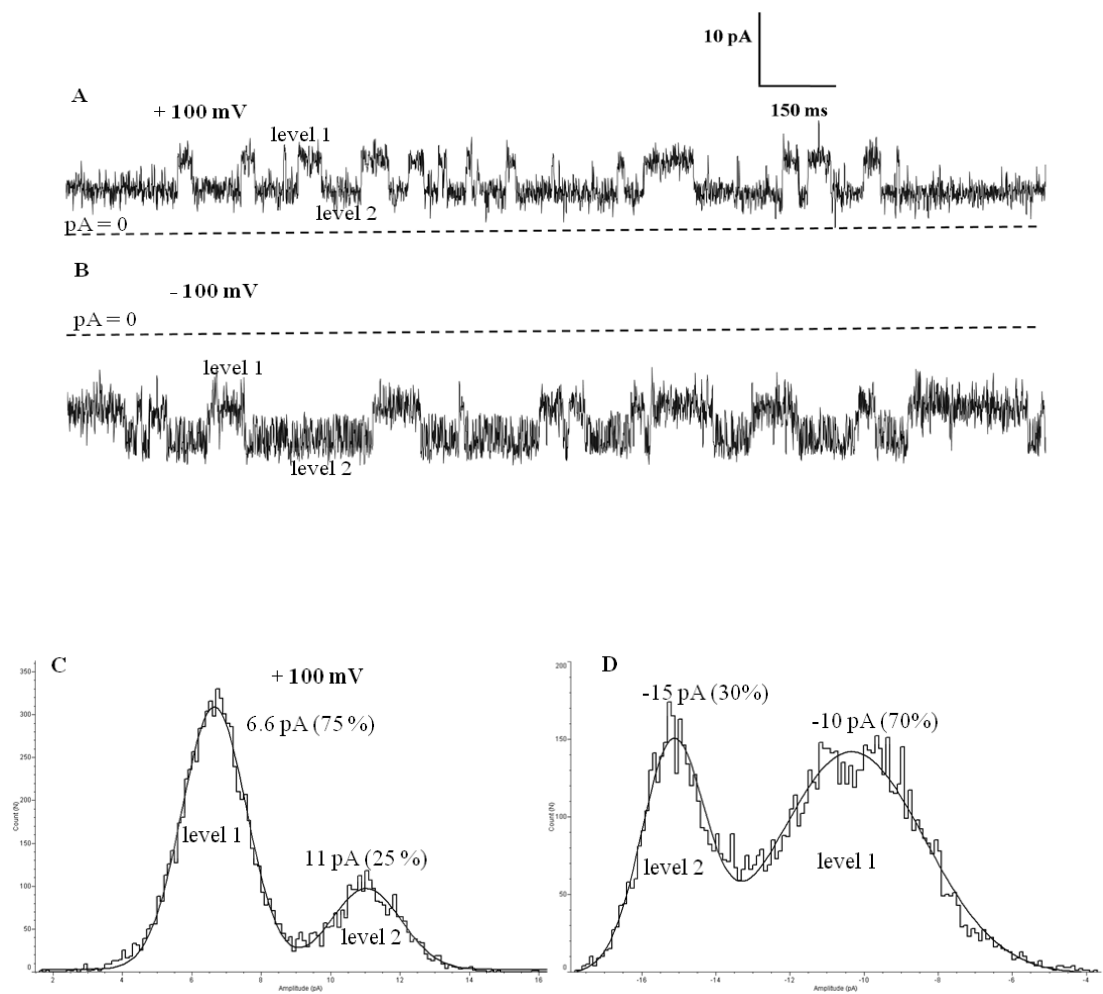

Figure 4.7: The two distinct sub-conductance states of single channel NanC in bilayer shown in symmetric 250 mM KCl, pH 7.0 at + and − 100 mV voltages in **A** and **B** respectively. The amplitudes of the two current levels shown in **C** and **D** are determined from the amplitude histogram analysis of the traces in **A** and **B**. Percentage values in parenthesis denote the probability the channel stays in the corresponding level for the duration the measurement is carried out. Sub-conductance states are native to the NanC and not induced by the experimental conditions i.e. in presence of buffer for example, HEPES/Phosphate or without buffer and high salt concentration. The occurrence of sub-conductance states is not so often (∼ < 5 %). The current traces shown are filtered digitally at 500 Hz. Dashed lines represent the zero current level.

# CHAPTER 5. SIALIC ACID TRANSPORT IN *ESCHERICHIA COLI*: ROLE OF OUTER-MEMBRANE PROTEIN NANC

## 5.1. <u>Introduction</u>

Sialic acids include more than 40 nine carbon negatively charged sugar moieties found primarily at the terminal positions of many eukaryotic surface-exposed glycoconjugates. Sialic acid plays diverse roles important for bacteria as it acts as a nutrient assisting bacterial colonization and interactions with the host (68). Under certain situations, the bacteria have to acquire sialic acid from its host environment and transport it into the cytoplasm where the sialic acid gets metabolized. The sialic acid has to be therefore, transported through the outer-membrane into the periplasm and from the periplasm into the cytoplasm via the inner membrane.

The transport of sialic acid through the inner membrane and its metabolism in *E. coli* is well characterized (69). Sialic acid in bacterial experiments is often used for Neu5Ac (67), the most abundant form of sialic acid. However, the mechanism of transport of Neu5Ac through the outer-membrane into the periplasm remains unresolved. It is widely believed that Neu5Ac is transported via the general porins OmpF/OmpC. However, there is no functional evidence to support their role in Neu5Ac transport. Neu5Ac can in fact induce a specific porin in the outer-membrane of *E.coli* for efficient uptake of Neu5Ac when the general porins OmpF/OmpC are absent. The Neu5Ac-specificity of NanC could not documented in the absence of functional measurements supporting its induction for a specific function (70).

The structure of NanC reveals characteristic features that support the specific role of NanC in Neu5Ac transport across the outer-membrane of *E.coli*(58). The pore region of NanC is predominately decorated by positively charged residues which are arranged such that they form two positively charged tracks facing each other across the pore. This particular arrangement of the positively charged residues in the pore region seems to be favorable for the passage of Neu5Ac through NanC.

In the earlier functional measurements of NanC carried out using patch-clamp experiments no significant change in the functioning of NanC was observed even in presence of upto 50 mM Neu5Ac (70). One of the possible reasons for the failure deduced from NanC's structure were the ionic conditions in which the experiments were performed for example, the concentration of salt solution and the pH buffering agent HEPES.

In this chapter, the goal is to identify the single channel functional behavior of NanC in presence of Neu5Ac and conduct *in vitro* measurement of Neu5Ac transport under the identified suitable ionic conditions in Chapter 4, through the reconstituted single channel NanC and compare its efficiency in transporting Neu5Ac by conducting similar measurements for OmpF.

## 5.2. <u>Methods</u>

### 5.2.1. NanC expression and purification

NanC is purified from the outer-membrane of *E.coli* as explained in the Methods section of Chapter 4.

### 5.2.2. Electrophysiology: Planar Lipid Bilayer Experiments

Planar lipid bilayer method is used to demonstrate and measure the effect of sialic acid on the functional properties of purified channel protein NanC. A single channel NanC is reconstituted in a preformed lipid bilayer in a set-up shown in Figure 3.1. A stock solution of ~ 1.1 $\mu$ g/ml of the purified NanC in 150 mM NaCl, 10 mM Tris, pH 8.0 containing 1 %( v/v) n-OPOE detergent is prepared. For reconstitution of a single NanC to the 'healthy' lipid bilayer we add ~0.1-0.2 $\mu$ l of the stock solution to the *cis* or *ground* side solution while stirring it. The Neu5Ac solution is prepared by dissolving Neu5Ac powder made commercially available by Sigma-Aldrich in Millipore water. The Neu5Ac solution is acidic (pH~2) when prepared. The pH of the Neu5Ac solution is adjusted to neutral pH with basic 1N KOH. The added increase in the concentration of potassium ions due to the asymmetric addition of Neu5Ac solution on one side of bilayer is calculated. The rest of the details concerning the measurement and analysis of data are similar as described in the Methods section of Chapter 4.

## 5.3. <u>Results</u>

### 5.3.1. Ionic Conditions for Sialic Acid Specific Activity of NanC in Bilayer

The specificity of NanC for sialic acid is deduced from its role observed in inducing NanC synthesis for transporting sialic acid into the periplasm in absence of the general porins OmpF/OmpC. However, the specificity of NanC to sialic acid could not been seen in the electrophysiology experiments (70).

From our previous work on the characterization of a single NanC's biophysical properties in bilayers we were able to identify the factors that might have contributed to NanC's insensitivity to sialic acid *i.e.* the ionic conditions (pH buffer and salt concentration) in which the experiments were conducted. We showed in bilayers that NanC is an anion selective channel, exhibits a large single channel conductance, and is voltage-gated. We identified that single channel of NanC shows high affinity to HEPES (a commonly used buffer in electrophysiology) that causes a decreased ion conductance and modulates its function. We were also able to identify from its high resolution structure and its function in bilayers that since, the pore region of NanC is relatively narrow full of positively charged residues; the concentrated salt solutions ($\geq$ 500 mM) may alter the 'fixed' charge distribution of the pore region due to screening effects thus affecting the sialic acid specific activity of NanC.

These key observations from the single channel planar lipid bilayer experiments of NanC suggested the ionic conditions that would be favorable for studying the sialic acid transport. We decided therefore, to carry out the sialic acid transport experiments in salt solutions of concentration ~ 250 mM (preferably KCl) adjusted to neutral pH without the buffer HEPES.

### 5.3.2. Single Channel Function of NanC in presence of Neu5Ac: Titration Experiment

We first carried out Neu5Ac a monomeric sialic acid titration experiment for determining the amount of Neu5Ac that triggers a significant change in the function of NanC. In this experiment we monitored the activity of single channel NanC in bilayer upon asymmetric addition (*i.e.* only in the ground side bath) of Neu5Ac. The control was 250 mM KCl, pH 7.0 with zero Neu5Ac present on both sides of the bilayer. The amount of Neu5Ac in 3.5 ml ground side bath filled with 250 mM KCl, pH 7.0 solution varied from 0 mM-55.37 mM.

In Figure 5.1 we show the single channel current traces recorded in response to the applied step voltages +/- 100 mV for the control experiment. In Figure 5.2 and Figure 5.3 we show the single channel current traces recorded at step voltages +/-100 mV in presence of 3.76 mM and 7.47 mM Neu5Ac. We noticed a obvious change in NanC's function at 7.47 mM Neu5Ac when suddenly, the current amplitude increased compared to its control and a distinct sub-level appeared. The increase in unit current was observed at both positive as well negative voltages. However, no significant increase in the unit current was observed as the amount of added Neu5Ac increased from 7.47 mM to 55.37 mM (Figure 5.4, Figure 5.5, and Figure 5.6). In addition to the increased unit current the gating of NanC changed also due to the addition of Neu5Ac. Distinct sub-levels in unitary current were noticed first at 7.47 mM Neu5Ac (panel A). These sub-levels kept appearing as the amount of added Neu5Ac varied from 7.47 mM to 55.37 mM. The appearance of sub-levels in unitary current was more prominent at negative voltages.

In Figure 5.7 we plot the unit current carried by NanC at +/- 100 mV in presence of Neu5Ac normalized by the control current. The normalized current is plotted vs. the Neu5Ac concentration in the ground side bath. We observe an increasing current ([Neu5Ac] = 7.47 mM − 9.89 mM) eventually saturating with further increase in Neu5Ac concentration. It is interesting to observe that the increase in the unitary current is greater at positive voltage than negative voltage.

We further show in Figure 5.8, the unitary current (I) - voltage (V) traces recorded in response to the ramp voltage protocol while adding Neu5Ac to the ground side bath. We show the I-V traces for the control and in presence of Neu5Ac (7.47 mM and 55.37 mM). It is obvious from these traces that the unitary current increases significantly from the control when 7.47 mM Neu5Ac is present. The slope conductance measurement is carried out over 60 mV range. The single channel slope conductance at 7.47 mM increases by 51% and by 74 % at 55.37 mM Neu5Ac compared to the control. The increase in slope conductance from 7.47 mM to 55.37 mM Neu5Ac is ~ 15 % indicating a sluggish increase in the ionic conductance compared to the drastic increase noticed at 7.47 mM Neu5Ac suggesting an eventual saturation to Neu5Ac sensitivity.  I-V traces appear to become linear as the amount of added Neu5Ac is increased from 0 mM to 55.37 mM (Figure 5.9, 3D plot).

In Figure 5.8, we also notice a shift in the reversal potential $V_{rev}$ from its zero value in control due to the asymmetric addition of Neu5Ac. The $V_{rev}$ shifts from zero towards the positive direction because of a change in the concentration of potassium ions contributed by the Neu5Ac potassium salt *i.e.* K-Neu5Ac. The amount of Cl⁻ remains equal on both sides. The positive shift of the $V_{rev}$ and the negative direction of the unit current at 0 mV are consistent with the direction of the gradient for potassium ions (more K⁺ on the ground side compared to the voltage side bath). However, since Neu5Ac is negatively charged and is present only on the ground side there is a gradient created for Neu5Ac as well and therefore, it also contributes to the current at 0 mV.

The titration experiment revealed the sensitivity of NanC to Neu5Ac and the amount of Neu5Ac at which NanC's function changes considerably. Neu5Ac appears to contribute to the ionic current resulting in an increased net unit current. The presence of Neu5Ac modulates the NanC's gating by inducing sub-current levels and increasing the frequency at which it opens or closes. The appearance of sub-levels seems to be because of a distinct steady current that overlaps with the control current (seen clearly when [Neu5Ac] ≥ 7.47 mM) at negative voltage and occasionally at positive voltage thus, resulting in an increased net ionic current. The action of Neu5Ac on NanC's function seems to be voltage-dependent. The shift in reversal potential due to Neu5Ac is significant suggesting its specific interaction with NanC.

### 5.3.3. Single Channel Function of NanC in presence of Neu5Ac: Symmetric Addition Experiment

The titration experiment was important for testing the sensitivity of NanC to Neu5Ac. It indicated the amount of Neu5Ac that induces a noticeable change in the function of NanC. However, since the permeation scenario of NanC is complicated due to the anions/cations and Neu5Ac that are conducted it becomes difficult to interpret the biophysical basis of permeation or transport of Neu5Ac through NanC. Therefore, the next series of experiments were carried out under symmetric conditions *i.e.* in presence of equal amounts of salt solution and Neu5Ac on both (ground and voltage) sides of the bilayer.

We monitored the single channel activity of NanC in presence of 20 mM Neu5Ac in 250 mM KCl, pH 7.0 on both sides of the bilayer and compared to the control 0 mM Neu5Ac in 250 mM KCl, pH 7.0. Figure 5.10-Figure 5.15, shows the unitary current measured at step voltages +/- 100 mV, +/- 150 mV, and +/- 200 mV in presence and absence of Neu5Ac. The channel under control conditions remains mostly open at +/- 100 mV but in presence of Neu5Ac the gating changes with rapid closures. A distinct steady current appears seen as a sub-level. A considerable increase in the unitary current amplitude is noticed. At higher voltages (+/-150 mV and +/-200 mV) the channel closes under control conditions whereas in presence of Neu5Ac the unitary current goes into multi-levels. The separate distinct sub-level in the measured unitary current appears to gate independently. These observations suggest that Neu5Ac seems to carry current contributing to the ionic current and changes the gating properties of NanC.

Figure 5.16 shows the measured I-V traces under control conditions vs. in the presence of Neu5Ac. The slope conductance is determined over 60 mV range and it turns out to be significantly larger in presence of Neu5Ac. The measured I-V trace appears to become linear in presence of Neu5Ac.

From these single channel experiments of NanC in presence of Neu5Ac we are therefore, able to demonstrate that the ionic conductance of NanC increases significantly due to the addition of Neu5Ac suggesting that Neu5Ac is actually conducted through NanC.

### 5.3.4. Single Channel Function of NanC in presence of Neu5Ac and HEPES

The presence of HEPES has been shown to interfere with the ionic conductance of NanC in bilayers by dramatically reducing its unit conductance. On the other hand, Neu5Ac tends to increase the unit conductance of NanC. In order to investigate the combined consequences of HEPES and Neu5Ac on the single channel function of NanC we carried out the experiments in presence of equal amounts of Neu5Ac (20 mM) and HEPES (5 mM) in 250 mM KCl, pH 7.4 on both sides of the bilayer. We monitored the single channel function of NanC and found out that Neu5Ac further decreased the unit current and modified the gating.

Figure 5.17 shows the I-V traces measured under control conditions (250 mM KCl, 0 mM HEPES, 0 mM Neu5Ac, pH 7.0 in black) vs. in presence of HEPES (250 mM KCl, 5 mM HEPES, 0 mM Neu5Ac, pH 7.4 in green), Neu5Ac (250 mM KCl, 0 mM HEPES, 20 mM Neu5Ac, pH 7.0 in red) and HEPES + Neu5Ac (250 mM KCl, 5 mM HEPES, 20 mM

HEPES, pH 7.4 in blue). The slope conductance is determined over 60 mV range. The slope conductance of NanC decreases by 38% in presence of HEPES (250 mM KCl, 5 mM HEPES, 0 mM Neu5Ac, pH 7.4) and increases by 98% in presence of Neu5Ac (250 mM KCl, 0 mM HEPES, 20 mM Neu5Ac, pH 7.0) in comparison to the control. Interestingly, in presence of HEPES and Neu5Ac (250 mM KCl, 5 mM HEPES, 20 mM HEPES, pH 7.4) the unit conductance drops by 42% from its control value. However, there is only ~ 6% decrease from its value in presence of just HEPES.

We compare the unit currents measured at step voltages (+/-100 mV) for the control Figure 5.18, HEPES Figure 5.19, Neu5Ac Figure 5.20 and HEPES + Neu5Ac Figure 5.21 experiments. Under control conditions and at these voltages Figure 5.18 the channel remains mostly open with a bare minimum activity. In presence of just HEPES Figure 5.19 the channel seems to gate slightly more with decreased amplitude. When Neu5Ac is present with zero HEPES Figure 5.20 the unit current is increased along with the gating. Distinct sub-levels in the unit current are noticed. However, when Neu5Ac and HEPES are present together the measured amplitude of the unit current decreases Figure 5.21 compared to the unit current measured in control, just HEPES, and Neu5Ac. Distinct sub-levels are induced as well but they stay open for longer in comparison to Figure 5.20 and gate independently noticed as "flickers" (mostly at positive voltages) or sub-levels (at negative voltages). This additional level in the measured unit current appears to be the current carried by Neu5Ac on top of the current carried by the ions. This observation comes from the previous results of bilayer experiments and the structure (58) of NanC that demonstrate the high affinity of HEPES in its pore region. HEPES seems to reduce the ion/Neu5Ac accessible region of NanC's pore thereby reducing the net current. Therefore, it is most likely that the separate level that we see at negative voltage in Figure 5.20 and Figure 5.21 is contributed by Neu5Ac.

### 5.4. <u>Discussion and Conclusions</u>

Sialic acid (family of nine carbon sugar acids) is known to be used by bacteria for a variety of purposes that play important roles necessary for their survival, colonize, and cause diseases. The sialic acid transport from the periplasm of the bacterium into the cytoplasm where its metabolism takes place is characterized in great detail (69). However, little was known about the acquisition and transport of sialic acid across the outer-membrane. The general belief was that the bacterium transports sialic acid from the external environment into the periplasm via the general non-specific porins OmpF/OmpC present in the outer-membrane.

Condemine *et.al*., (70) demonstrated from their bacterial growth experiments that when Neu5Ac (monomeric form of sialic acid) was the sole carbon source and the general porins OmpF/OmpC were absent a specific porin NanC was induced to promote the uptake of Neu5Ac required by the bacteria for growth. However, their functional studies gave negative results. Therefore, the specificity of NanC over the general porins for Neu5Ac transport could not be established.

It became even more puzzling when the high resolution structure of NanC revealed structural features suggesting its specific role in Neu5Ac transport(58). We were therefore, able to successfully identify the factors that may have altered the pore architecture responsible for facilitating the Neu5Ac transport. Under the identified suitable conditions we have demonstrated the role of NanC in conducting Neu5Ac efficiently compared to the general porin OmpF (Figure 5.22, Figure 5.23, and Figure 5.24). Sialic acid binds and closes the three pores of the OmpF fully or partially at times in bilayers inhibiting the movement of sialic acid through the single  trimeric channel.

We have thus, presented the biophysical evidence of specificity of NanC to Neu5Ac. Further characterization of the underlying mechanism of sialic acid transport will require a diverse range of experiments where the different forms of sialic acid like dimeric, trimeric and mixture of homopolymer of sialic acid *i.e.* colominic acid will be tested on NanC under different ionic conditions. We believe that these experiments will be of enormous importance in understanding the biophysical, physiological and pathological role of sialic acid transport across the outer-membrane of *E.coli*.

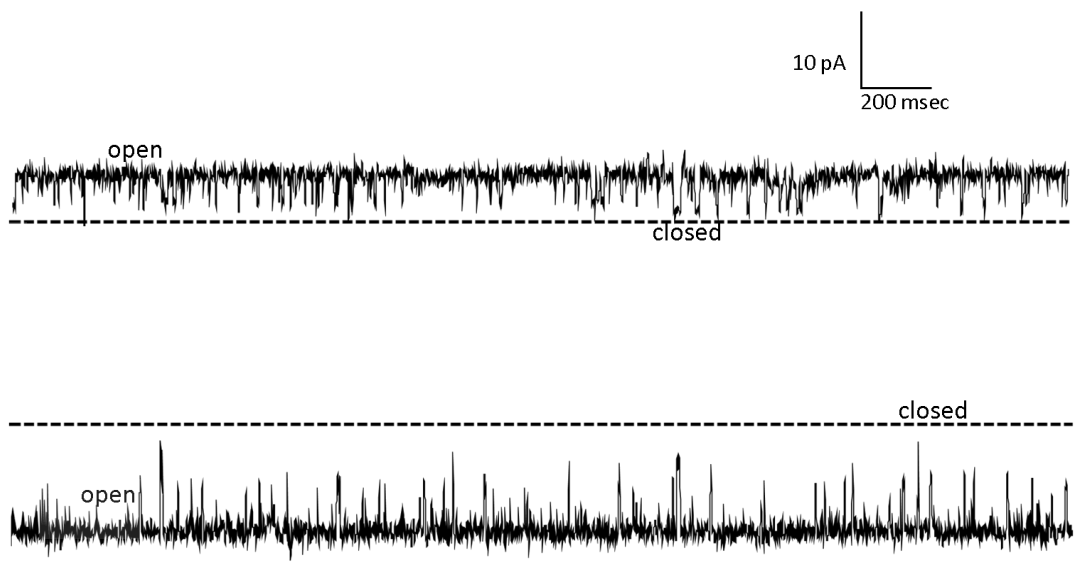

Figure 5.1: Control experiment: Unitary current traces recorded without sialic acid under conditions, ground (cis)/voltage (trans) 250 mM KCl, 0 mM Neu5Ac, pH 7.0 at + 100 mV and -100 mV respectively. The current traces in this figure and the subsequent figures (Figure 4b-4f) shown at step voltages (+/- 100 mV) are sampled at 5KHz and are filtered (analog) at 2KHz with low pass 8-pole Bessel filter. For analysis these current traces are filtered digitally at 300 Hz using low pass 8-pole Bessel filter and are corrected for leakage. The current amplitudes are determined from amplitude histogram analysis of these filtered traces. The unitary current amplitudes are $I_{+100\ mV}$ = 6.14 pA and $I_{-100\ mV}$ = -14.56 pA. Dashed lines represent the zero current level.

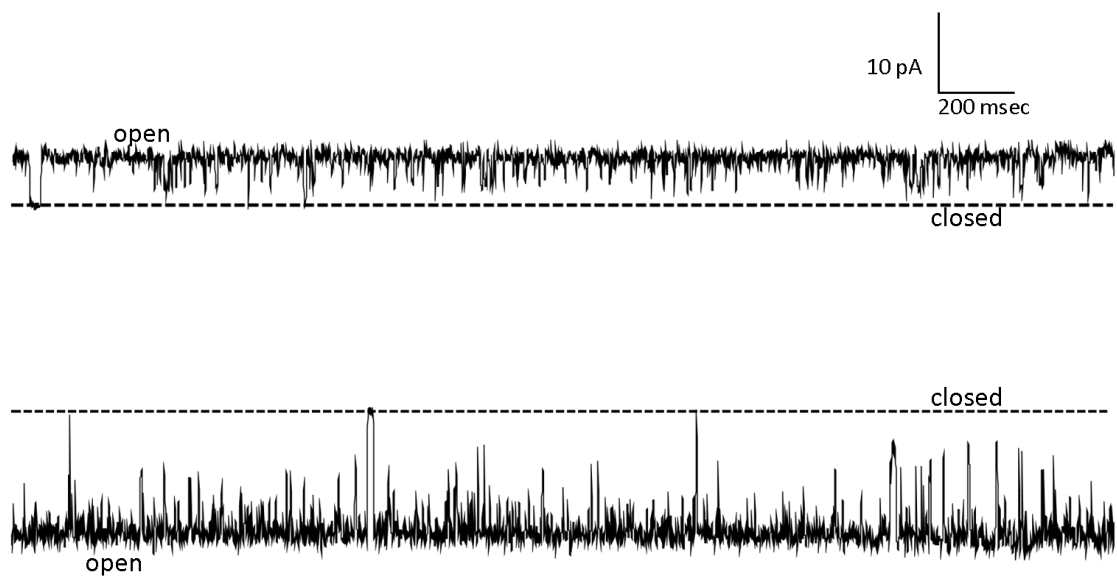

Figure 5.2: Sialic acid titration experiment: Unitary current traces recorded when sialic acid (monomeric Neu5Ac) is added 'only' to the ground side (cis) - 250 mM KCl, 3.76 mM Neu5Ac, pH 7.0 and voltage side (trans)- 250 mM KCl, 0 mM Neu5Ac, pH 7.0   at + 100 mV and -100 mV respectively. The unitary current amplitudes are $I_{+100\ mV}$ = 6.09 pA and $I_{-100\ mV}$ = -15.14 pA. Dashed lines represent the zero current level.

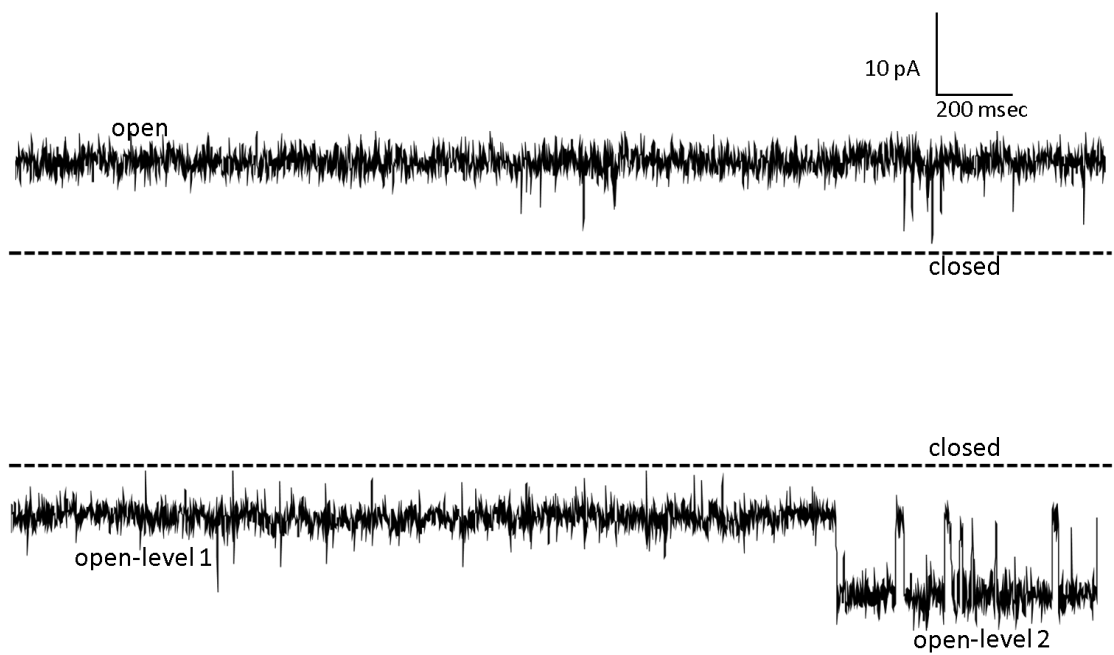

Figure 5.3: Sialic acid titration experiment: Unitary current traces recorded when sialic acid (monomeric Neu5Ac) is added 'only' to the ground side (cis) - 250 mM KCl, 7.47 mM Neu5Ac, pH 7.0 and voltage side (trans)- 250 mM KCl, 0 mM Neu5Ac, pH 7.0 at + 100 mV and -100 mV respectively. The unitary current amplitudes are $I_{+100\ mV}$ = 12.25 pA, $I_{-100\ mV}$ (level-1) = -8.23 pA, and $I_{-100\ mV}$ (level-2) = -18.04 pA .Dashed lines represent the zero current level.

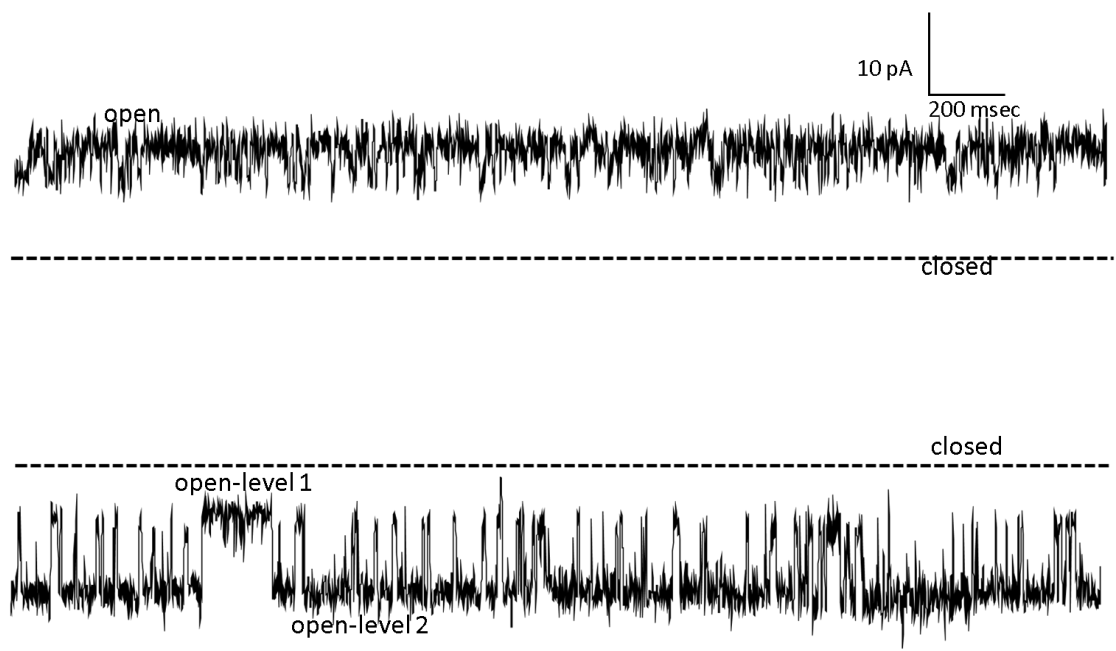

Figure 5.4: Sialic acid titration experiment: Unitary current traces recorded when sialic acid (monomeric Neu5Ac) is added 'only' to the ground side (cis) - 250 mM KCl, 9.89 mM Neu5Ac, pH 7.0 and voltage side (trans)- 250 mM KCl, 0 mM Neu5Ac, pH 7.0 at + 100 mV and -100 mV respectively. The unitary current amplitudes are $I_{+100\ mV}$ = 15.74 pA, $I_{-100\ mV}$ (level-1) = -9.63 pA, and $I_{-100\ mV}$ (level-2) = -18.18 pA. Dashed lines represent the zero current level.

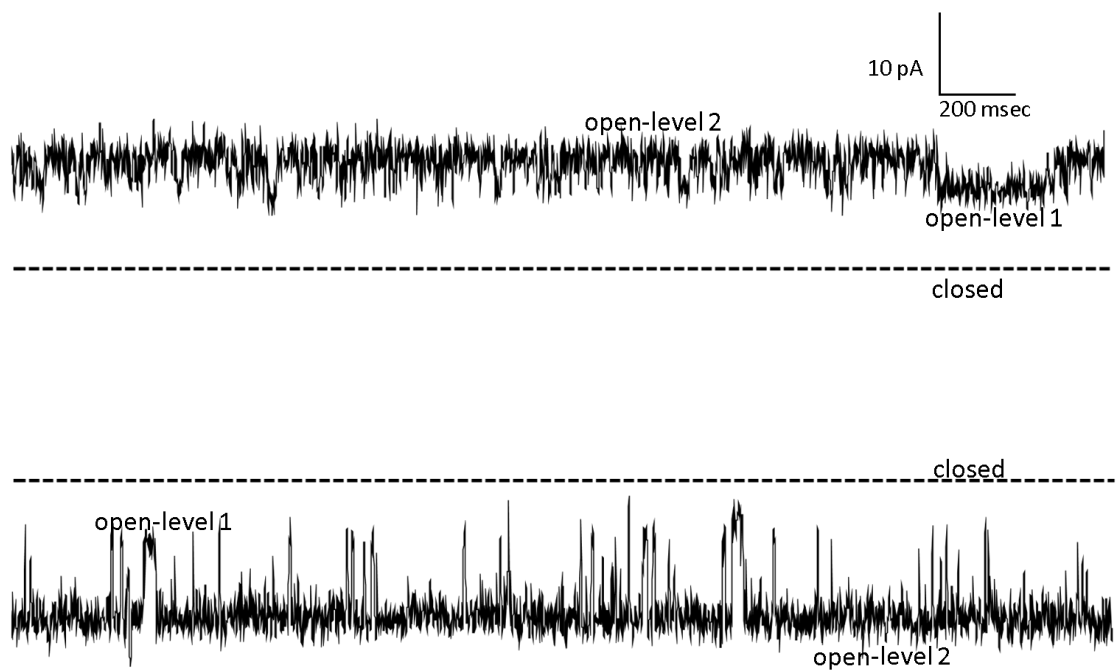

Figure 5.5: Sialic acid titration experiment: Unitary current traces recorded when sialic acid (monomeric Neu5Ac) is added 'only' to the ground side (cis) - 250 mM KCl, 23.94 mM Neu5Ac, pH 7.0 and voltage side (trans)- 250 mM KCl, 0 mM Neu5Ac, pH 7.0 at + 100 mV and -100 mV respectively. The unitary current amplitudes are $I_{+100 \text{ mV}}$ (level-1) = 11.11 pA, $I_{+100 \text{ mV}}$ (level-2) = 14.72 pA, $I_{-100 \text{ mV}}$ (level-1) = -8.82 pA, and $I_{-100 \text{ mV}}$ (level-2) = -19.08 pA. Dashed lines represent the zero current level.

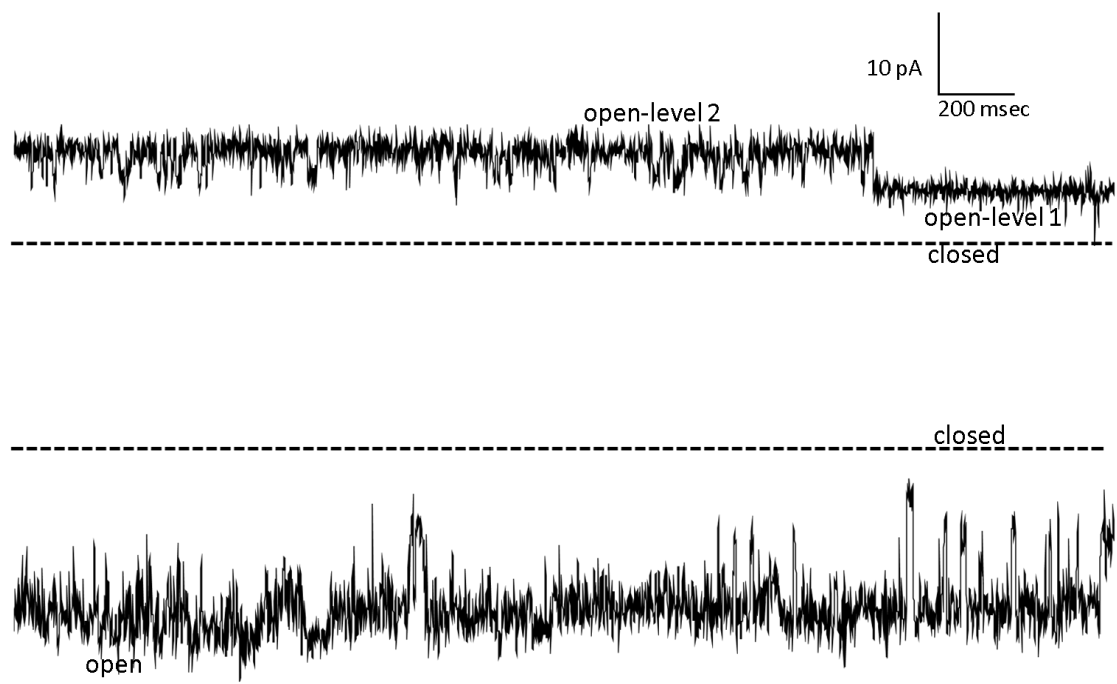

Figure 5.6: Sialic acid titration experiment: Unitary current traces recorded when sialic acid (monomeric Neu5Ac) is added 'only' to the ground side (cis) - 250 mM KCl, 55.37 mM Neu5Ac, pH 7.0 and voltage side (trans)- 250 mM KCl, 0 mM Neu5Ac, pH 7.0 at + 100 mV and -100 mV respectively. The unitary current amplitudes are $I_{+100\,mV}$ (level-1) = 7.58 pA, $I_{+100\,mV}$ (level-2) = 12.72 pA, and $I_{-100\,mV}$ = -20.13 pA. Dashed lines represent the zero current level.

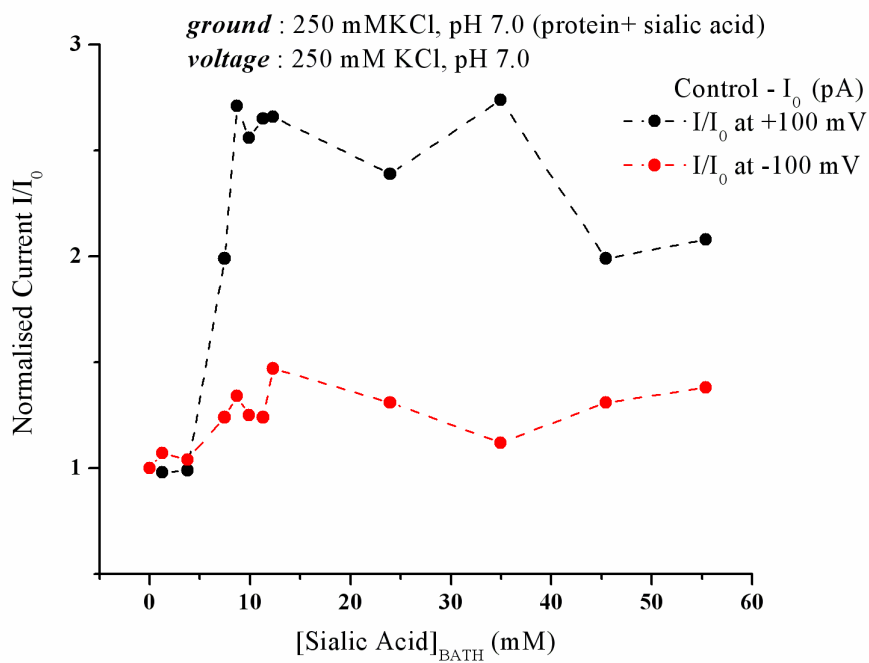

Figure 5.7: Normalized unit current plotted as varying concentration of sialic acid (0 mM -55.37 mM) in the ground side bath. The unit current recorded at +/-100 mV is normalized by the corresponding unit current under control (zero sialic acid) conditions. The normalized unit current increases as the amount of added sialic acid increases and seems to saturate.  At + 100 mV (black), the increase in the unit current appears to be larger than at - 100 mV (red). Dashed lines in black and red are drawn to join points and carry no physical meaning.

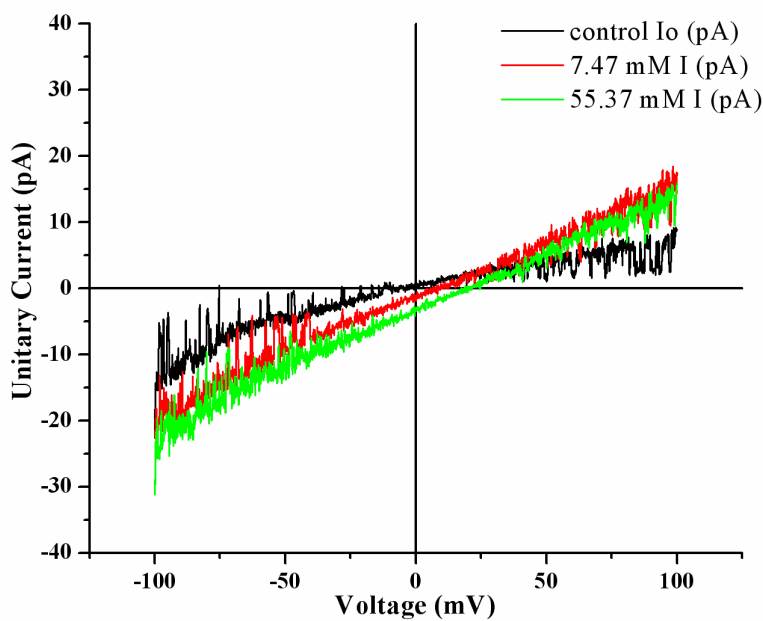

Figure 5.8: Single channel current (I)-voltage (V) traces plotted for the control experiment $I_o$ (black) in 250 mM KCl, pH 7.0 (ground side) and 250 mM KCl, 0 mM Neu5Ac, pH 7.0 (voltage side). vs. the added sialic acid experiment in presence of 7.47 mM $I_{7.47\ mM}$ (red) and 55.37 mM $I_{55.37\ mM}$ (green) in 250 mM KCl, x mM Neu5Ac, pH 7.0 (ground side) and 250 mM KCl, 0 mM Neu5Ac, pH 7.0 (voltage side). The slope conductances measured over 60 mV range under control conditions $g_0 = 96.23$ pS, in presence of 7.47 mM $g_{7.47\ mM} = 144.89$ pS, and $g_{55.37\ mM} = 166.54$ pS. Significant increase in slope conductance occurs as the amount of sialic acid increases. In presence of sialic acid, the unit current at 0 mV is no longer zero and the $V_{rev}$ shifts towards right from zero.

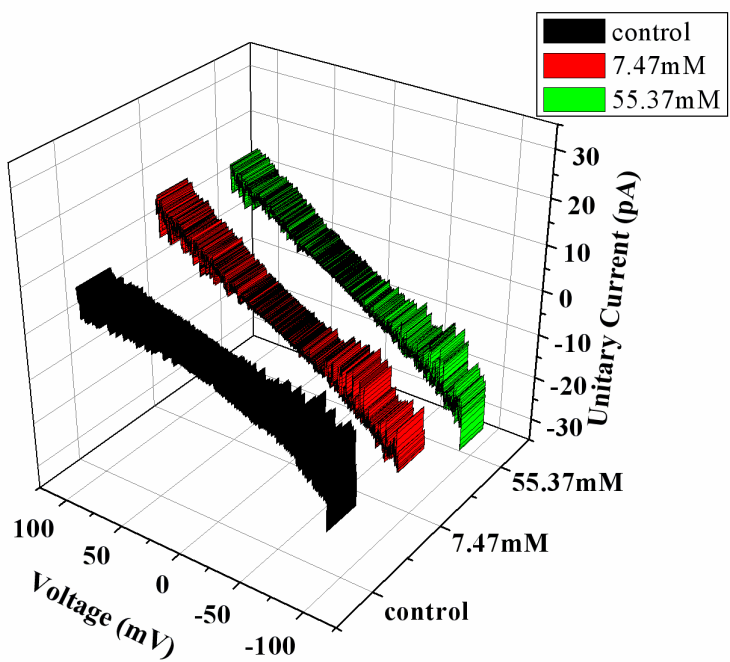

Figure 5.9: Three dimensional plot of unit current (I)-voltage (V) traces plotted vs. concentration of sialic acid for the control experiment $I_o$ (black) and the added sialic acid experiment in presence of 7.47 mM $I_{7.47\ mM}$ (red) and 55.37 mM $I_{55.37\ mM}$ (green) in 250 mM KCl, pH 7.0 (ground side) and 250 mM KCl, 0 mM Neu5Ac, pH 7.0 (voltage side). The I-V traces appear to become linear with increasing sialic acid.

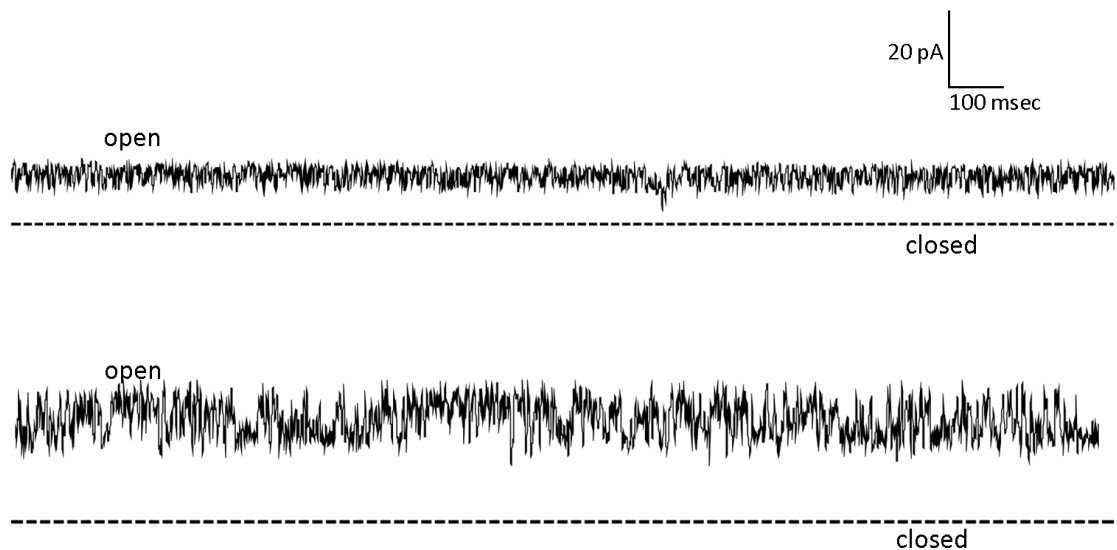

Figure 5.10: Control experiment (top): Unitary current traces recorded without sialic acid under conditions, ground (cis)/voltage (trans) 250 mM KCl, 0 mM Neu5Ac, pH 7.0 at + 100 mV. Sialic acid addition experiment (bottom): Unitary current traces recorded with equal amount of sialic acid on both sides, ground (cis)/voltage (trans) 250 mM KCl, 20 mM Neu5Ac, pH 7.0 at + 100 mV . Unitary current increases in presence of sialic acid (bottom) compared to the control(top). The sub-conductances become distinct with an increase in activity of the channel. The current traces in this figure for the control and the sialic acid experiment in the subsequent figures (Figure 8b-8f) shown at step voltages (+/- 100-200 mV) are sampled at 5KHz and are filtered (analog) at 2KHz with low pass 8-pole Bessel filter. For analysis these current traces are filtered digitally at 250 Hz (control) and 300 Hz (sialic acid) using low pass 8-pole Bessel filter and are corrected for leakage. Dashed lines represent the zero current level.

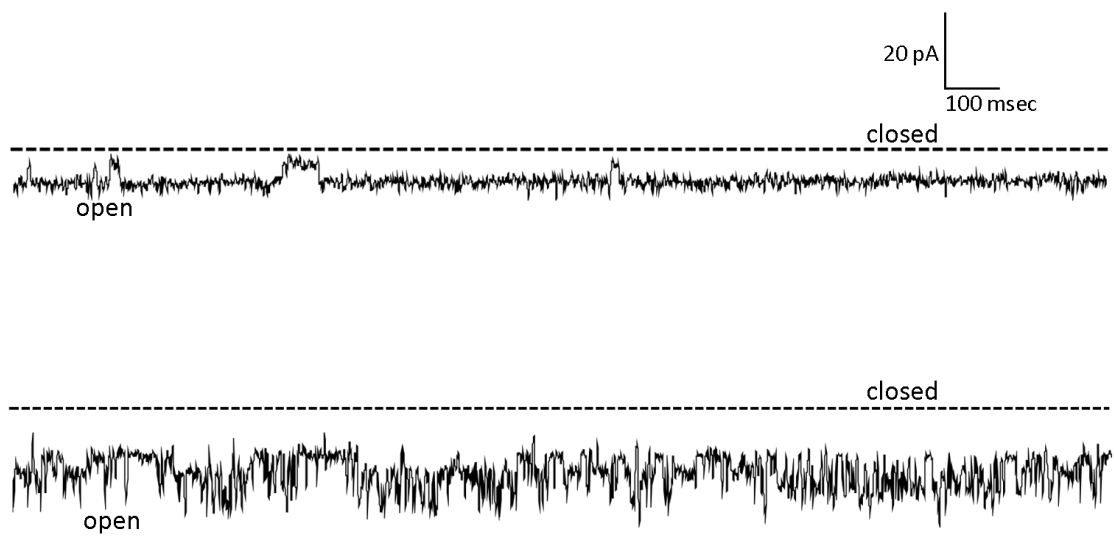

Figure 5.11: Control experiment (top): Unitary current traces recorded without sialic acid under conditions, ground (cis)/voltage (trans) 250 mM KCl, 0 mM Neu5Ac, pH 7.0 at - 100 mV. Sialic acid addition experiment (bottom): Unitary current traces recorded with equal amount of sialic acid on both sides, ground (cis)/voltage (trans) 250 mM KCl, 20 mM Neu5Ac, pH 7.0 at - 100 mV . Unitary current increases in presence of sialic acid (bottom) compared to the control(top). The sub-conductances become distinct with an increase in activity of the channel.

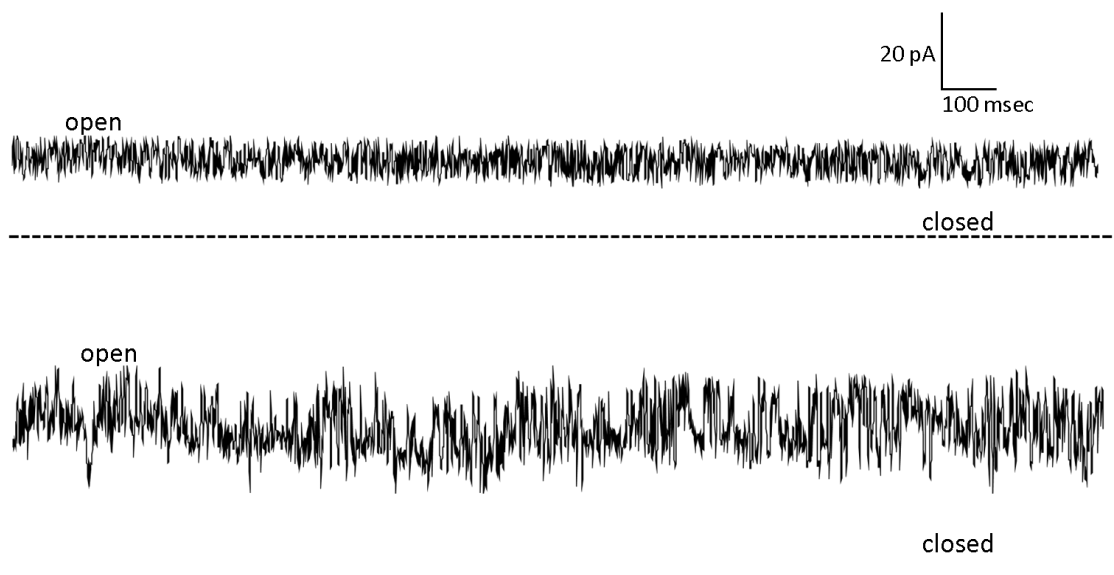

Figure 5.12: Control experiment (top): Unitary current traces recorded without sialic acid under conditions, ground (cis)/voltage (trans) 250 mM KCl, 0 mM Neu5Ac, pH 7.0 at + 150 mV. Sialic acid addition experiment (bottom): Unitary current traces recorded with equal amount of sialic acid on both sides, ground (cis)/voltage (trans) 250 mM KCl, 20 mM Neu5Ac, pH 7.0 at + 150 mV. Unitary current increases in presence of sialic acid (bottom) compared to the control (top). The sub-conductances become distinct with an increase in activity of the channel.

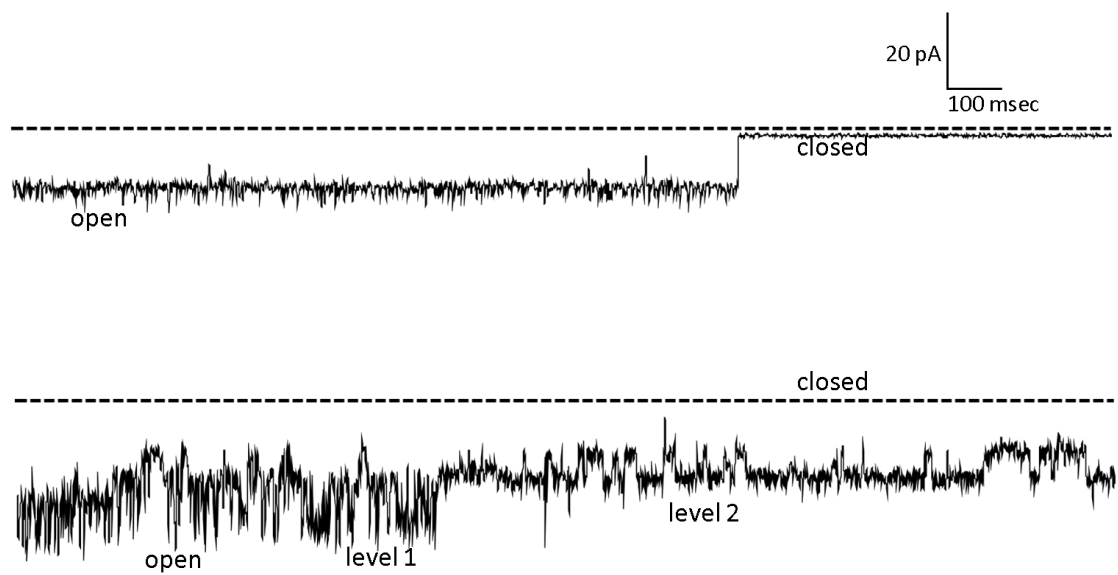

Figure 5.13: Control experiment (top): Unitary current traces recorded without sialic acid under conditions, ground (cis)/voltage (trans) 250 mM KCl, 0 mM Neu5Ac, pH 7.0 at -150 mV. Sialic acid addition experiment (bottom): Unitary current traces recorded with equal amount of sialic acid on both sides, ground (cis)/voltage (trans) 250 mM KCl, 20 mM Neu5Ac, pH 7.0 at -150 mV. Unitary current increases in presence of sialic acid (bottom) compared to the control (top). The sub-conductances become distinct with an increase in activity of the channel.

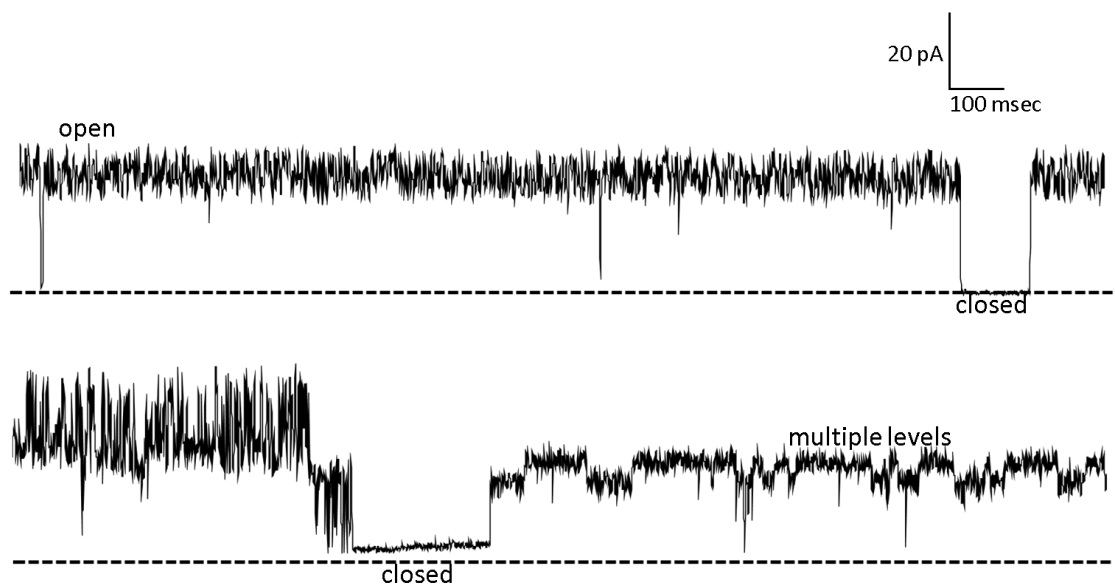

Figure 5.14: Control experiment (top): Unitary current traces recorded without sialic acid under conditions, ground (cis)/voltage (trans) 250 mM KCl, 0 mM Neu5Ac, pH 7.0 at + 200 mV. Sialic acid addition experiment (bottom): Unitary current traces recorded with equal amount of sialic acid on both sides, ground (cis)/voltage (trans) 250 mM KCl, 20 mM Neu5Ac, pH 7.0 at + 200 mV. Unitary current increases in presence of sialic acid (bottom) compared to the control (top). The sub-conductances become distinct with an increase in activity of the channel. Channel closes and opens again at this large applied voltage.

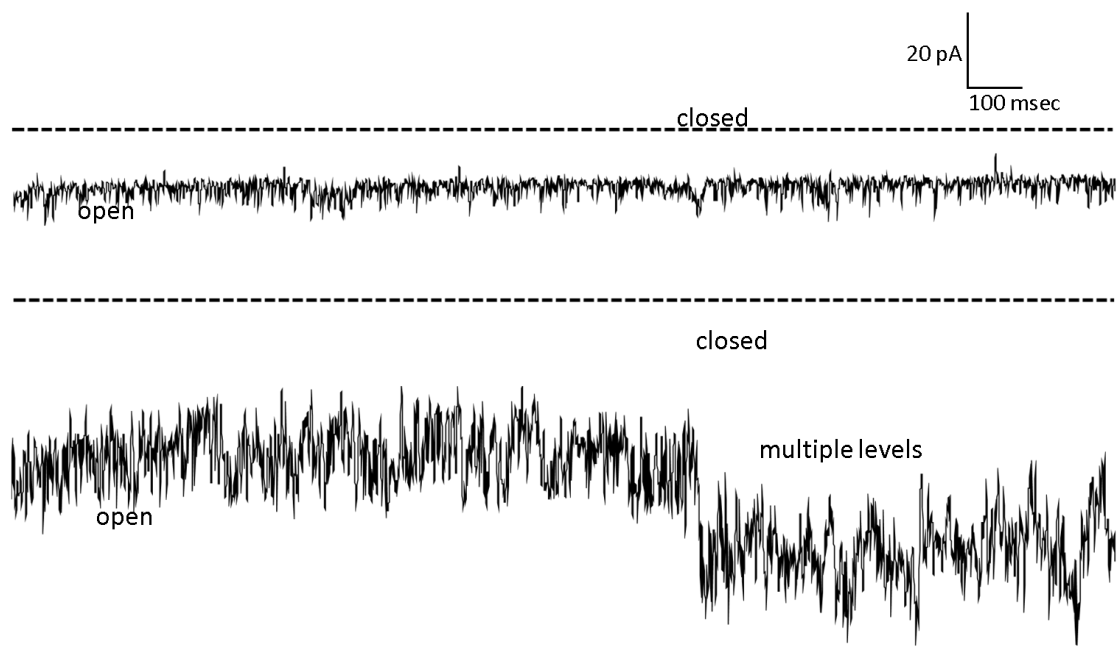

Figure 5.15: Control experiment (top): Unitary current traces recorded without sialic acid under conditions, ground (cis)/voltage (trans) 250 mM KCl, 0 mM Neu5Ac, pH 7.0 at - 200 mV. Sialic acid addition experiment (bottom): Unitary current traces recorded with equal amount of sialic acid on both sides, ground (cis)/voltage (trans) 250 mM KCl, 20 mM Neu5Ac, pH 7.0 at - 200 mV. Unitary current increases in presence of sialic acid (bottom) compared to the control (top). Channel goes into multiple sub-conductance levels. The sub-conductances become distinct with an increase in activity of the channel.

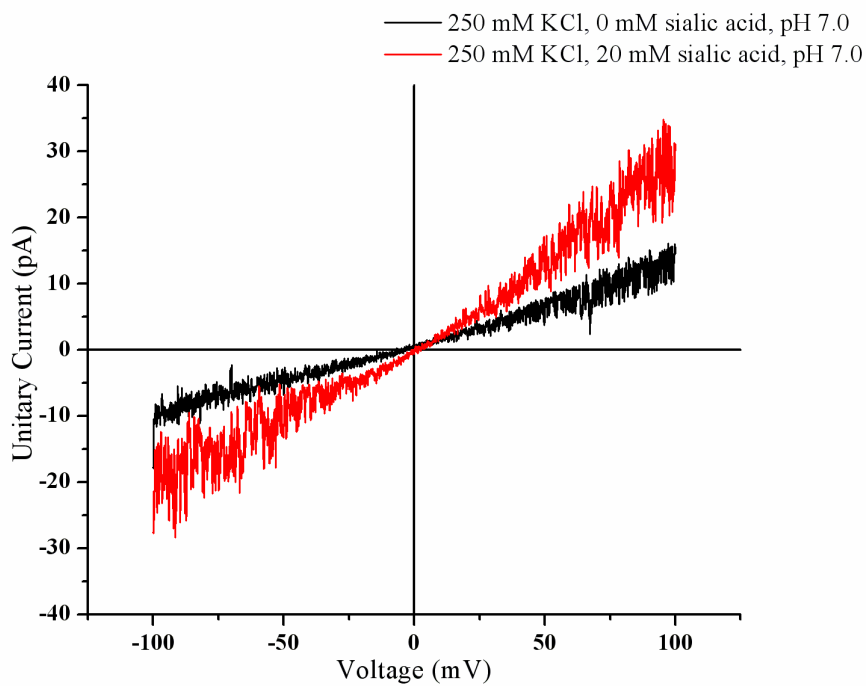

Figure 5.16: Single channel current-voltage traces recorded under control (without sialic acid, black) condition vs. in presence of sialic acid (red) on both sides of the bilayer. The current traces shown are sampled at 5KHz and filtered (analog) at 2KHz with low pass 8-pole Bessel filter. For analysis these current traces are filtered at 250 Hz (control) / 300 Hz (with sialic acid) and are corrected for leakage and offset. The unit slope conductances measured over 60 mV range under control conditions $g_0 = 114.09 \pm 4.16$ pS (n = 12) and in presence of sialic acid $g = 226.15 \pm 9.95$ pS (n = 29), n is the number of measurements. The single channel conductance increases considerably due to sialic acid.

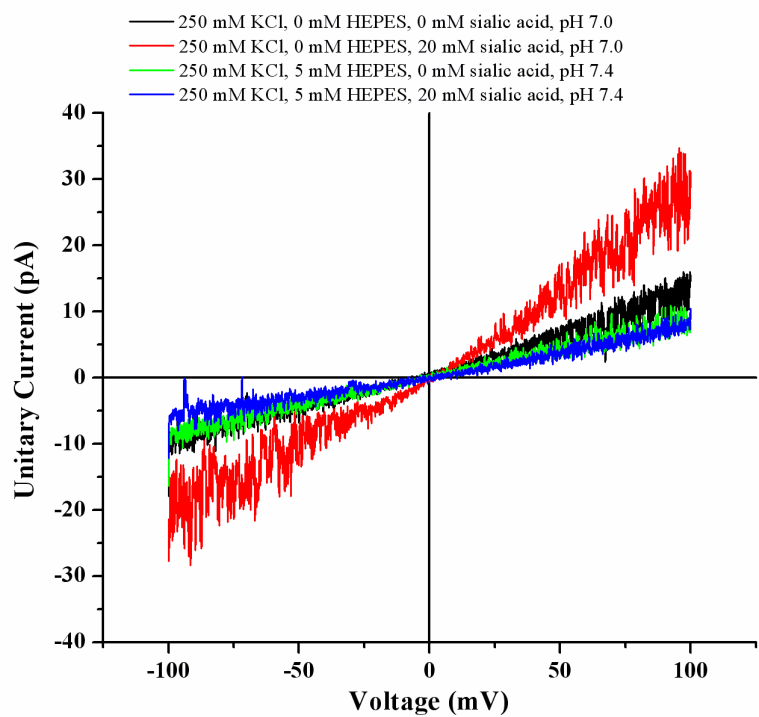

Figure 5.17: Single channel I-V traces shown for control experiment (cis/trans: 250 mM KCl, 0 mM HEPES, 0 mM Neu5Ac, pH 7.0 in black) vs. HEPES experiment (cis/trans: 250 mM KCl, 5 mM HEPES, 0 mM Neu5Ac, pH 7.4 in green) , sialic acid experiment (cis/trans: 250 mM KCl, 0 mM HEPES, 20 mM Neu5Ac, pH 7.0 in red) , and HEPES+ Neu5Ac (cis/trans: 250 mM KCl, 5 mM HEPES, 20 mM Neu5Ac, pH 7.4 in blue). The unit slope conductance measured over 60 mV range under control conditions is $g_0 = 114.09 \pm 4.16$ pS (n = 12), in presence of HEPES without Neu5Ac, $g_{HEPES} = 70.70 \pm 5.17$ pS (n = 15), in presence of Neu5Ac without HEPES $g_{Neu5Ac} = 226.15 \pm 9.95$ pS (n = 29), and in presence of HEPES and Neu5Ac $g_{HEPES+Neu5Ac} = 66.21 \pm 0.74$ pS (n = 26). Significant decrease in unit ionic conductance in comparison to the control takes place in presence of 5 mM HEPES (0 mM Neu5Ac) whereas the unit ionic conductance increases considerably from its control value in presence of 20 mM Neu5Ac (0 Mm HEPES). However, the presence of 5 mM HEPES and 20 mM Neu5Ac considerably decreases the unit ionic conductance from its control value but not a significant change occurs in comparison to the HEPES experiment. Effect of Neu5Ac on the unit ionic conductance of NanC is not so dramatic when HEPES is around.

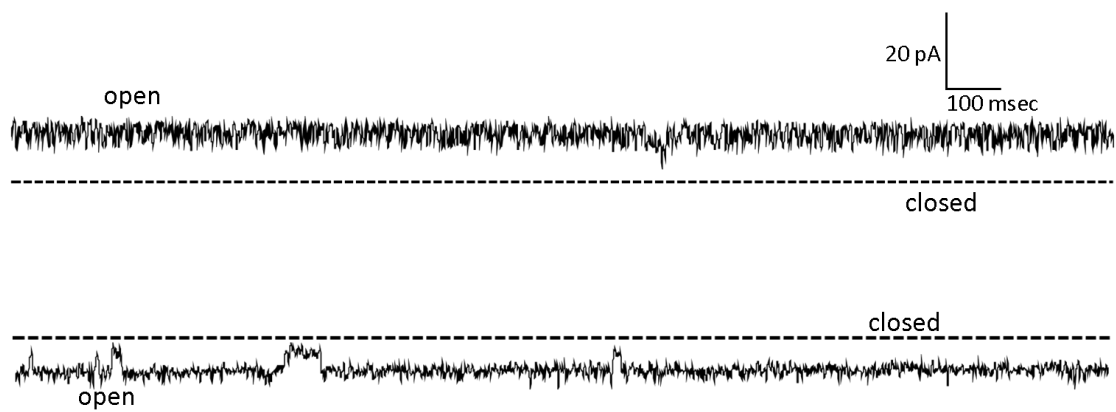

Figure 5.18: Control experiment: Unitary current traces recorded at +/- 100 mV (top and bottom respectively) under symmetric conditions, cis/trans: 250 mM KCl, 0 mM HEPES, 0 mM Neu5Ac, pH 7.0. Channel remains mostly open at these voltages. Dashed lines represent the zero current levels.

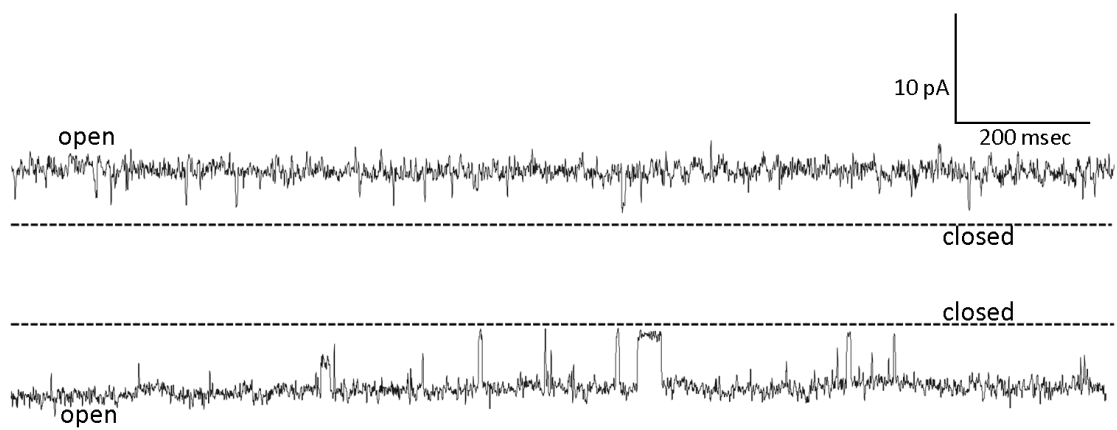

Figure 5.19: HEPES experiment: Unitary current traces recorded at +/- 100 mV (top and bottom respectively) under symmetric conditions, cis/trans: 250 mM KCl, 5 mM HEPES, 0 mM Neu5Ac, pH 7.4. Channel closes occasionally at −100 mV and shows gating. Dashed lines represent the zero current levels.

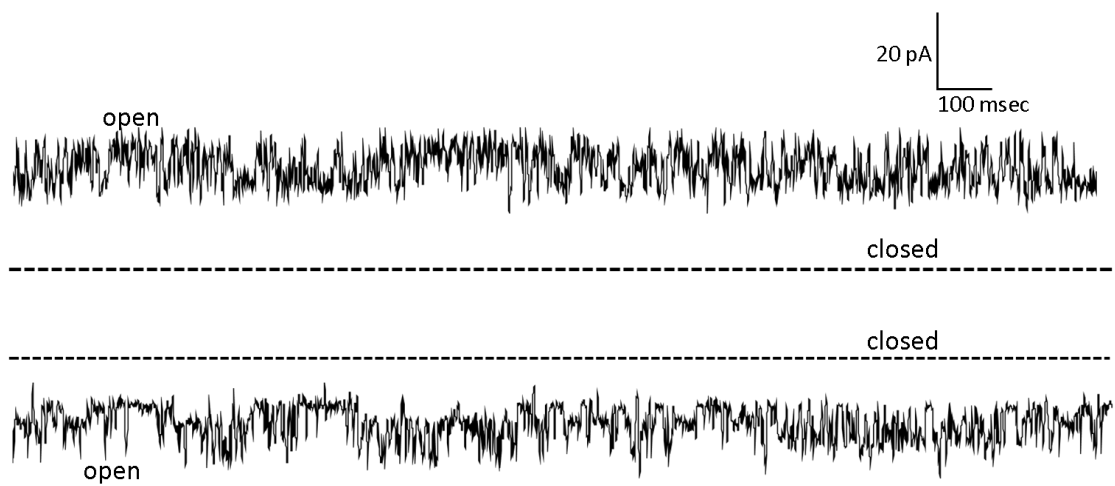

Figure 5.20: Sialic acid experiment: Unitary current traces recorded at +/- 100 mV (top and bottom respectively) under symmetric conditions, cis/trans: 250 mM KCl, 0 mM HEPES, 20 mM Neu5Ac, pH 7.0. Channel shows an increased activity and goes into sub-conductances. Dashed lines represent the zero current levels.

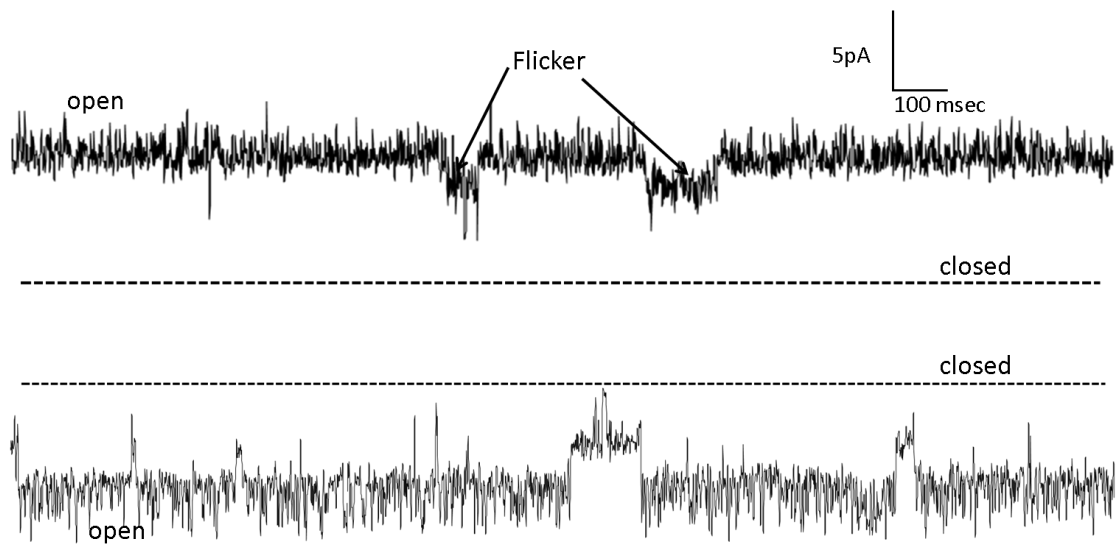

Figure 5.21: HEPES and Sialic acid experiment: Unitary current traces recorded at +/- 100 mV (top and bottom respectively) under symmetric conditions, cis/trans: 250 mM KCl, 5 mM HEPES, 20 mM Neu5Ac, pH 7.0. Channel shows 'flickering' at + 100 mV and at -100 mV, goes into sub-conductances that seem to gate independently. Dashed lines represent the zero current levels.

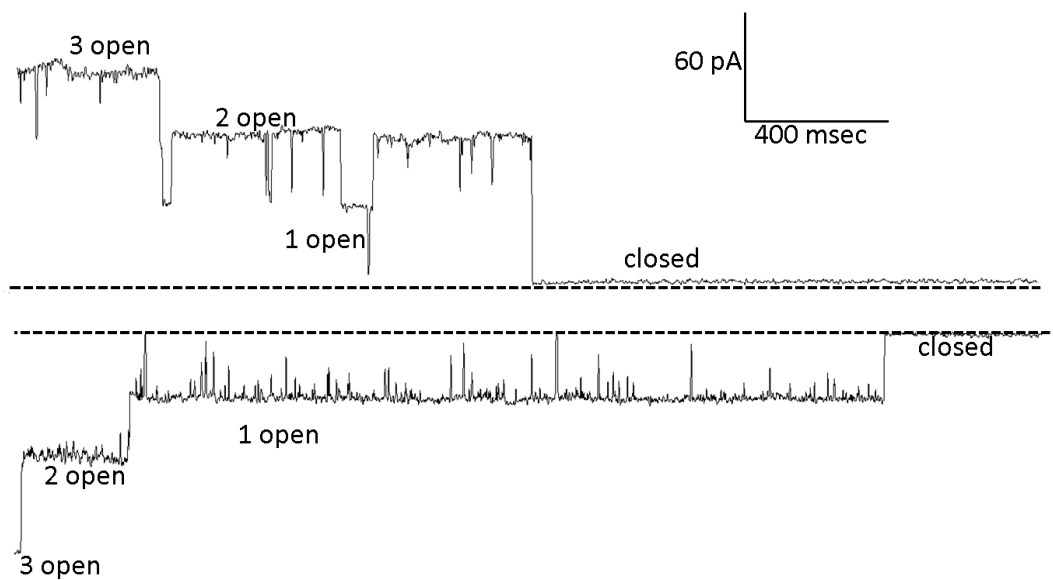

Figure 5.22: Trimeric current traces measured for OmpF (wild type) under symmetric cis/trans : 250 Mm KCl, 0 mM Neu5Ac, pH 7.0 at step voltages + 100 mV (top) and -100 mV (bottom) respectively. A single trimer opens and closes in 3 levels as shown. Dashed lines represent the zero current levels.

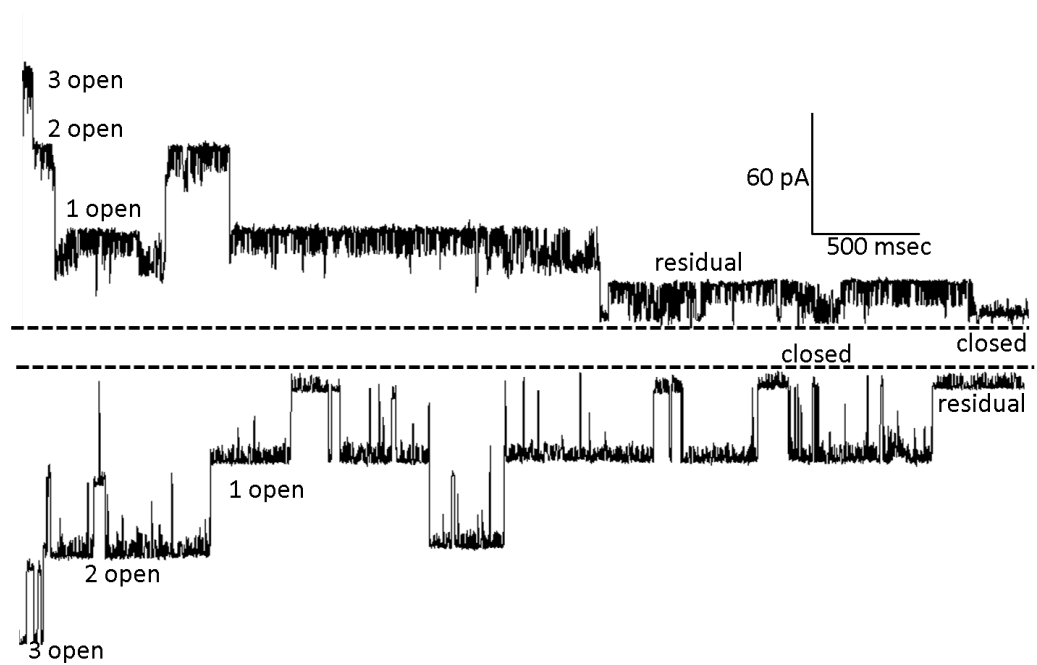

Figure 5.23: Trimeric current traces measured for OmpF (wild type) under symmetric cis/trans: 250 Mm KCl, 8 mM Neu5Ac, pH 7.0 conditions at step voltages + 100 mV (top) and -100 mV (bottom) respectively. A single trimer opens and closes in 3 levels as shown. At + 100 mV, the trimer further opens/closes from the $2^{nd}$ level to the $1^{st}$ level to fully close and later on opens at only one level that gates with a reduced amplitude compared to the $1^{st}$ level in the beginning that eventually closes with some residual current. Dashed lines represent the zero current levels.

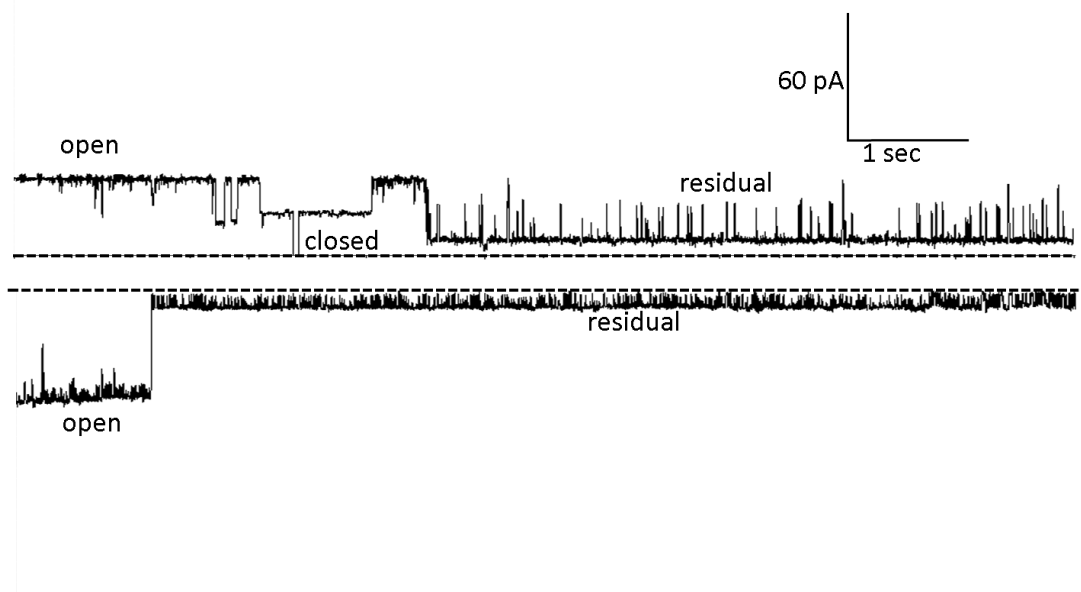

Figure 5.24: Trimeric current traces measured for OmpF (wild type) under symmetric cis/trans: 250 mM KCl, 59.5 mM Neu5Ac, pH 7.0 conditions at step voltages + 100 mV (top) and -100 mV (bottom) respectively. The characteristic trimeric behavior vanishes at larger Neu5Ac concentrations. The three independently functioning pores of a unit OmpF appears to be blocked partially/fully at times and therefore, appear with a reduced amplitude showing a unique gating very different from its control. Dashed lines represent the zero current levels.

# CITED LITERATURE

1.      Alberts, B., et al., Essential Cell Biology. Third ed. 1998, New York: Garland. 630.

2.      Alberts, B., et al., Molecular Biology of the Cell. Third ed. 1994, New York: Garland. 1294.

3.      Gumery, P.Y., et al., An adaptive detector of genioglossus EMG reflex using Berkner transform for time latency measurement in OSA pathophysiological studies. IEEE Trans Biomed Eng, 2005. **52**(8): p. 1382-1389.

4.      Hille, B., Ionic Channels of Excitable Membranes. 3rd ed. 2001, Sunderland: Sinauer Associates Inc. 1-814.

5.      Ashcroft, F.M., Ion Channels and Disease. 1999, New York: Academic Press. 481.

6.      Mikami A, I.K., Tanabe T, Niidome T, Mori Y, et al, Primary structure and functional expression of the cardiac dihydropyridine-sensitive calcium channel. Nature, 1989. **340**: p. 230-233.

7.      Tanabe T, T.H., Mikami A, Flockerzi V, Takahashi H, et al, Primary structure of the receptor for calcium channel blockers from skeletal muscle. Nature, 1987. **328**: p. 313-318.

8.      Wu, X.S., H.D. Edwards, and W.A. Sather, Side chain orientation in the selectivity filter of a voltage-gated Ca2+ channel. J.Biol.Chem., 2000. **275**: p. 31778-31785.

9.      Yang, J., et al., Molecular determinants of Ca2+ selectivity and ion permeation in L-type Ca2+ channels. Nature, 1993. **366**: p. 158-161.

10.     Favre, I., E. Moczydlowski, and L. Schild, On the structural basis for ionic selectivity among Na+, K+, and Ca2+ in the voltage-gated sodium channel. Biophys J, 1996. **71**(6): p. 3110-3125.

11.     Sun, Y.M., et al., On the structural basis for size-selective permeation of organic cations through the voltage-gated sodium channel. Effect of alanine mutations at the DEKA locus on selectivity, inhibition by Ca2+ and H+, and molecular sieving. J Gen Physiol, 1997. **110**(6): p. 693-715.

12.     Clay, J.R., Potassium current in the squid giant axon. Int Rev Neurobiol, 1985. **27**: p. 363-384.

13.     Hodgkin, A.L., The Conduction of the Nervous Impulse. 1971, Liverpool: Liverpool University Press. 1-108.

14.     Hodgkin, A.L., Chance and Design. 1992, New York: Cambridge University Press. 401.

15.     Benham, C.D., P. Hess, and R.W. Tsien, Two types of calcium channels in single smooth muscle cells from rabbit ear artery studied with whole-cell and single-channel recordings. Circ Res, 1987. **61**(4 Pt 2): p. I10-16.

16.     Dimitrov, R.A. and R.R. Crichton, Self-consistent field approach to protein structure and stability. I: pH dependence of electrostatic contribution. Proteins, 1997. **27**(4): p. 576-596.

17.   Ellinor, P.T., et al., Ca2+ channel selectivity at a single locus for high-affinity Ca2+ interactions. Neuron, 1995. **15**: p. 1121-1132.

18.   Friel, D.D. and R.W. Tsien, Voltage-gated calcium channels: direct observation of the anomalous mole fraction effect at the single-channel level. Proceedings of the National Academy of Sciences, 1989. **86**: p. 5207-5211.

19.   Rosenberg, R.L., P. Hess, and R.W. Tsien, Cardiac calcium channels in planar lipid bilayers. L-type channels and calcium-permeable channels open at negative membrane potentials. J Gen Physiol, 1988. **92**(1): p. 27-54.

20.   Sather, W.A. and E.W. McCleskey, Permeation and selectivity in calcium channels. Annu Rev Physiol, 2003. **65**: p. 133-159.

21.   Tsien, R.W. and R.Y. Tsien, Calcium channels, stores, and oscillations. Annu Rev Cell Biol, 1990. **6**: p. 715-760.

22.   Tsien, R.W., Calcium channels in excitable cell membranes. Annu Rev Physiol, 1983. **45**: p. 341-358.

23.   Diamond, J.M. and E.M. Wright, Biological membranes: the physical basis of ion and nonelectrolyte selectivity. Annu Rev Physiol, 1969. **31**: p. 581-646.

24.   Doyle, D.A., et al., The Structure of the Potassium Channel: Molecular Basis of K+ Conduction and Selectivity. Science, 1998. **280**: p. 69-77.

25.   Dutzler, R., et al., X-ray structure of a ClC chloride channel at 3.0 A reveals the molecular basis of anion selectivity. Nature, 2002. **415**(6869): p. 287-294.

26.   Dutzler, R., E.B. Campbell, and R. MacKinnon, Gating the selectivity filter in ClC chloride channels. Science, 2003. **300**(5616): p. 108-112.

27.   Dutzler, R., The structural basis of ClC chloride channel function. Trends Neurosci, 2004. **27**(6): p. 315-320.

28.   Dutzler, R., Structural basis for ion conduction and gating in ClC chloride channels. FEBS Lett, 2004. **564**(3): p. 229-233.

29.   Dutzler, R., A structural perspective on ClC channel and transporter function. FEBS Lett, 2007. **581**(15): p. 2839-2844.

30.   Morais-Cabral, J.H., Y. Zhou, and R. MacKinnon, Energetic optimization of ion conduction rate by the K+ selectivity filter. Nature, 2001. **414**(6859): p. 37-42.

31.   Mullins, L.J., An analysis of pore size in excitable membranes. J Gen Physiol, 1960. **43**: p. 105-117.

32.   Mullins, L.J., Ion selectivity of carriers and channels. Biophys J, 1975. **15**(9): p. 921-931.

33.   Park, C.S. and R. MacKinnon, Divalent cation selectivity in a cyclic nucleotide-gated ion channel. Biochemistry, 1995. **34**(41): p. 13328-13333.

34.   Valiyaveetil, F.I., et al., Glycine as a D-amino acid surrogate in the K(+)-selectivity filter. Proc Natl Acad Sci U S A, 2004. **101**(49): p. 17045-17049.

35.   Valiyaveetil, F.I., et al., Structural and functional consequences of an amide-to-ester substitution in the selectivity filter of a potassium channel. J Am Chem Soc, 2006. **128**(35): p. 11591-11599.

36.     Zhou, Y. and R. MacKinnon, The occupancy of ions in the K+ selectivity filter: charge balance and coupling of ion binding to a protein conformational change underlie high conduction rates. J Mol Biol, 2003. **333**(5): p. 965-975.

37.     Zhou, M. and R. MacKinnon, A mutant KcsA K(+) channel with altered conduction properties and selectivity filter ion distribution. J Mol Biol, 2004. **338**(4): p. 839-846.

38.     Eisenman, G. and R. Horn, Ionic selectivity revisited: The role of kinetic and equilibrium processes in ion permeation through channels. J. Membrane Biology, 1983. **76**: p. 197-225.

39.     Hille, B., Ionic Selectivity, saturation, and block in sodium channels. A four barrier model. J Gen Physiol., 1975. **66**: p. 535-560.

40.     Andreev, S., D. Reichman, and G. Hummer, Effect of flexibility on hydrophobic behavior of nanotube water channels. J Chem Phys, 2005. **123**(19): p. 194502.

41.     Corry, B., et al., Mechanisms of permeation and selectivity in calcium channels. Biophys J, 2001. **80**(1): p. 195-214.

42.     Corry, B., et al., A model of calcium channels. Biochim Biophys Acta, 2000. **1509(1-2)**(.): p. 1-6.

43.     Corry, B. and S.H. Chung, Mechanisms of valence selectivity in biological ion channels. Cell Mol Life Sci, 2006. **63**(3): p. 301-315.

44.     Bardhan, J.P., R.S. Eisenberg, and D. Gillespie, Discretization of the induced-charge boundary integral equation. Phys Rev E Stat Nonlin Soft Matter Phys, 2009. **80**(1 Pt 1): p. 011906.

45.     Boda, D., et al., Computing induced charges in inhomogeneous dielectric media: application in a Monte Carlo simulation of complex ionic systems. Phys Rev E Stat Nonlin Soft Matter Phys, 2004. **69**(4 Pt 2): p. 046702.

46.     Boda, D., et al., Monte Carlo simulations of ion selectivity in a biological Na+ channel: charge-space competition. Physical Chemistry Chemical Physics (PCCP), 2002. **4**: p. 5154-5160.

47.     Boda, D., et al., Steric selectivity in Na channels arising from protein polarization and mobile side chains. Biophys J, 2007. **93**(6): p. 1960-1980.

48.     Boda, D., et al., Volume exclusion in calcium selective channels. Biophys J, 2008. **94**(9): p. 3486-3496.

49.     Boda, D., et al., The effect of protein dielectric coefficient on the ionic selectivity of a calcium channel. J Chem Phys, 2006. **125**(3): p. 34901.

50.     Pauling, L., Nature of the Chemical Bond. Third Edition ed. 1960, New York: Cornell University Press. 664.

51.     Gillespie, D., et al., (De)constructing the ryanodine receptor: modeling ion permeation and selectivity of the calcium release channel. J Phys Chem B, 2005. **109**(32): p. 15598-15610.

52.     Gillespie, D., Energetics of divalent selectivity in a calcium channel: the ryanodine receptor case study. Biophys J, 2008. **94**(4): p. 1169-1184.

53. Gillespie, D. and M. Fill, Intracellular calcium release channels mediate their own countercurrent: the ryanodine receptor case study. Biophys J, 2008. **95**(8): p. 3706-3714.

54. Gillespie, D., J. Giri, and M. Fill, Reinterpreting the anomalous mole fraction effect: the ryanodine receptor case study. Biophys J, 2009. **97**(8): p. 2212-2221.

55. Miedema, H., et al., Permeation properties of an engineered bacterial OmpF porin containing the EEEE-locus of Ca2+ channels. Biophys J, 2004. **87**(5): p. 3137-3147.

56. Miedema, H., et al., $Ca^{2+}$ selectivity of a chemically modified OmpF with reduced pore volume. Biophysical Journal, 2006. **91**: p. 4392-4440.

57. Vrouenraets, M., et al., Chemical modification of the bacterial porin OmpF: gain of selectivity by volume reduction. Biophys J, 2006. **90**(4): p. 1202-1211.

58. Wirth, C., et al., NanC crystal structure, a model for outer-membrane channels of the acidic sugar-specific KdgM porin family. J Mol Biol, 2009. **394**(4): p. 718-731.

59. Achouak, W., T. Heulin, and J.M. Pagees, Multiple facets of bacterial porins. FEMS Microbiol Lett, 2001. **199**: p. 1-7.

60. Benz, R., ed. Bacterial and Eukaryotic Porins: Structure, Function, Mechanism. 2004, Wiley-Vch.

61. Buchanan, S.K., Beta Barrel proteins from bacterial outer membranes: structure, function and refolding. Curr Opin Struct Biol, 1999. **9**: p. 455-461.

62. Delcour, A.H., Solute uptake through general porins. Front. Biosci., 2003. **8**: p. 1055-1071.

63. Nikaido, H., Porins and specific channels of bacterial outer membranes. Mol Microbiol, 1992. **6**(4): p. 435-442.

64. Schirmer, T., General and specific porins from bacterial outer membranes. Journal of Structural Biology, 1998. **121**: p. 101-109.

65. Schulz, G.E., Beta-Barrel membrane proteins. Curr Opin Struct Biol, 2000. **10**: p. 443-447.

66. Benz, R., A. Schmid, and G.H. Vos-Scheperkeuter, Mechanism of sugar transport through the sugar-specific LamB channel of Escherichia coli outer membrane. J Membr Biol, 1987. **100**(1): p. 21-29.

67. Comb, D.G. and S. Roseman, I. The structure and enzymatic synthesis of N-acetylneuraminic acid. The Journal of Biological Chemistry, 1960. **235**(9): p. 2529-2537.

68. Severi, E., D.W. Hood, and G.H. Thomas, Sialic acid utilization by bacterial pathogens. Microbiology, 2007. **153**: p. 2817-2822.

69. Vimr, E.R., et al., Diversity of microbial sialic acid metabolism. Microbiol and Mol Bio Rev, 2004. **68**(1): p. 132-153.

70. Condemine, G., et al., Function and expression of an N-acetylneuraminic acid-inducible outer membrane channel in Escherichia coli. J Bacteriol, 2005. **187**(6): p. 1959-1965.

71. Chazalviel, J.-N., Coulomb Screening by Mobile Charges. 1999, New York: Birkhäuser. 355.

72. Roux, B., Perspectives on: Molecular dynamics and computational methods. Journal of General Physiology, 2010. **135**(6): p. 547-548.

73. Roux, B., Exploring the Ion Selectivity Properties of a Large Number of Simplified Binding Site Models. Biophysical Journal, 2010. **98**(12): p. 2877-2885.

74. Almers, W. and E.W. McCleskey, Non-selective conductance in calcium channels of frog muscle: calcium selectivity in a single-file pore. J.Physiol., 1984. **353**: p. 585-608.

75. Byerly, L. and S. Hagiwara, Calcium currents in internally perfused nerve cell bodies of Limnea stagnalis. J Physiol, 1982. **322**: p. 503-528.

76. Corvalan, V., et al., Neuronal modulation of calcium channel activity in cultured rat astrocytes. Proc Natl Acad Sci U S A, 1990. **87**(11): p. 4345-4348.

77. Gerasimov, V.D., P.G. Kostyuk, and V.A. Maiskii, Action potential production in giant neurons of mollusks. Fed Proc Transl Suppl, 1965. **24**(5): p. 763-767.

78. Gordon, S.E., Perspectives on: Local calcium signaling. The Journal of General Physiology, 2010. **136**(2): p. 117.

79. Hagiwara, S. and K.I. Naka, The Initiation of Spike Potential in Barnacle Muscle Fibers under Low Intracellular Ca++. J Gen Physiol, 1964. **48**: p. 141-162.

80. Hagiwara, S., Differentiation of Na and Ca channels during early development. Soc Gen Physiol Ser, 1979. **33**: p. 189-197.

81. Hagiwara, S. and H. Ohmori, Studies of calcium channels in rat clonal pituitary cells with patch electrode voltage clamp. J Physiol, 1982. **331**: p. 231-252.

82. Hagiwara, S. and K. Kawa, Calcium and potassium currents in spermatogenic cells dissociated from rat seminiferous tubules. J Physiol, 1984. **356**: p. 135-149.

83. Kirischuk, S., et al., Calcium signalling in granule neurones studied in cerebellar slices. Cell Calcium, 1996. **19**(1): p. 59-71.

84. Kostyuk, P.G., et al., Ionic currents in the neuroblastoma cell membrane. Neuroscience, 1978. **3**(3): p. 327-332.

85. Kostyuk, P.G., Diversity of calcium ion channels in cellular membranes. Neuroscience, 1989. **28**(2): p. 253-261.

86. Kostyuk, P. and A. Verkhratsky, Calcium stores in neurons and glia. Neuroscience, 1994. **63**(2): p. 381-404.

87. Lee, K.S. and R.W. Tsien, High selectivity of calcium channels as determined by reversal potential measurements in single dialyzed heart cells of the guinea pig. Journal of Physiology (London), 1984. **354**: p. 253-272.

88. McCleskey, E.W., Calcium channels: cellular roles and molecular mechanisms. Curr Opin Neurobiol, 1994. **4**(3): p. 304-312.

89. Rodríguez-Contreras, A. and E.N. Yamoah, Effects of Permeant Ion Concentrations on the Gating of L-Type Ca2+ Channels in Hair Cells. Biophysical Journal, 2003. **84**(5): p. 3457-3469.

90.     Tsien, R.W., et al., Calcium channels: mechanisms of selectivity, permeation, and block. Annual Review of Biophysics and Biophysical Chemistry, 1987. **16**: p. 265-290.

91.     Yamashita, N., S. Ciani, and S. Hagiwara, Effects of internal Na+ on the Ca channel outward current in mouse neoplastic B lymphocytes. J Gen Physiol, 1990. **96**(3): p. 559-579.

92.     Zhang, J.F., et al., Multiple structural elements in voltage-dependent Ca2+ channels support their inhibition by G proteins. Neuron, 1996. **17**(5): p. 991-1003.

93.     Boda, D., et al., Ionic selectivity in L-type calcium channels by electrostatics and hard-core repulsion. J Gen Physiol, 2009. **133**(5): p. 497-509.

94.     Antosiewicz, J., J.A. McCammon, and M.K. Gilson, The determinants of pKas in proteins. Biochemistry, 1996. **35**(24): p. 7819-7833.

95.     Baker, N.A. and J.A. McCammon, Electrostatic interactions. Methods Biochem Anal, 2003. **44**: p. 427-440.

96.     Davis, M.E. and J.A. McCammon, Electrostatics in biomolecular structure and dynamics. Chem. Rev., 1990. **90**: p. 509–521.

97.     Dominy, B.N. and C.L. Brooks, Development of Generalized Born Model Parameterization for Proteins and Nucleic Acids. J Phys Chem B, 1999. **103**(18): p. 3675-3773.

98.     Dzubiella, J., J.M. Swanson, and J.A. McCammon, Coupling hydrophobicity, dispersion, and electrostatics in continuum solvent models. Phys Rev Lett, 2006. **96**(8): p. 087802.

99.     Gilson, M.K. and B. Honig, The dielectric constant of a folded protein. Biopolymers, 1985. **25**: p. 2097-2119.

100.    Gilson, M.K. and B. Honig, Calculation of the total electrostatic energy of a macromolecular system: solvation energies, binding energies, and conformational analysis. Proteins, 1988. **4**(1): p. 7-18.

101.    Gilson, M.K., K.A. Sharp, and B.H. Honig, Calculating the Electrostatic Potential of Molecules in Solution: Method and Error Assessment. Journal of Computational Chemistry, 1988. **9**: p. 327-335.

102.    Gilson, M.K., Theory of electrostatic interactions in macromolecules. Curr Opin Struct Biol, 1995. **5**(2): p. 216-223.

103.    Head-Gordon, T. and C.L. Brooks, The role of electostatics in the binding of small ligands to enzymes. J Phys Chem B, 1987. **91**(12): p. 3342-3349.

104.    Honig, B., K. Sharp, and M. Gilson, Electrostatic interactions in proteins. Prog Clin Biol Res, 1989. **289**: p. 65-74.

105.    Honig, B. and K. Sharp, Macroscopic models of aqueous solutions: biological and chemical applications. Journal of Physical Chemistry, 1993. **97**: p. 1101-1109.

106.    Honig, B. and A. Nichols, Classical electrostatics in biology and chemistry. Science, 1995. **268**: p. 1144-1149.

107.    Im, W., M.S. Lee, and C.L. Brooks III, Generalized Born Model With A Simple Smoothing Function. J Comput Chem, 2003. **12**(9): p. 1894-1901.

108.    Nielsen, J.E. and J.A. McCammon, Calculating pKa values in enzyme active sites. Protein Sci, 2003. **12**(9): p. 1894-1901.

109.    Rahin, A.A. and B. Honig, Reevaluation of the Born model of ion hydration. J.Phys.Chem. B, 1985. **89**(26): p. 5588-5593.

110.    Roux, B., Implicit solvent models, in Computational Biophysics, O. Becker, et al., Editors. 2001, Marcel Dekker Inc: New York. p. p. 133-155.

111.    Russell, S.T. and A. Warshel, Calculations of electrostatic energies in proteins. The energetics of ionized groups in bovine pancreatic trypsin inhibitor. J Mol Biol, 1985. **185**(2): p. 389-404.

112.    Schutz, C.N. and A. Warshel, What are the dielectric "constants" of proteins and how to validate electrostatic models? Proteins, 2001. **44**(4): p. 400-417.

113.    Sharp, K., A. Jean- Charles, and B. Honig, A local dielectric constant model for solvation free energies which accounts for solute polarizability. J Phys Chem B, 1992. **96**(9): p. 3822-3828.

114.    Simonson, T. and C.L. Brooks, Charge Screening and the Dielectric Constant of Proteins: Insights from Molecular Dynamics. Journal of the American Chemical Society, 1996. **118**(35): p. 8452-8458.

115.    Swanson, J.M., J. Mongan, and J.A. McCammon, Limitations of atom-centered dielectric functions in implicit solvent models. J Phys Chem B Condens Matter Mater Surf Interfaces Biophys, 2005. **109**(31): p. 14769-14772.

116.    Warshel, A. and S.T. Russell, Calculations of electrostatic interactions in biological systems and in solutions. Quarterly Review of Biophysics, 1984. **17**: p. 283-422.

117.    Warshel, A., Electrostatic origin of the catalytic power of enzymes and the role of preorganized active sites. J Biol Chem, 1998. **273**(42): p. 27035-27038.

118.    Zhu, J., E. Alexov, and B. Honig, Comparative Study of Generalized Born Models: Born Raddi and Peptide Folding. J Phys Chem B, 2005. **109**(7): p. 3008-3022.

119.    Henderson, D., Attractive Energy and Entropy or Particle Size: the Yin and Yang of Physical and Biological Science. Interdisciplinary Sciences: Computational Life Sciences, 2009. **1**: p. 1-11    available on arXiv http://arxiv.org/ with PaperID 0901.3641.

120.    Eisenberg, B., Y. Hyon, and C. Liu, Energy Variational Analysis EnVarA of Ions in Water and Channels: Field Theory for Primitive Models of Complex Ionic Fluids. Preprint# 2317 of the reprint series of the Institute for Mathematics and its Applications    (IMA,    University    of    Minnesota,    Minneapolis) http://www.ima.umn.edu/preprints/jun2010/jun2010.html, 2010.

121.    Hyon, Y., B. Eisenberg, and C. Liu, A mathematical model for the hard sphere repulsion in ionic solutions. Preprint# 2318 of the reprint series of the Institute for Mathematics and its Applications (IMA, University of Minnesota, Minneapolis) http://www.ima.umn.edu/preprints/jun2010/jun2010.html, 2010.

122. Barreiro, G., C.R. Guimaraes, and R.B. de Alencastro, A molecular dynamics study of an L-type calcium channel model. Protein Eng, 2002. **15**: p. 109-122.

123. Lipkind, G.M. and H.A. Fozzard, Modeling of the outer vestibule and selectivity filter of the L-type Ca2+ channel. Biochemistry, 2001. **40**(23): p. 6786-6794.

124. Malasics, A., et al., Protein structure and ionic selectivity in calcium channels: Selectivity filter size, not shape, matters. Biochim Biophys Acta, 2009.

125. Nonner, W., L. Catacuzzeno, and B. Eisenberg, Binding and Selectivity in L-type Ca Channels: a Mean Spherical Approximation. Biophysical Journal, 2000. **79**: p. 1976-1992.

126. Rutkai, G., D. Boda, and T. Kristof, Relating binding affinity to dynamical selectivity from dynamic Monte Carlo simulations of a model calcium channel. J.Phys.Chem.Lett., 2010. **1**: p. 2179-2184.

127. Eisenberg, B., Living Transistors: a Physicist's View of Ion Channels (version 2). http://arxiv.org/ q-bio.BM: arXiv:q-bio/0506016v2  2005.

128. Markowich, P.A., C.A. Ringhofer, and C. Schmeiser, Semiconductor Equations. New York, Prentice Hall, 1990.

129. Shur, M., Physics of Semiconductor Devices. 1990, New York: Prentice Hall. 680.

130. Koch, S.E., et al., Architecture of Ca2+ Channel Pore-lining Segments Revealed by Covalent Modification of Substituted Cysteines. Journal of Biological Chemistry, 2000.

131. McCleskey, E.W., Ion channel selectivity using an electric stew. Biophys J, 2000. **79**(4): p. 1691-1692.

132. Boda, D., et al., Combined effect of pore radius and protein dielectric coefficient on the selectivity of a calcium channel. Phys Rev Lett, 2007. **98**(16): p. 168102.

133. Metropolis, N., et al., Equation of State Calculations by Fast Computing Machines. The Journal of Chemical Physics, 1953. **21**(6): p. 1087-1092.

134. Boda, D., D. Henderson, and D.D. Busath, Monte Carlo study of the selectivity of calcium channels: improved geometrical mode. Molecular  Physics, 2002. **100**: p. 2361-2368.

135. Valleau, J.P. and L.K. Cohen, Primitive model electrolytes. I. Grand canonical Monte Carlo computations. The Journal of Chemical Physics, 1980. **72**(11): p. 5935-5941.

136. Boda, D., D. Henderson, and D. Busath, Monte Carlo study of the selectivity of calcium channels: improved geometrical mode. Molecular  Physics, 2002. **100**: p. 2361-2368.

137. Malasics, A., D. Gillespie, and D. Boda, Simulating prescribed particle densities in the grand canonical ensemble using iterative algorithms. Journal of Chemical Physics, 2008. **128**: p. 124102.

138. Malasics, A. and D. Boda, An efficient iterative grand canonical Monte Carlo algorithm to determine individual ionic chemical potentials in electrolytes. J. Chem. Phys., 2010. **132**: p. 244103.

139. Almers, W., E.W. McCleskey, and P.T. Palade, Non-selective cation conductance in frog muscle membrane blocked by micromolar external calcium ions. J. Physiol., 1984. **353**: p. 565-583.

140. Almers, W. and E.W. McCleskey, Non-Selective conductance in calcium channels of frog muscle: calcium selectivity in a single-file pore. J.Physiol. , 1984. **353**: p. 585-608.

141. Gillespie, D. and D. Boda, The Anomalous Mole Fraction Effect in Calcium Channels: A Measure of Preferential Selectivity. Biophys. J., 2008. **95**(6): p. 2658-2672.

142. Nonner, W. and B. Eisenberg, Ion Permeation and Glutamate Residues Linked by Poisson-Nernst-Planck Theory in L-type Calcium Channels. Biophys. J., 1998. **75**: p. 1287-1305.

143. Varma, S., D. Sabo, and S.B. Rempe, K+/Na+ selectivity in K channels and valinomycin: over-coordination versus cavity-size constraints. J Mol Biol, 2008. **376**(1): p. 13-22.

144. Chen, D., et al., Rate Constants in Channology. Biophys. J., 1997. **73**(3): p. 1349-1354.

145. Cooper, K., E. Jakobsson, and P. Wolynes, The theory of ion transport through membrane channels. Prog. Biophys. Molec. Biol., 1985. **46**: p. 51–96.

146. Cooper, K.E., P.Y. Gates, and R.S. Eisenberg, Surmounting barriers in ionic channels. Quarterly Review of Biophysics, 1988. **21**: p. 331–364.

147. Cooper, K.E., P.Y. Gates, and R.S. Eisenberg, Diffusion theory and discrete rate constants in ion permeation. J. Membr. Biol., 1988. **109**: p. 95–105.

148. Eisenberg, R.S., M.M. Klosek, and Z. Schuss, Diffusion as a chemical reaction: Stochastic trajectories between fixed concentrations. J. Chem. Phys., 1995. **102**: p. 1767–1780.

149. Eisenberg, R.S., Computing the field in proteins and channels. J. Membrane Biol., 1996. **150**: p. 1–25. Also available on http:\\arxiv.org as Paper arXiv 1009.2857v1001.

150. Eisenberg, R.S., Atomic Biology, Electrostatics and Ionic Channels., in New Developments and Theoretical Studies of Proteins, R. Elber, Editor. 1996, World Scientific: Philadelphia. p. 269-357. Published in the Physics ArXiv as arXiv:0807.0715.

151. Eisenberg, R.S., From Structure to Function in Open Ionic Channels. Journal of Membrane Biology, 1999. **171**: p. 1-24.

152. Eisenberg, B., Permeation as a Diffusion Process, in Biophysics Textbook On Line "Channels, Receptors, and Transporters" http://www.biophysics.org/btol/channel.html#5, L.J. DeFelice, Editor. 2000. p. Published in ArXiv as arXiv:0807.0721.

153. Fleming, G. and P. Hänggi, Activated Barrier Crossing: applications in physics, chemistry and biology. 1993, River Edge, New Jersey: World Scientific.

154. Hänggi, P., P. Talkner, and M. Borokovec, Reaction-rate theory: fifty years after Kramers. Reviews of Modern Physics, 1990. **62**: p. 251-341.

155. Dang, T.X. and E.W. McCleskey, Ion channel selectivity through stepwise changes in binding affinity. J Gen Physiol, 1998. **111**: p. 185-193.

156. Cibulsky, S.M. and W.A. Sather, The EEEE locus is the sole high-affinity Ca2+ binding structure in the pore of a voltage-gated Ca2+ channel: block by Ca2+ entering from the intracellular pore entrance. . Journal of General Physiology, 2000. **116**: p. 349-362.

157. Wu XS, E.H., and Sather WA, Side chain orientation in the selectivity filter of a voltage-gated Ca2+ channel. J.Biol.Chem., 2000. **275**: p. 31778-31785.

158. Christian, P.R. and G. Casella, Monte Carlo Statistical Methods 2004.

159. Gubernatis, J.E., ed. The Monte Carlo Method in the Physical Sciences, Celebrating the 50th Anniversary of the Metropolis Algorithm. AIP Conference Proceedings: 690. 2003, American Institute of Physics: Melville, NY. 416.

160. Gillespie, D., et al., Synthetic Nanopores as a Test Case for Ion Channel Theories: The Anomalous Mole Fraction Effect without Single Filing. Biophys. J., 2008. **95**(2): p. 609-619.

161. Gillespie, D., W. Nonner, and R.S. Eisenberg, Coupling Poisson-Nernst-Planck and Density Functional Theory to Calculate Ion Flux. Journal of Physics (Condensed Matter), 2002. **14**: p. 12129-12145.

162. Gillespie, D., J. Giri, and M. Fill, Reinterpreting the Anomalous Mole Fraction Effect. The ryanodine receptor case study. Biophyiscal Journal, 2009. **97**(8): p. pp. 2212 - 2221

163. Gillespie, D., Analytic Theory for Dilute Colloids in a Charged Slit. The Journal of Physical Chemistry B, 2010. **114**(12): p. 4302-4309.

164. Eisenberg, B., Crowded Charges in Ion Channels. Advances in Chemical Physics (in the press), 2010. also available at http:\\arix.org as Paper arXiv 1009.1786v1

165. Eisenberg, B., Y. Hyon, and C. Liu, Energy Variational Analysis EnVarA of Ions in Water and Channels: Field Theory for Primitive Models of Complex Ionic Fluids. Preprint# 2317 of the reprint series of the Institute for Mathematics and its Applications 2010. (IMA, University of Minnesota, Minneapolis) http://www.ima.umn.edu/preprints/jun2010/jun2010.html.

166. Gillespie, D., M. Valisko, and D. Boda, Density functional theory of the electrical double layer: the RFD functional. Journal of Physics: Condensed Matter 2005. **17**: p. 6609-6626.

167. Gillespie, D., et al., (De)construcing the Ryanodine Receptor: modeling ion permeation and selectivity of the calcium release channel. Journal of Physical Chemistry, 2005. **109**: p. 15598-15610.

168. Hyon, Y., B. Eisenberg, and C. Liu, A mathematical model for the hard sphere repulsion in ionic solutions Preprint# 2318 of the reprint series of the Institute for Mathematics and its Applications 2010. (IMA, University of Minnesota, Minneapolis) http://www.ima.umn.edu/preprints/jun2010/jun2010.html.

169. Boda, D., D. Henderson, and D.D. Busath, Monte Carlo Study of the Effect of Ion and Channel Size on the Selectivity of a Model Calcium Channel. Journal of Physical Chemistry B, 2001. **105**(47): p. 11574-11577.

170. Koshland, D., Application of a Theory of Enzyme Specificity to Protein Synthesis. Proc. Natl. Acad. Sci. , 1958. **44**(2): p. 98–104.

171. Boda, D., et al., Monte Carlo Simulations of the Mechanism of Channel Selectivity: the competition between Volume Exclusion and Charge Neutrality. Journal of Physical Chemistry B, 2000. **104**: p. 8903-8910.

172. Boda, D., et al., Monte Carlo simulations of ion selectivity in a biological Na+ channel: charge-space competition. Physical Chemistry Chemical Physics (PCCP), 2002. **4**: p. 5154-5160.

173. Kokubo, H., et al., Molecular Basis of the Apparent Near Ideality of Urea Solutions. Biophys. J., 2007. **93**(10): p. 3392-3407.

174. Yu, H. and B. Roux, On the utilization of energy minimization to the study of ion selectivity. Biophys J, 2009. **97**(8): p. L15-17.

175. Boulton, A.A., G.B. Baker, and W. Walz, eds. Patch Clamp Applications and Protocols. 1995, Humana Press: Totowa, NJ. 316.

176. Miller, C., ed. Ion Channel Reconstitution. 1986, Plenum Press: New York.

177. Molleman, A., Patch Clamping: An Introductory Guide to Patch Clamp Electrophysiology. 2003: Wiley.

178. Sattelle, D.B., ed. Planar Lipid Bilayers-Methods and Applications. Biological Techniques Series, ed. W. Hanke and W.-R. Schlue. 1993, Academic Press Harcourt Brace and Company.

179. Tien, H.T. and A. Ottova-Lietmannova, eds. Membrane Science and Technology Series. Planar Lipid Bilayers (BLMs) and Their Applications. 2003, Elsevier.

180. Hamill, O.P., et al., Improved patch-clamp techniques for high-resolution current recording from cells and cell-free membrane patches. Pflugers Arch, 1981. **391**(2): p. 85-100.

181. Neher, E. and B. Sakmann, Single channel currents recorded from the membrane of denervated muscle fibers. Nature, 1976. **260**: p. 799-802.

182. Mueller, P., et al., Reconstitution of cell membrane structure in vitro and its transformation into an excitable system. Nature, 1962. **194**: p. 979-980.

183. Colquhoun, D. and F.J. Sigworth, Single-Channel Recording. 1983, New York: Plenum Press, New York.

184. Sakmann, B. and E. Neher, Single Channel Recording. Second ed. 1995, New York: Plenum. 700.

185. Levis, R.A. and J.L. Rae, Constructing a patch clamp setup., in Methods in Enzymology, L. Iverson and B. Rudy, Editors. 1992, Academic Press: NY. p. 14-66.

186. Levis, R.A. and J.L. Rae, Technology of patch clamp recording electrodes, in Patch-clamp Applications and Protocols, W. Walz, A. Boulton, and G. Baker, Editors. 1995, Humana Press.: Totowa, NJ.

187.    Levis, R.A. and J.L. Rae, Low noise patch clamp techniques. . Methods in Enzymology. Vol. 293. 1998.

188.    Rae, J.L. and R.A. Levis, Patch Clamp Recordings from the Epithelium of the Lens Obtained using Glasses Selected for Low Noise and Improved Sealing Properties. Biophysical Journal, 1984. **45**(1): p. 144-146.

189.    Rae, J.L. and R.A. Levis, Glass Technology for Patch Clamp Electrodes, in Methods in Enzymology, L. Iverson and B. Rudy, Editors. 1992, Academic Press: NY. p. 66-92.

190.    Rae, J.L. and R.A. Levis, Single-cell electroporation. Pflugers Archives, 2002. **443**(4): p. 664-670.

191.    Rae, J.L. and R.A. Levis, Fabrication of patch pipets., in Current Protocol Neuroscience. 2004, John Wiley and Sons, Inc.

192.    Milton, R.L. and J.H. Caldwell, Na current in membrane blebs: implications for channel mobility and patch clamp recording. Journal of Neuroscience 1990. **10**: p. 885-893.

193.    Suchyna, T.M., V.S. Markin, and F. Sachs, Biophysics and structure of the patch and the gigaseal. Biophysical Journal, 2009. **97**: p. 738-747.

194.    Tang, J.M., et al., Perfusing pipettes. Pflugers Arch, 1990. **416**(3): p. 347-350.

195.    Tang, J.M., J. Wang, and R.S. Eisenberg, Perfusing patch pipettes. Methods Enzymol, 1992. **207**: p. 176-181.

196.    Montal, M. and P. Mueller, Formation of bimolecular membranes from lipid monolayers and a study of their electrical properties. Proceedings of National Academy of Sciences, U.S.A, 1972. **69**: p. 3561-3566.

197.    WarnerInstruments, Electrophysiology Equipment.

198.    Bard, A.J. and L.R. Faulkner, Electrochemical Methods: Fundamentals and Applications. 2 ed. 2000, New York: John Wiley & Sons.

199.    Axopatch, A Guide to Electrophysiology and Biophysics Laboratory Techniques. Axon Guide. **2500-012 Rev. C**.

200.    Barry, P.H. and J.M. Diamond, Junction potential, electrode standard potentials, and other problems in interpreting electrical properties of membranes. Journal of Membrane Biology, 1970. **3**: p. 931-122.

201.    Barry, P.H., Permeation mechanisms in epithelia: Biionic potentials, dilution potentials, conductances and streming potentials, in Methods in Enzymology, Biomembranes, Part M : Biological Transport. 1989. p. 678-715.

202.    Barry, P.H. and J.W. Lynch, Topical Review: Liquid junction potentials and small cell effects in patch clamp analysis. Journal of Membrane Biology, 1991. **121**: p. 101-117.

203.    Neher, E., Correction for liquid junction potentials in patch-clamp experiments, in Ion Channels, Methdods Enzymology 1991. p. 123-131.

204.    Miedema, H., et al., Ca2+ selectivity of a chemically modified OmpF with reduced pore volume. Biophys J, 2006. **91**(12): p. 4392-4400.

205.  Barry, P.H., JPCalc, a software package for calculating liquid junction potential corrections in patch-clamp, intracellular, epithelial and bilayer measurements and for correcting junction potential measurements. Journal of Neuroscience Methods, 1994. **51**(1): p. 107-116.

206.  Barry, P.H. JPCalc for Windows (JPCalcW) Junction Potential Calculator Users' Manual. 1996-2009; Available from: http://web.med.unsw.edu.au/phbsoft/JPCalcWManual-web-2009.pdf.

207.  Barry, P.H., et al., Further analysis of counterion permeation through anion-selective glycine receptor channels. Channels, 2010. **4**(3): p. 142-149.

208.  Sugiharto, S., et al., Anion-cation permeability correlates with hydrated counter-ion size in glycine receptor channels. Biophysical Journal, 2008. **95**: p. 4698-4715.

209.  Sugiharto, S., et al., External divalent cations increase anion-cation permeability ratio in glycine receptor channels. Pflugers Archive, 2010. **460**: p. 131-152.

210.  Bockris, J. and A.M.E. Reddy, Modern Electrochemistry. 1970, New York: Plenum Press. 1432.

211.  Kunz, W. and R. Neueder, An Attempt at an Overview, in Specific Ion Effects. 2010, World Scientific Singapore. p. 11.

212.  Dickinson, E.J.F., L. Freitag, and R.G. Compton, Dynamic Theory of Liquid Junction Potentials. The Journal of Physical Chemistry B, 2009. **114**(1): p. 187-197.

213.  Eisenberg, B., Y. Hyon, and C. Liu, Energy variational analysis of ions in water and channels: Field theory for primitive models of complex ionic fluids. The Journal of Chemical Physics, 2010. **133**(10): p. 104104-104123.

214.  Cowan, S.W., et al., Crystal structures explain functional properties of two E coli porins. Nature, 1992. **358**: p. 727-733.

215.  Baslé A, et al., Crystal structure of osmoporin OmpC from E. coli at 2.0 A. J Mol Biol., 2006. **362**(5): p. 933-942.

216.  Schirmer, T., et al., Structural basis for sugar translocation through maltoporin channels at 3.1Å resolution. Science, 1995. **267**: p. 512-514.

217.  Koronakis, V., et al., Crystal structure of the bacterial membrane protein TolC central to mutlidrug efflux and protein export. Nature, 2000. **405**: p. 914-919.

218.  Garavito, R.M. and J.P. Rosenbusch, Three-dimensional crystals of an integral membrane protein: an initial X-ray analysis. J Cell Biol, 1980. **86**(327-329).

219.  Delcour, A.H., Function and modulation of bacterial porins: insights from electrophysiology. FEMS Microbiology Letters, 1997. **151**(2): p. 115-123.

220.  Bayley, H., Engineered Nanopores, in NanoBiotechnology, C.M. Niemeyer and C.A. Mirkin, Editors. 2005, Wiley-VCH, Weinheim.

221.  Miedema, H., et al., A Biological Porin Engineered into a Molecular, Nanofluidic Diode. Nano Lett., 2007. **7**(9): p. 2886-2891.

222.  Nikaido, H., Molecular basis of bacterial outer membrane permeability revisited. Microbiology and Molecular Biology Reviews, 2003. **67**(4): p. 593-656.

223.  Wandersman, C., M. Schwartz, and T. Ferenci, Escherichia coli Mutants Impaired in Maltodextrin Transport. Journal of Bacteriology. **140**(1): p. 1-13.

224.  Luckey, M. and H. Nikaido, Specificity of diffusion channels produced by λ phage receptor protein of Escherichia coli. PNAS, 1980. **77**(1): p. 167-171.

225.  Schindler, H. and J.P. Rosenbusch, Matrix protein from Escherichia coli outer membranes form voltage controlled channels in lipid bilayers. Proceedings of the National Academy of Science USA, 1978. **75**: p. 3751-3755.

226.  Cohen, F.S., J. Zimmerberg, and A. Finkelstein, Fusion of Phospholipid Vesicles with Planar Phospholipid Bilayer Membranes.  II.  Incorporation of a Vesicular Membrane Marker into the Planar Membrane. Journal of General Physiology, 1980. **75**: p. 251-270.

227.  Miller, C. and R. Racker, Calcium induced fusion of fragmented scarcoplasmic reticulum with artificial planar bilayers. Journal of Membrane Biology, 1976. **30**: p. 283-300.

228.  Schindler, H., Concepts and techniques for membrane transport reconstitution, in Model systems and reconstitution, R. Antolini, A. Gliozzi, and A. Gorio, Editors. 1982, Raven Press: New York. p. 75-85.

229.  Schindler, H., Planar lipid protein membranes: Strategies of formation and of detecting dependencies of ion transport function on membrane conditions. Methods Enzymology, 1989. **171**: p. 225-253.

230.  Amman, D., Ion-selective Microelectrodes. 1986, Berlin: Springer-Verlag.

231.  Morf, W.E., The Principles of Ion-Selective Electrodes and of Membrane Transport. 1981, Amsterdam, New York: Elsevier.

232.  Hanrahan, J.W. and J.A. Tabcharani, Inhibition of outwardly rectifying anion channel by HEPES and related buffers. Journal of Membrane Biology, 1990. **116**: p. 65-77.

233.  Yamamoto, D. and N. Suzuki, Blockage of chloride channels by HEPES buffer. Proc Roy Soc B, 1987. **230**(1258): p. 93-100.